\documentclass[prd,preprint,tightenlines,floatfix,showpacs,preprintnumbers,nofootinbib,eqsecnum]{revtex4}

 \usepackage[dvips,final]{graphicx}
  \usepackage{amssymb}
   \usepackage{amsmath}
    \usepackage{amsfonts}
     \usepackage{epsfig}
      \usepackage{bm}

\usepackage[section]{placeins}

\usepackage{multirow}
\usepackage{ctable}
\usepackage{booktabs}
\usepackage{array}
\usepackage{tabularx}
\usepackage{xcolor}
\usepackage{pstricks}

\newcommand{\GeV}{\textrm{GeV}}

\newcommand{\dif}{\mathrm{d}}
\newcommand{\diff}[1]{\frac{\mathrm{d}#1}{#1}}

\begin{document}

\vfill
\title{Open charm production at the LHC - \bm{$k_{t}$}-factorization approach}

\author{Rafa{\l} Maciu{\l}a}
\email{rafal.maciula@ifj.edu.pl} \affiliation{Institute of Nuclear
Physics PAN, PL-31-342 Cracow, Poland}

\author{Antoni Szczurek}
\email{antoni.szczurek@ifj.edu.pl} \affiliation{Institute of Nuclear
Physics PAN, PL-31-342 Cracow,
Poland and\\
University of Rzesz\'ow, PL-35-959 Rzesz\'ow, Poland}

\date{\today}

\begin{abstract}
We discuss inclusive production of open charm in proton-proton scattering at LHC.
The calculation is performed within the $k_t$-factorization approach.
Different models of unintegrated gluon distributions (UGDF) from
the literature are used. The theoretical transverse momentum as well as (pseudo)rapidity
distributions of charmed mesons are compared with recent experimental
data of ATLAS, ALICE and LHCb collaborations.
Only the calculation with Kimber-Martin-Ryskin (KMR) UGDF gives results
comparable to experimental ones. All other popular models of UGDF significantly
underpredict experimental data. Several sources of uncertainties of the theoretical predictions are also studied in details. In addition we discuss correlations
between $D$ and $\bar D$ mesons. Good description of experimental 
distribution in invariant mass and in relative azimuthal angle between 
$D$ and $\bar D$ mesons is achieved for the KMR UGDF. 
The considered correlation observables measured by the LHCb experiment were not discussed in other approaches in the literature.
\end{abstract}


\maketitle

\section{Introduction}

At high energy hadronic scattering the gluon-gluon fusion is known to be the dominant
mechanism of open charm production. Even at RHIC the contribution from quark-antiquark anihilation
constitutes only a small fraction of the cross section.
Usually in the studies of heavy quark production the main efforts concentrate on inclusive distributions.
The transverse momentum distribution of charmed mesons is the best
example. Standard collinear NLO approach \cite{NLOPM} as well as its improved schemes e.g. FONLL \cite{FONLL} or GM-VFNS \cite{GM-VFNS} are states of art in this
respect. These approaches cannot be, however, used  when transverse
momenta of charm quark and antiquark are not equal. This means in practice 
that it cannot be used for studies of correlation observables 
for charmed meson pairs or for meson-nonphotonic electron modes.

The $k_t$-factorization approach seems much more efficient tool in this
respect \cite{Teryaev,Baranov00,BLZ,Shuvaev,JKLZ2011,JKLZ2012,Saleev2009,Saleev2012}. Different unintegrated gluon distributions in the proton (UGDF)
have been used in the literature in this context \cite{KMR,KMS,Kutak-Stasto,Jung,GBW}.
Recently we have applied this formalism to the description of
inclusive distributions of so-called nonphotonic electrons \cite{LMS09} 
and electron-positron correlations \cite{MSS2011} at RHIC. Rather good
description of correlation observables has been achieved there.

The quark mass is sufficiently large to apply perturbative calculation,
but still small enough that interesting low-x effects may appear too. In the 
$k_t$-factorization approach the latter effects are contained in
the unintegrated gluon distributions -- the buildning blocks of the
formalism. In principle a comparison of experimental data and predictions with the UGDFs which include such 
effects may tell us more about footprints of the saturation effects --
the topic extensively discussed in recent years.

Recently ATLAS \cite{ATLASincD}, ALICE \cite{ALICEincD,ALICEincDs} 
and LHCb \cite{LHCbincD} collaborations have measured inclusive 
distributions (mainly transverse momentum distributions) of different charmed mesons.
The LHCb collaboration has measured in addition  
a few correlation observables for charmed mesons for the first time 
in the history in the forward rapidity region \cite{LHCb-DPS-2012}.
Before the STAR collaboration at RHIC has measured correlation of charmed mesons
and nonphotonic electrons \cite{Mischke}. At RHIC a study of meson-meson
correlations was not possible due to limited statistics caused by
relatively small cross sections. It was accessible only at Tevatron where first midrapidity measurements
of azimuthal angle correlations between charmed mesons have been performed by the CDF experiment \cite{Tevatron-DD}.

In the present paper we wish to concentrate first on inclusive distributions of
charmed mesons in order to test different models of unintegrated
gluon distributions from the literature. Next we wish to focus on $D \bar D$ meson correlations. Conclusions will close our paper.

\section{Sketch of the formalism}

The cross section for the production of a pair of charm quark -- 
charm antiquark can be written as:
\begin{eqnarray}
\frac{d \sigma(p p \to c \bar c X)}{d y_1 d y_2 d^2 p_{1t} d^2 p_{2t}} 
&& = \frac{1}{16 \pi^2 {\hat s}^2} \int \frac{d^2 k_{1t}}{\pi} \frac{d^2 k_{2t}}{\pi} \overline{|{\cal M}^{off}_{g^{*}g^{*} \to c\; \bar c}|^2} \nonumber \\
&& \times \;\; \delta^2 \left( \vec{k}_{1t} + \vec{k}_{2t} - \vec{p}_{1t} - \vec{p}_{2t}
\right)
{\cal F}_g(x_1,k_{1t}^2,\mu^2) {\cal F}_g(x_2,k_{2t}^2,\mu^2).
\end{eqnarray}
The main ingredients in the formula are off-shell matrix elements for $g^{*}g^{*} \rightarrow c \;\bar{c}$ subprocess
and unintegrated gluon distributions (UGDF). The relevent matrix
elements are known and can be found in Refs.~\cite{CCH91,CE91,BE01}. 
The unintegrated gluon distributions are functions of
longitudinal momentum fraction $x_1$ or $x_2$ of gluon with respect to its parent nucleon and of gluon transverse momenta $k_{t}$.
Some of them depend in addition on the
factorization scale $\mu$.
The longitudinal momentum fractions can be calculated as:
\begin{eqnarray}
x_1 = \frac{m_{1t}}{\sqrt{s}}\exp( y_1) 
     + \frac{m_{2t}}{\sqrt{s}}\exp( y_2),\nonumber \\
x_2 = \frac{m_{1t}}{\sqrt{s}}\exp(-y_1)
     + \frac{m_{2t}}{\sqrt{s}}\exp(-y_2),
\end{eqnarray}
where $m_{it} = \sqrt{p_{it}^2 + m_Q^2}$ is the transverse mass of produced quark/antiquark.

Various unintegrated gluon distributions have been discussed in the literature
\cite{KMR,KMS,Kutak-Stasto,Jung,GBW}.
In contrast to the collinear gluon distributions (PDFs) they differ
considerably among themselves. 
One may expect that they will lead to different
production rates of $c \bar c$ pairs at the LHC. Since the production of charm quarks
is known to be dominated by the gluon-gluon fusion, the charm production at the LHC
can be used to verify the quite different models of UGDFs.

Below we wish to concentrate for a while on the Kimber-Martin-Ryskin (KMR)
unintegrated gluon distribution which, as will be discussed in this paper, gives the best
description of the LHC experimental data, taking into account also correlation observables.

According to the KMR approach the unintegrated gluon distribution is given
by the following formula
\begin{eqnarray} \label{eq:UPDF}
  f_g(x,k_t^2,\mu^2) &\equiv& \frac{\partial}{\partial \log k_t^2}\left[\,g(x,k_t^2)\,T_g(k_t^2,\mu^2)\,\right]\nonumber \\ &=& T_g(k_t^2,\mu^2)\,\frac{\alpha_S(k_t^2)}{2\pi}\,\sum_{b }\,\int_x^1\! d z\,P_{gb}(z)\,b \left (\frac{x}{z}, k_t^2 \right).
\end{eqnarray}
This definition is fully satisfied for $k_t > \mu_0$, where $\mu_0\sim 1$ GeV
is the minimum scale for which DGLAP evolution of the conventional
collinear gluon distributions, $g(x,\mu^2)$, is valid.

The virtual (loop) contributions may be resummed to all orders by the Sudakov form factor,
\begin{equation} \label{eq:Sudakov}
  T_g (k_t^2,\mu^2) \equiv \exp \left (-\int_{k_t^2}^{\mu^2}\!\diff{\kappa_t^2}\,\frac{\alpha_S(\kappa_t^2)}{2\pi}\,\sum_{b}\,\int_0^1\!\dif{z}\;z \,P_{b g}(z) \right ),
\end{equation}
which gives the probability of evolving from a scale $k_t$ to a scale $\mu$ without parton emission.

The exponent of the gluon Sudakov form factor can be simplified using the following identity: $P_{qg}(1-z)=P_{qg}(z)$.  Then the gluon Sudakov form factor is
\begin{equation}
  T_g(k_t^2,\mu^2) = \exp\left(-\int_{k_t^2}^{\mu^2}\!\diff{\kappa_t^2}\,\frac{\alpha_S(\kappa_t^2)}{2\pi}\,\left( \int_{0}^{1-\Delta}\!\dif{z}\;z \,P_{gg}(z) + n_F\,\int_0^1\!\dif{z}\,P_{qg}(z)\right)\right),
\end{equation}
where $n_F$ is the quark--antiquark active number of flavours into which the gluon may split and $\Delta = k_{t}/(k_{t} + \mu)$ which introduces a restricton of the phase space for gluon emmision due to the angular-ordering condition.
Due to the presence of the Sudakov form factor in the KMR prescription only last emission generates transverse momentum of incoming gluons. This scheme is the direct analogy to the techniques usually applied in all standard parton shower Monte Carlo generators. The unique feature of the KMR model of UGDF is that it provides possibility for the emission of at most one additional gluon. Therefore one can expect that the KMR model may include in an effective way NLO corrections to heavy quark production cross section.

In the literature often somewhat differently defined UGDFs are used. 
They differ by the following transformation:
\begin{eqnarray}
{\cal F}_g(x,k_{t}^2,\mu^2) \equiv \frac{1}{k_t^2}\,f_g(x,k_t^2,\mu^2)
\; .
\end{eqnarray}
The normalisation condition for unintegrated distributions
\begin{equation} \label{eq:norm}
  g(x,\mu^2) = \int_0^{\mu^2}\! d k_t^2\,f_g(x,k_t^2,\mu^2)
\end{equation}
is exactly satisfied if we define
\begin{equation} \label{eq:smallkt}
  \left.\frac{1}{k_t^2}\,f_g(x,k_t^2,\mu^2)\right\rvert_{k_t<\mu_0} = \frac{1}{\mu_0^2}\,g(x,\mu_0^2)\,T_g(\mu_0^2,\mu^2),
\end{equation}
so that the density of gluons in the proton is constant for $k_t<\mu_0$
at fixed $x$ and $\mu$.

The precise expression for the unintegrated gluon distribution reads
\begin{eqnarray} \label{eq:a12}
f_g(x,k_t^2,\mu^2) 
&& = T_g(k_t^2,\mu^2)\,\frac{\alpha_S(k_t^2)}{2\pi}\, \times \nonumber \\  
&& \int_x^1\! d z \left[\sum_q P_{gq}(z)\frac{x}{z}q\left(\frac{x}{z},k_t^2\right) + P_{gg}(z)\frac{x}{z}g\left(\frac{x}{z},k_t^2\right)\Theta\left(\frac{\mu}{\mu+k_t}-z\right)\right]. \;\;\;\;\;
\end{eqnarray}
%
\section{Charm quark/anti-quark production at LHC}

In this section we wish to concentrate on the production of charm
quarks and antiquarks. Thus this section has rather theoretical
character. The cross sections for production of charmed mesons will 
be discussed in the next section. Before we go to the presentation of differential distributions let us
summarize integrated cross sections for $c \bar c$ production.

Using the KMR model of unintegrated gluon distributions, the total cross section
for charm quark/antiquark production at $\sqrt{s} = 7$ TeV
is obtained to be 
$\sigma^{KMR}_{tot} (pp\rightarrow c\bar{c}X) =
7.36^{+2.34}_{-1.77}(\mu)^{+6.03}_{-2.94}(m_{c})$ mb. The predicted value has large uncertainties related to the choice of factorization/renormalization scales $\mu$ and due to the charm quark mass $m_{c}$. 
The obtained cross section is very large, of the same order as e.g. cross 
section for elastic scattering
or single diffraction. This means that in practice charm
quark/antiquarks appear in almost each inelastic event. This is
a rather new situation which requires more detailed studies.

Taking into account acceptance of 
ATLAS, LHCb and ALICE detectors we get 
$\sigma^{KMR}_{ATLAS}(pp\rightarrow c\bar{c}X) =
2.53^{+0.83}_{-0.60}(\mu)^{+1.66}_{-0.90}(m_{c})$ mb, 
$\sigma^{KMR}_{LHCb}(pp\rightarrow c\bar{c}X) = 
1.54^{+0.50}_{-0.37}(\mu)^{+1.27}_{-0.62}(m_{c})$ mb and 
$\sigma^{KMR}_{ALICE}(pp\rightarrow c\bar{c}X) =
0.91^{+0.30}_{-0.23}(\mu)^{+0.68}_{-0.35}(m_{c})$ mb, 
respectively. These numbers together with theoretical uncertainties are consistent 
with recent LHC measurements as well as with the recent FONLL
\cite{Cacciari} and GM-VFNS \cite{Kniehl2012} predictions of charm cross section.

\begin{figure}[!h]
\begin{minipage}{0.32\textwidth}
 \centerline{\includegraphics[width=1.0\textwidth]{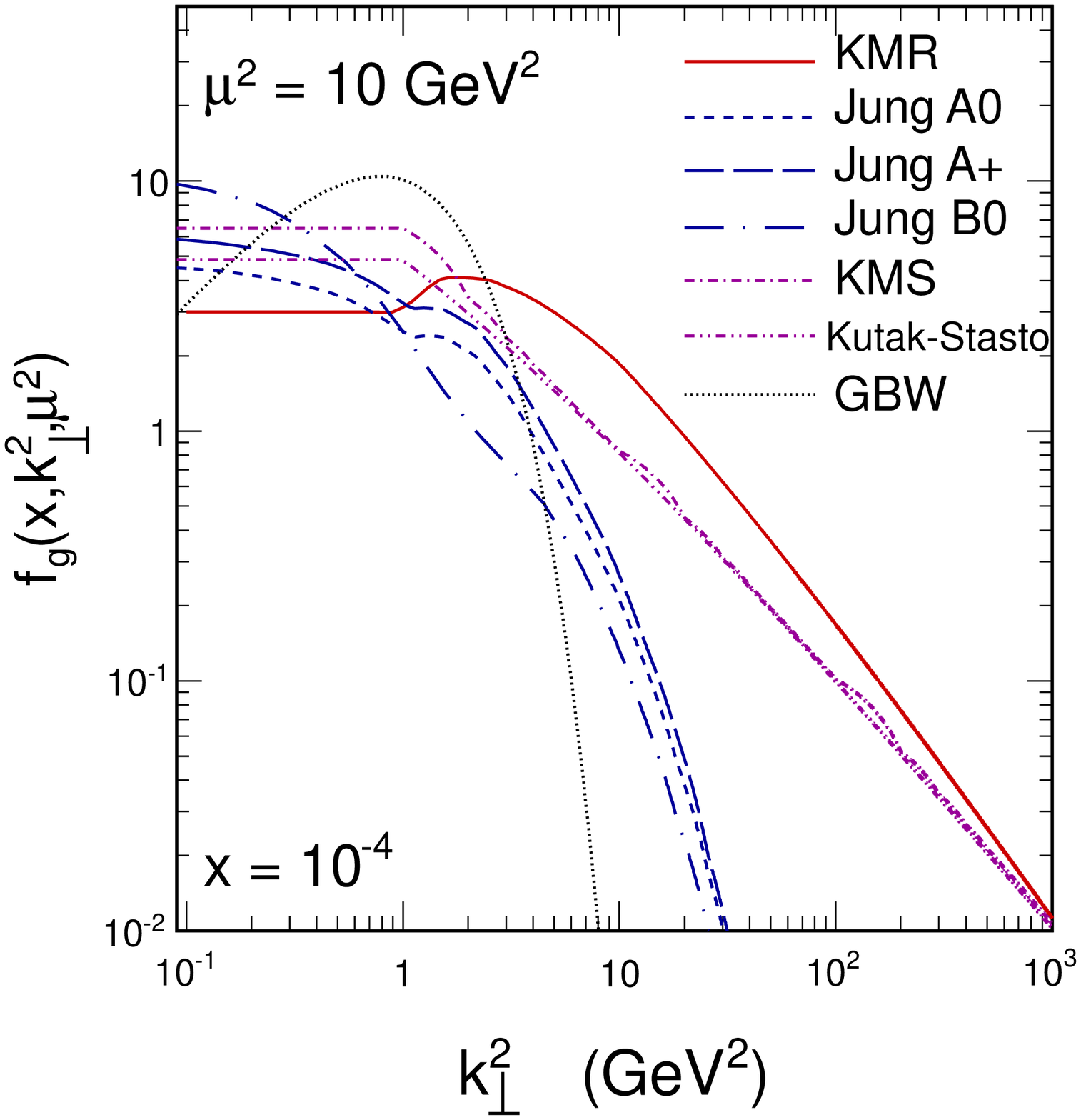}}
\end{minipage}
\hspace{0.05cm}
\begin{minipage}{0.32\textwidth}
 \centerline{\includegraphics[width=1.0\textwidth]{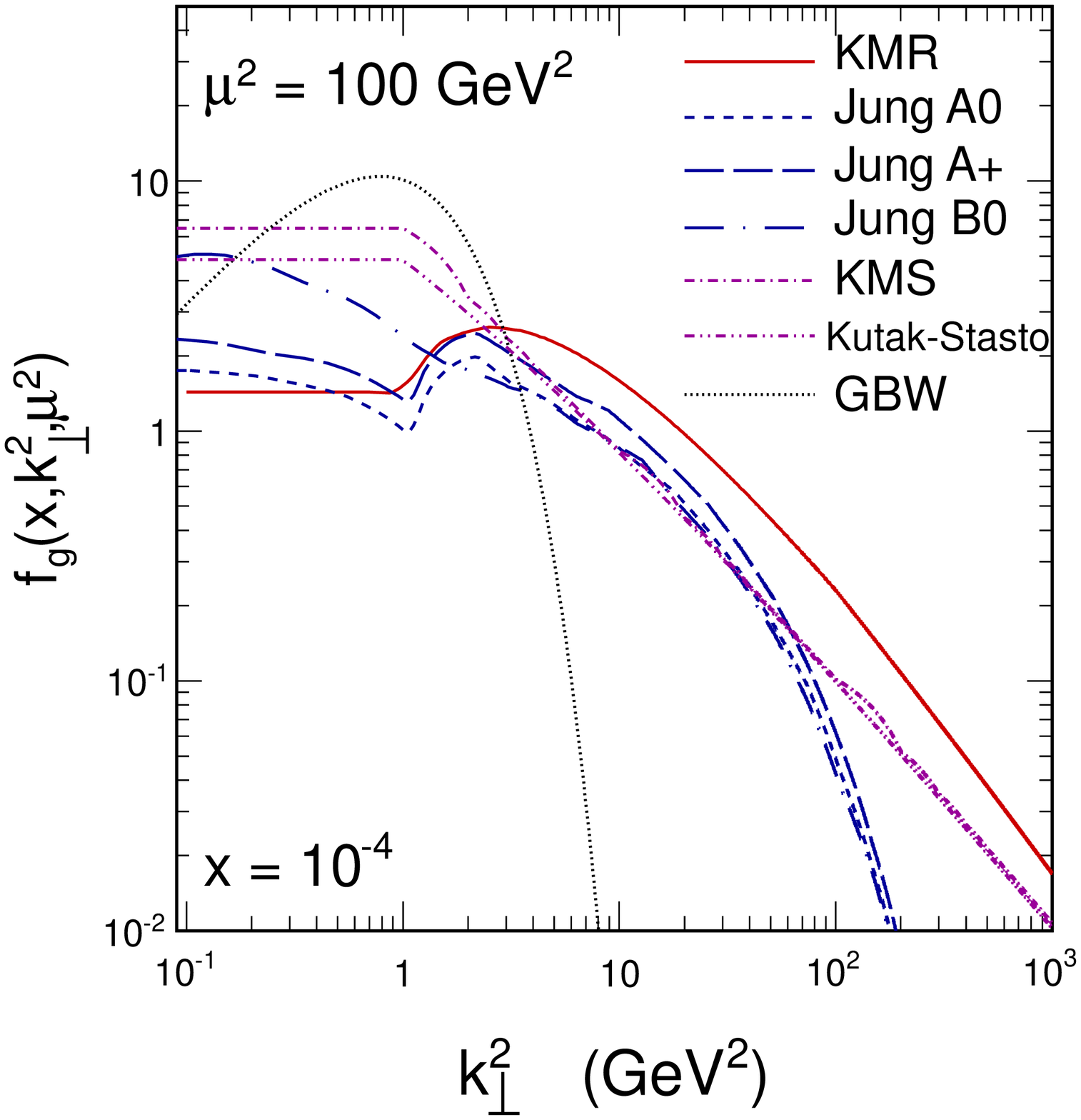}}
\end{minipage}
\hspace{0.05cm}
\begin{minipage}{0.32\textwidth}
 \centerline{\includegraphics[width=1.0\textwidth]{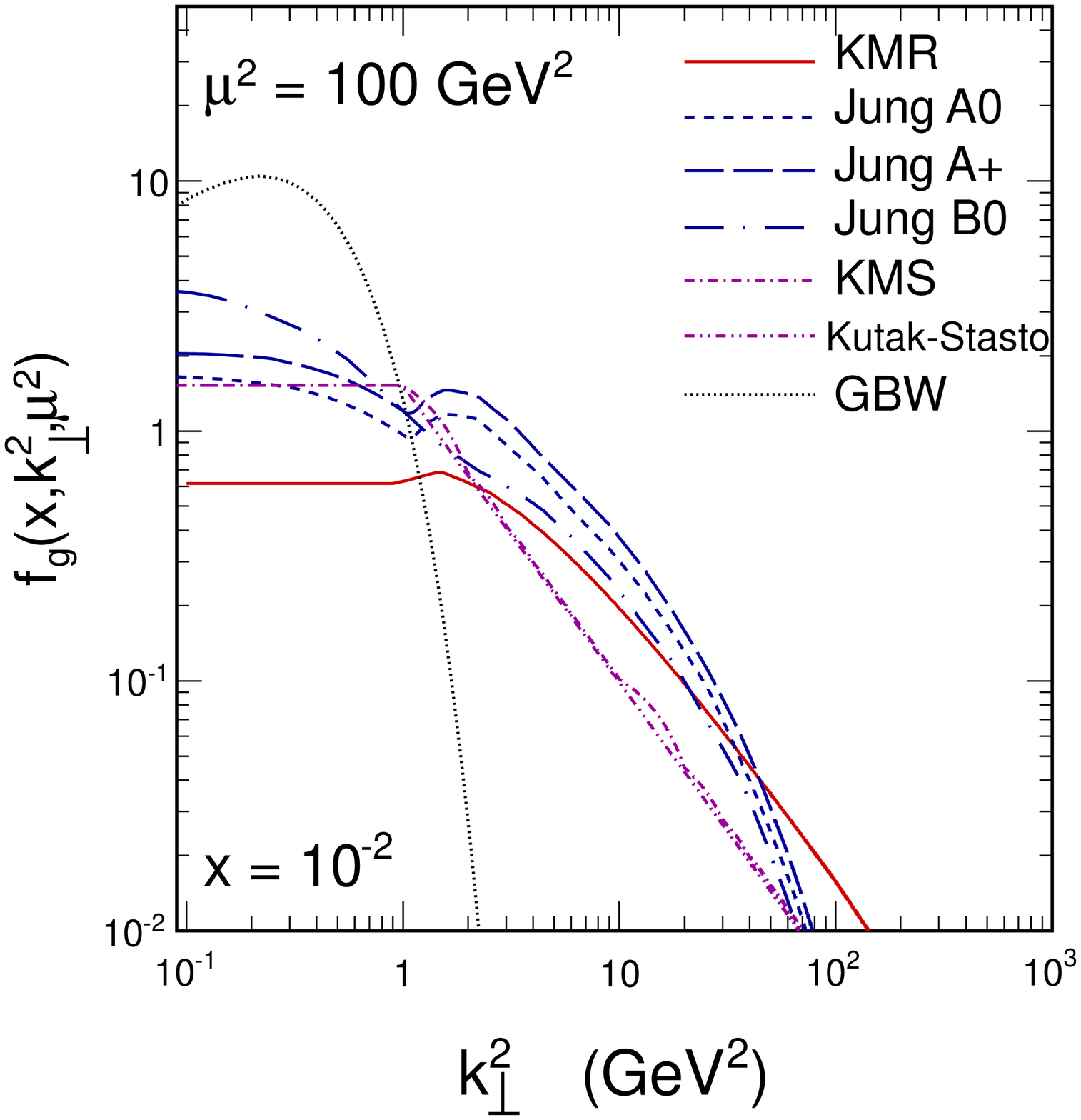}}
\end{minipage}
   \caption{
\small Different unintegrated gluon distributions from the literature as
a function of gluon transverse momentum squared $k_{t}^{2}$ for different values of longitudinal
momentum fraction $x$ of the gluon initiating the hard process and for different factorization scale $\mu$. }
 \label{fig:qt2-ugdfs}
\end{figure}

As it was mentioned in the previous section our predictions are very sensitive to the choice of unintegrated gluon distributions. 
Different UGDFs are very often based on quite different theoreticl assumptions. This has a crucial meaning for their kinematical characteristics.
In Fig.~\ref{fig:qt2-ugdfs} we show dependence of the unintegrated
gluon distributions functions on gluon transverse momentum squared $k_{t}^{2}$
for several values of $x$ relevant 
for the production of charm quarks and antiquarks at LHC energy. Differences in shapes in $k_{t}^{2}$ of the plotted functions are significant.
One can also see different dependence on $x$ of the different considered UGDFs. Changing the value of $x$, the mutual trends between them also change what makes the overall picture more complicated.
Especially the KMR model seems to reveal the strongest $x$-dependence.

\begin{figure}[!h]
\begin{minipage}{0.35\textwidth}
 \centerline{\includegraphics[width=1.0\textwidth]{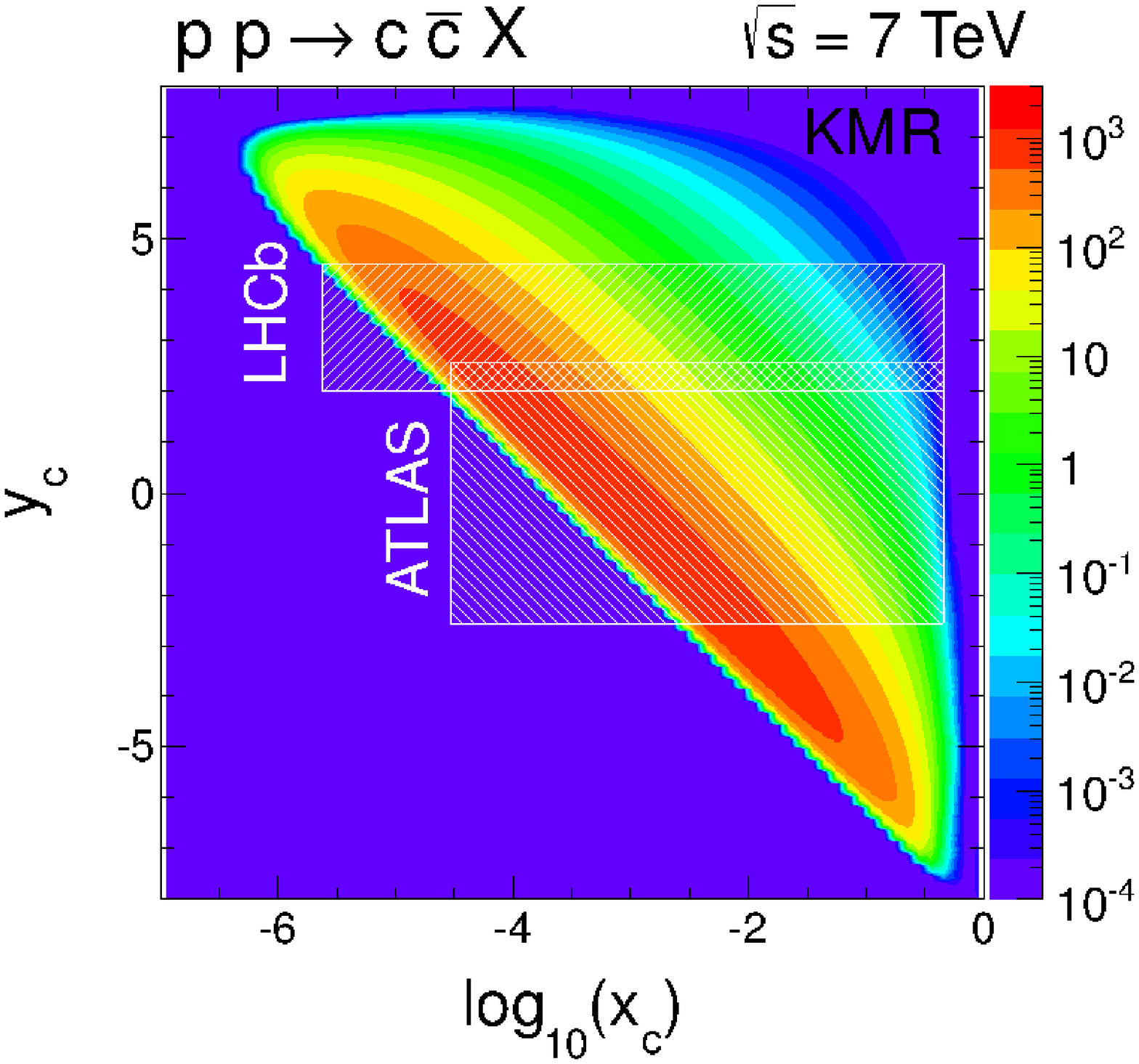}}
\end{minipage}
\hspace{0.5cm}
\begin{minipage}{0.35\textwidth}
 \centerline{\includegraphics[width=1.0\textwidth]{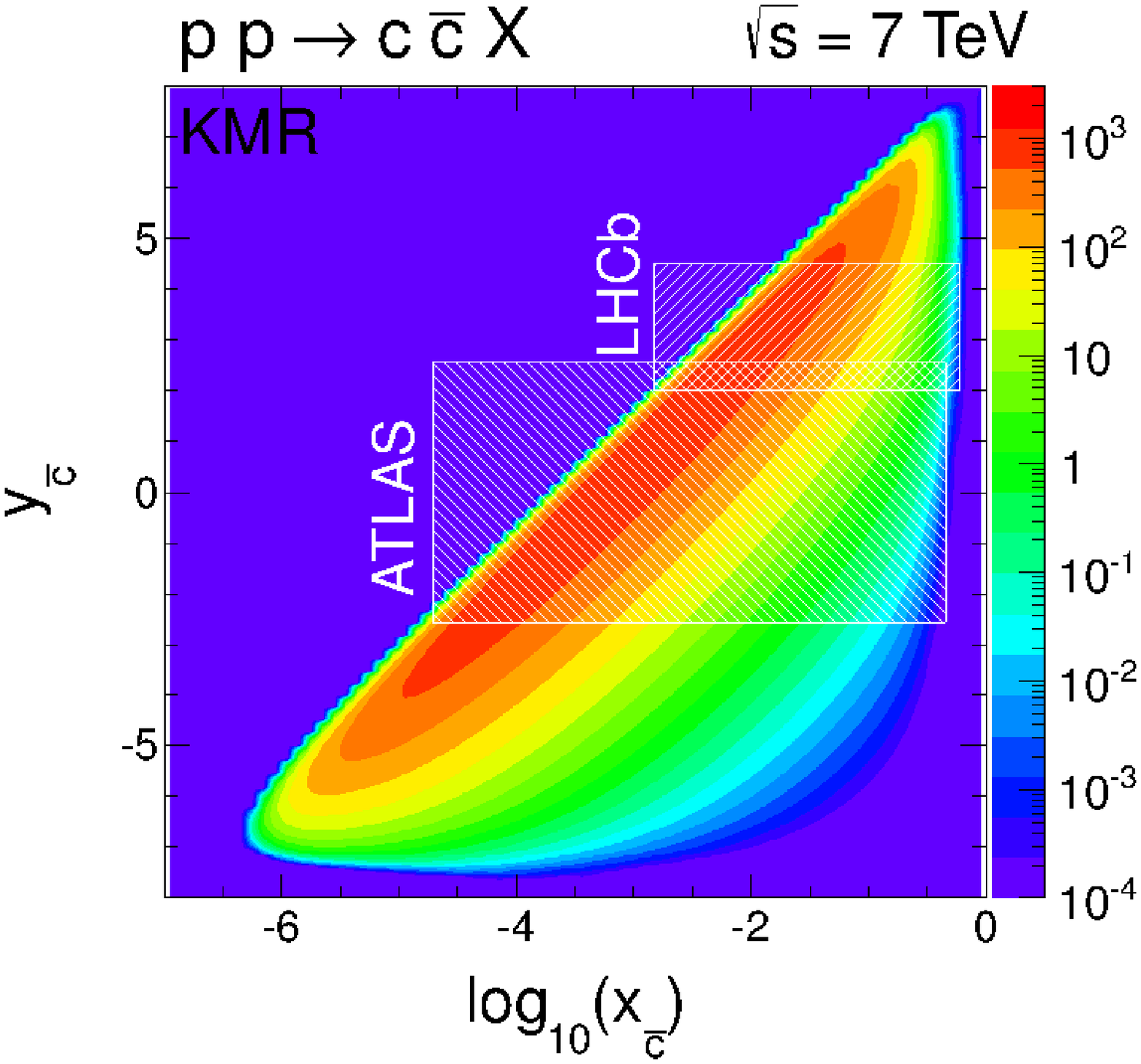}}
\end{minipage}
   \caption{
\small The range of longitudinal momentum fraction of gluons and its
correlation to the rapidity of charm quark (left) or antiquark (right).
In addition regions of the coverage for the ATLAS and LHCb experiments are shown.}
 \label{fig:ylogx-kmr}
\end{figure}

The rapidity of the quark or antiquark are strongly correlated with
longitudinal momentum fractions of gluons initiating the hard process.
This is shown in Fig.~\ref{fig:ylogx-kmr} for the KMR UGDF.
At rapidities $|y| >$ 5 one starts to probe longitudinal momentum
fractions smaller than 10$^{-4}$. This is a new situation compared to earlier 
measurements at RHIC or Tevatron. The unintegrated gluon distributions (UGDFs) as well as standard collinear ones (PDFs) were not
tested so far in this region.

\begin{figure}[!h]
\begin{minipage}{0.32\textwidth}
 \centerline{\includegraphics[width=1.0\textwidth]{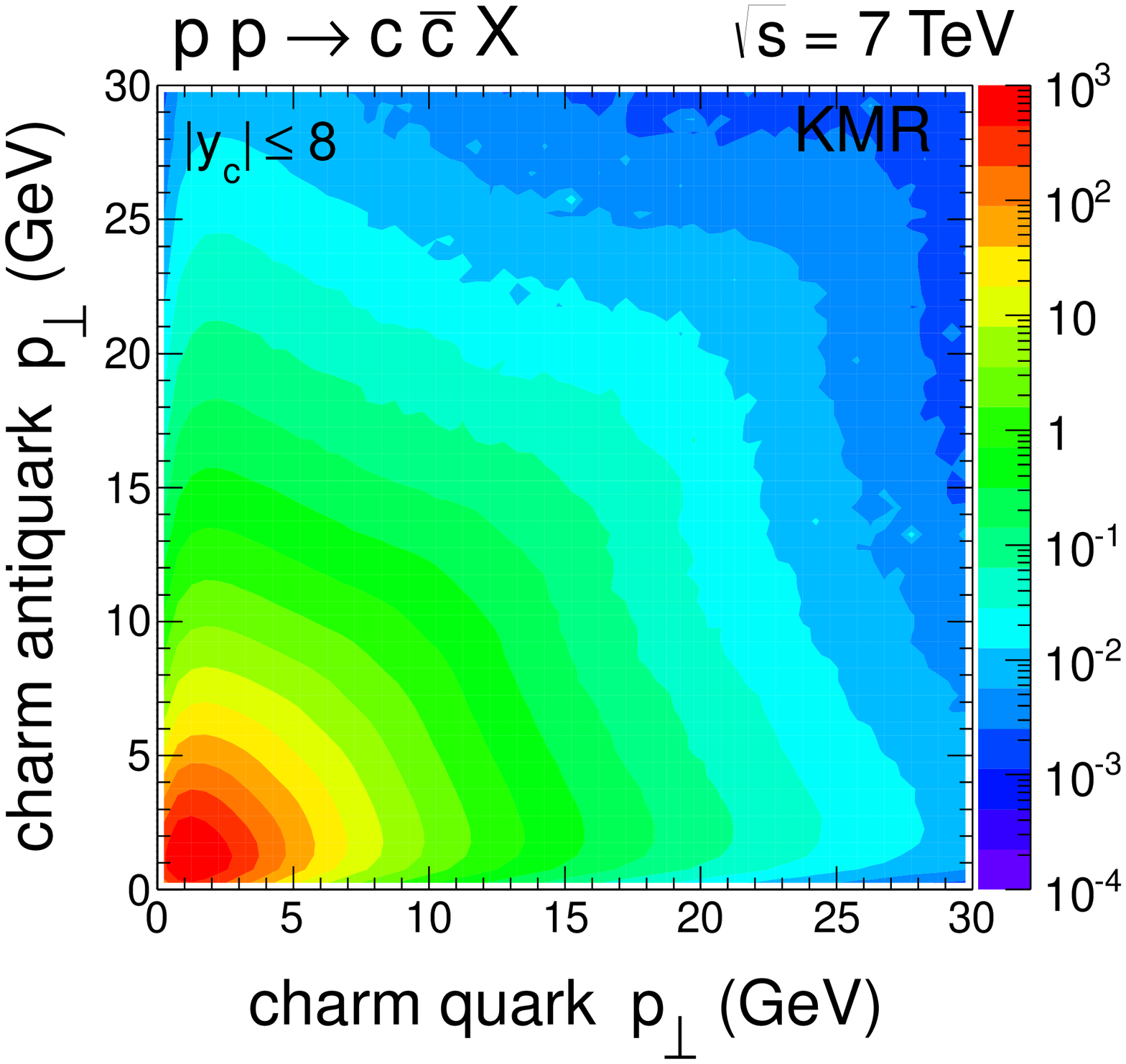}}
\end{minipage}
\hspace{0.05cm}
\begin{minipage}{0.32\textwidth}
 \centerline{\includegraphics[width=1.0\textwidth]{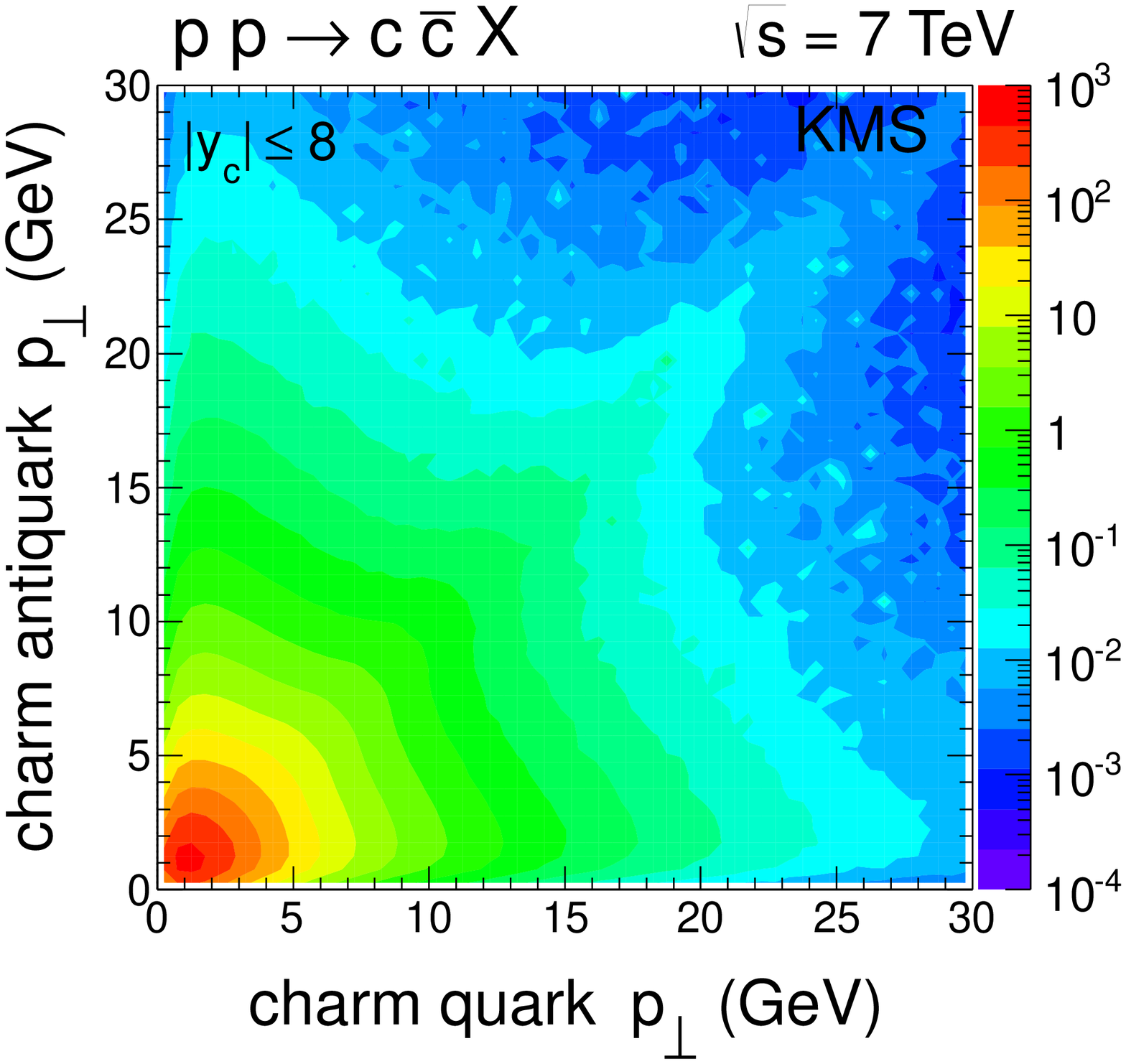}}
\end{minipage}
\hspace{0.05cm}
\begin{minipage}{0.32\textwidth}
 \centerline{\includegraphics[width=1.0\textwidth]{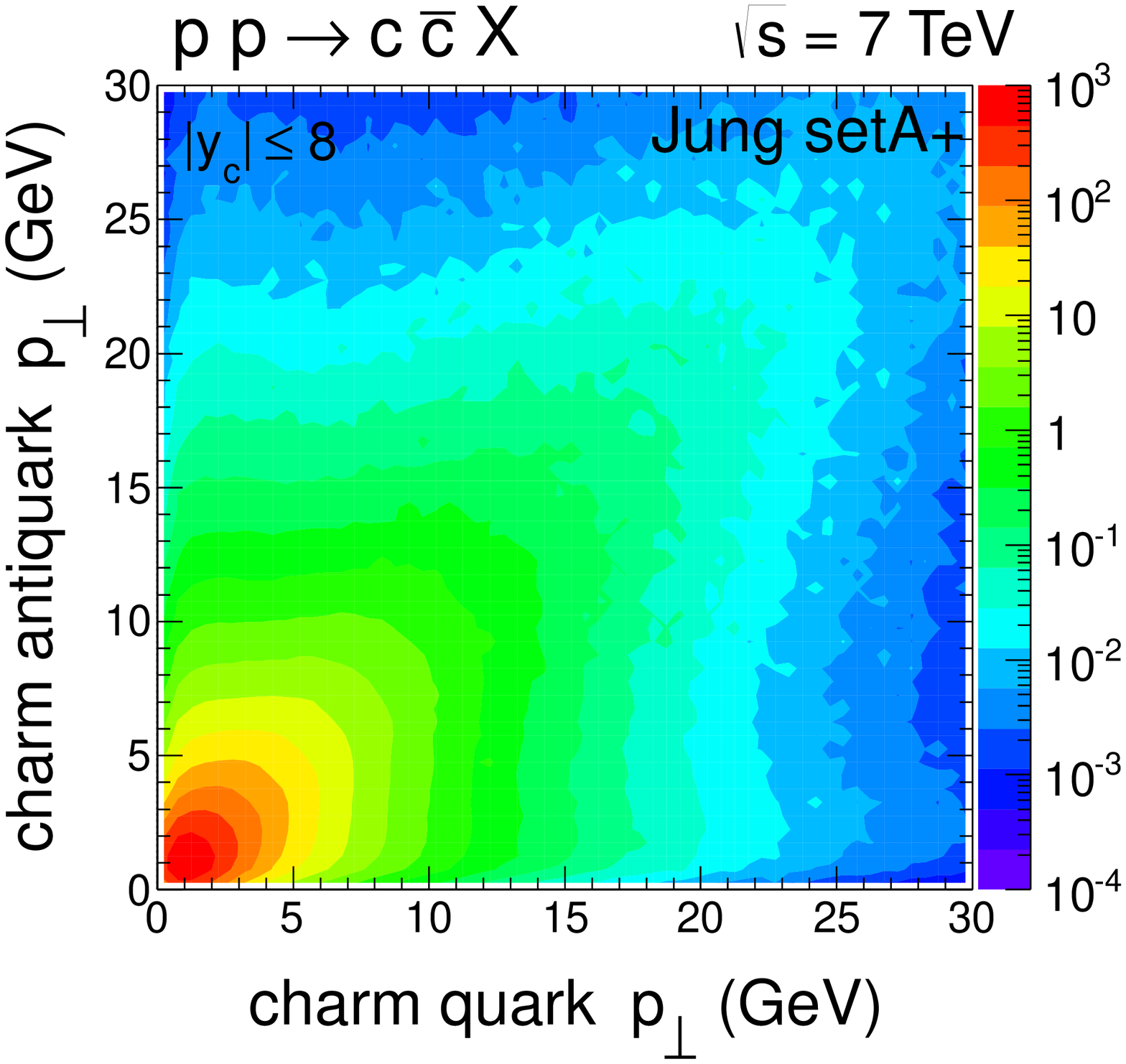}}
\end{minipage}

\begin{minipage}{0.32\textwidth}
 \centerline{\includegraphics[width=1.0\textwidth]{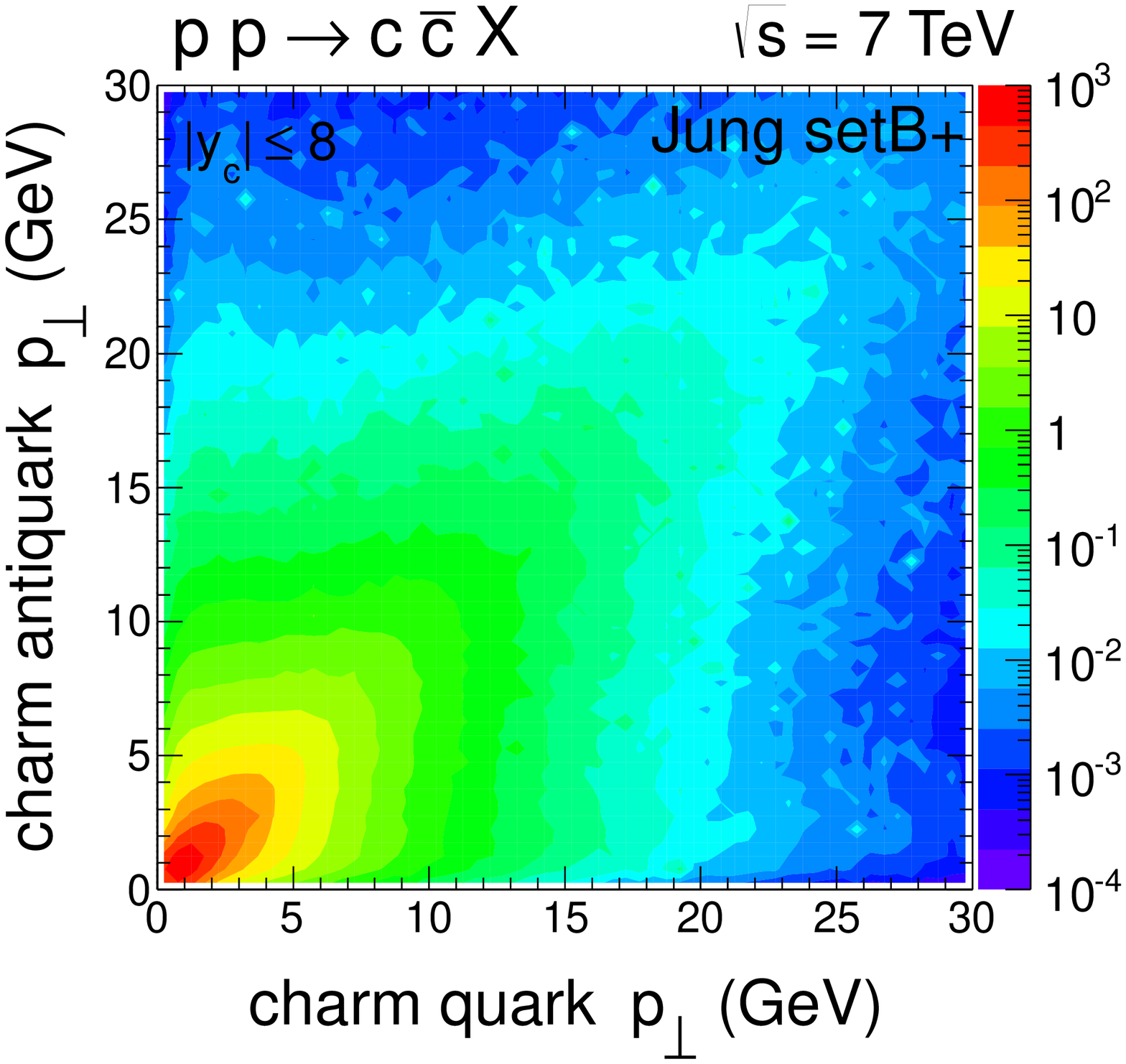}}
\end{minipage}
\hspace{0.05cm}
\begin{minipage}{0.32\textwidth}
 \centerline{\includegraphics[width=1.0\textwidth]{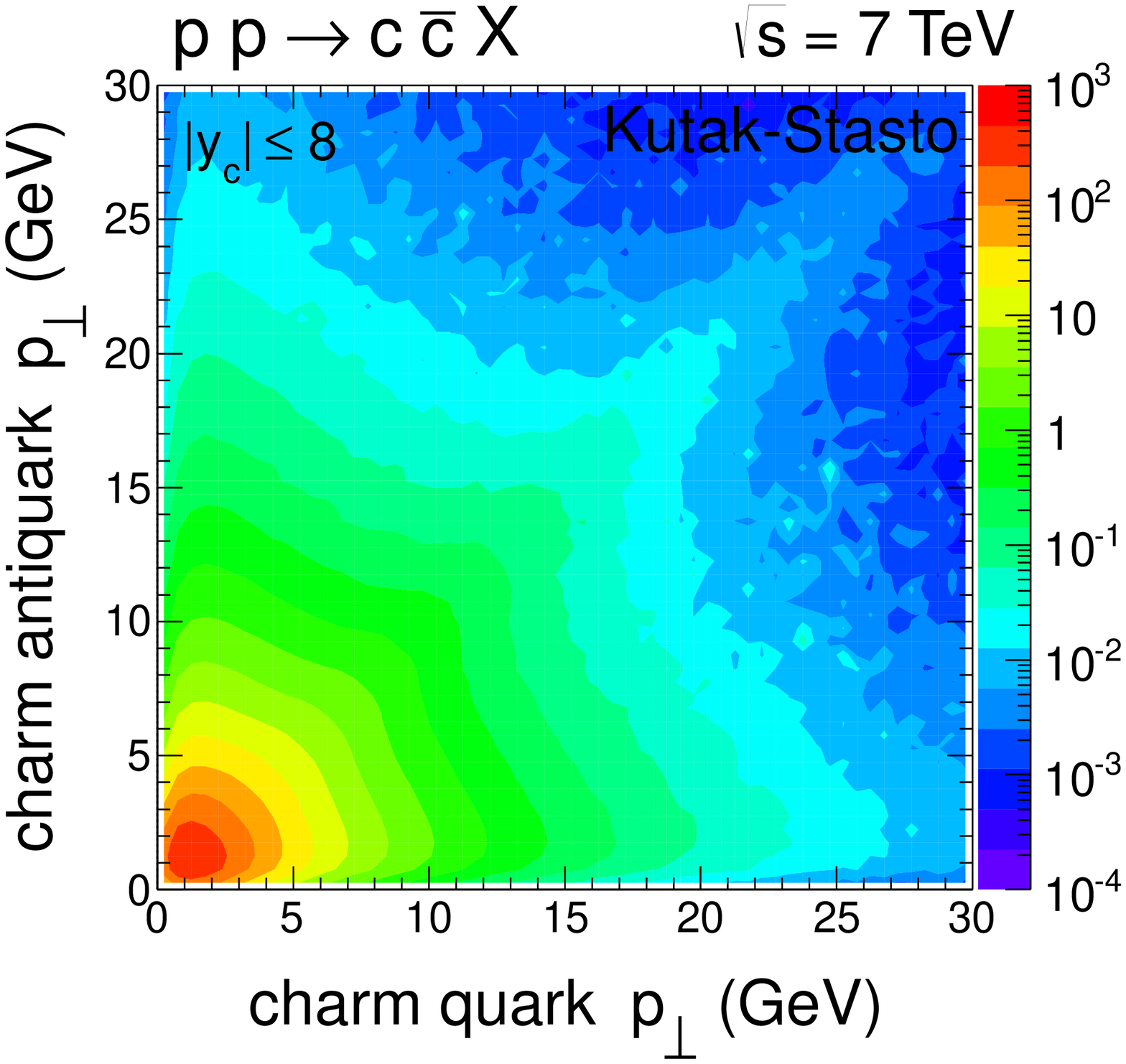}}
\end{minipage}
\hspace{0.05cm}
\begin{minipage}{0.32\textwidth}
 \centerline{\includegraphics[width=1.0\textwidth]{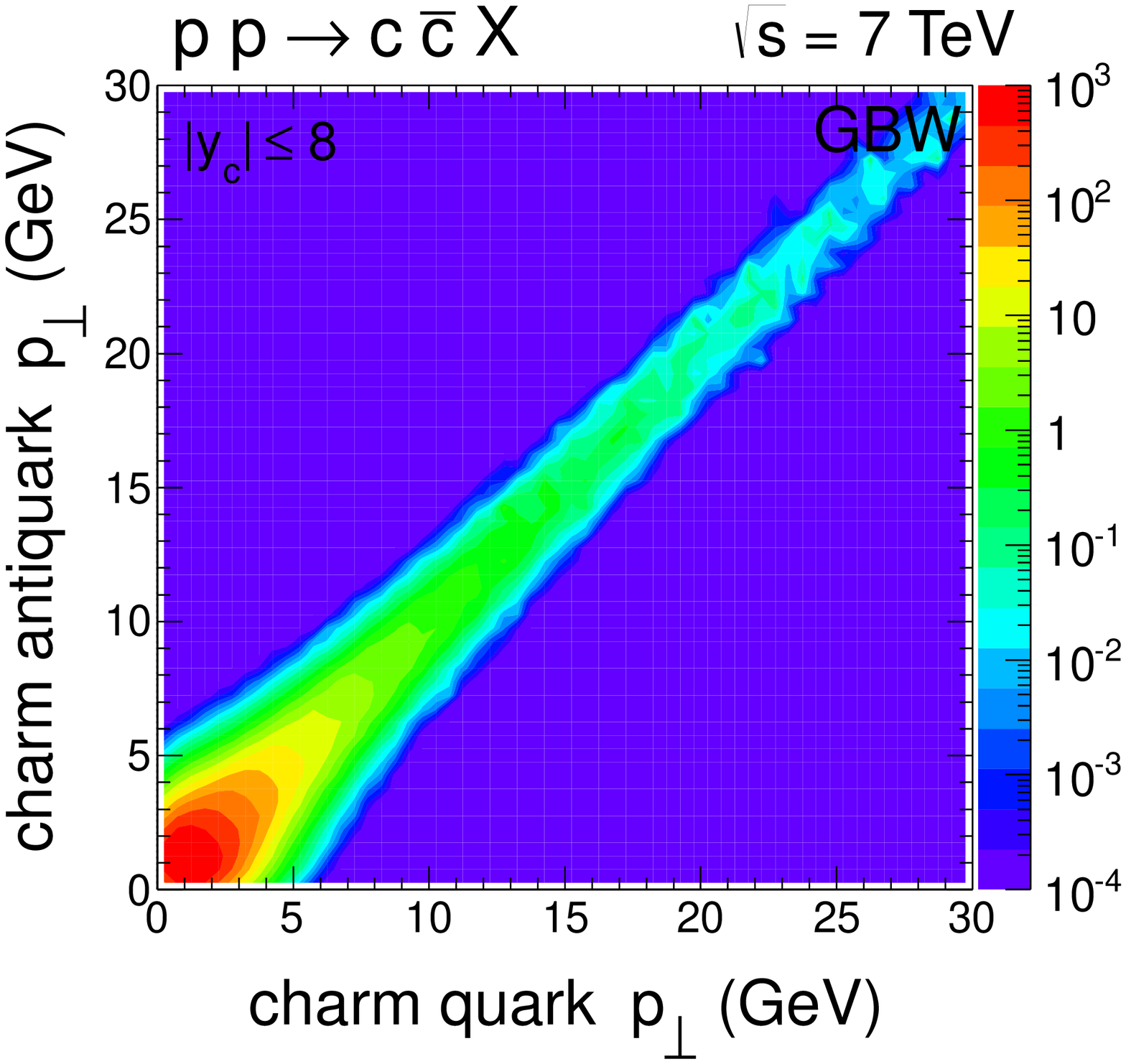}}
\end{minipage}
   \caption{
\small Two dimensional maps in transverse momentum of charm quark and
transverse momentum of the charm antiquark for different unintegrated
gluon distributions.
}
 \label{fig:pt1pt2-ugdfs}
\end{figure}

It was advocated in Ref.~\cite{LS2006} that the
two-dimensional distribution in transverse momentum of charm quark
and charm antiquark can be a good "theoretical observable" to study
unintegrated gluon distributions. In Fig.~\ref{fig:pt1pt2-ugdfs}
we show such distributions for different UGDFs from the literature. We use here KMR \cite{KMR}, KMS \cite{KMS}, Kutak-Stasto \cite{Kutak-Stasto},
Jung setA+, setB+ \cite{Jung} and GBW \cite{GBW} parametrizations.
Quite different pattern is obtained for different UGDFs.
This may have direct consequences for correlation observables for mesons
or/and nonphotonic electrons. Moreover, events when one $p_{t}$ is small and second one is large correspond to the region relevant for higher order collinear corrections.
It is clear from this $p_{1t}p_{2t}$-plane that effects of an effective inclusion of NLO diagrams in the $k_{t}$-factorization approach strongly depend on the construction of UGDFs.

\begin{figure}[!h]
\begin{minipage}{0.47\textwidth}
 \centerline{\includegraphics[width=1.0\textwidth]{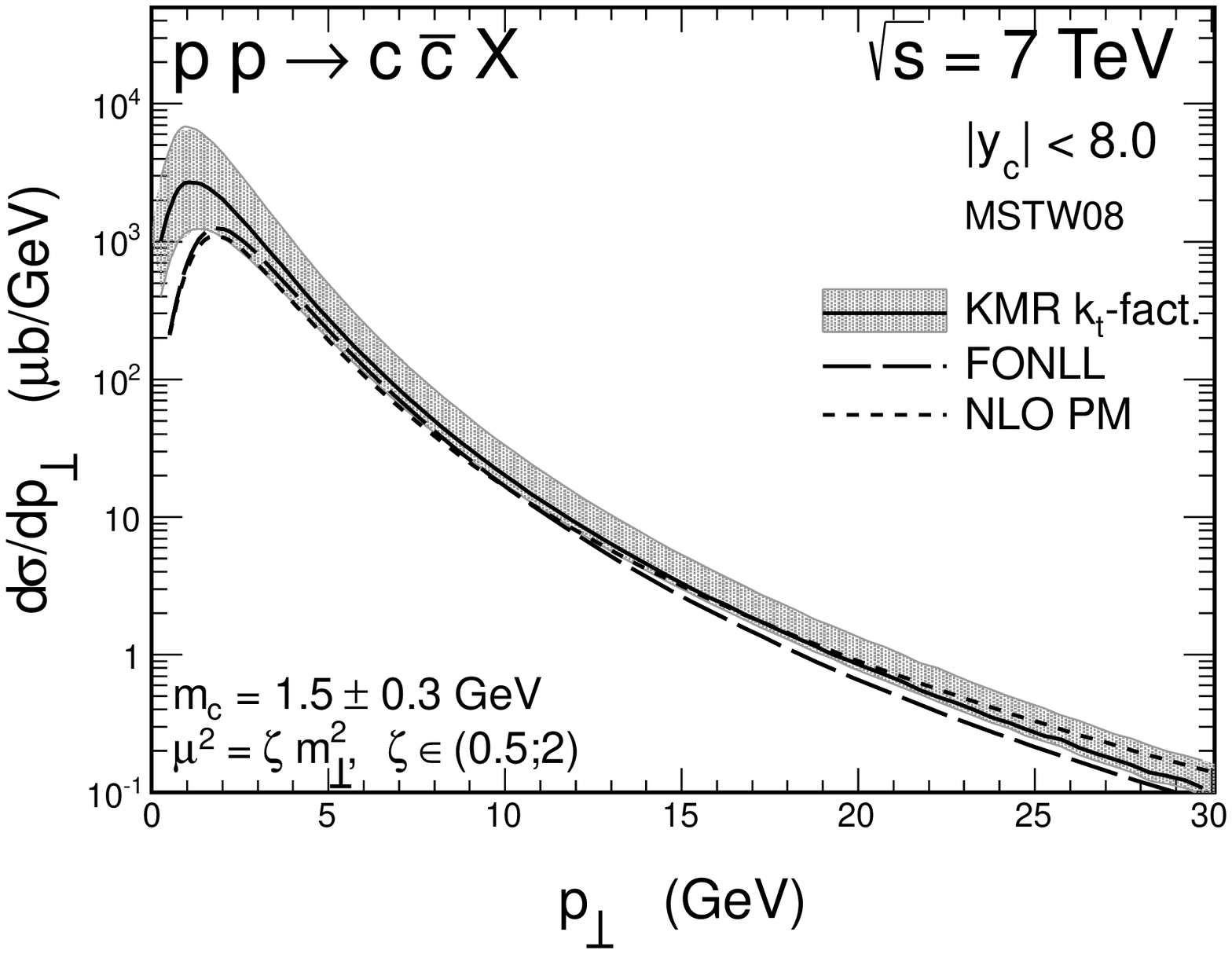}}
\end{minipage}
\hspace{0.5cm}
\begin{minipage}{0.47\textwidth}
 \centerline{\includegraphics[width=1.0\textwidth]{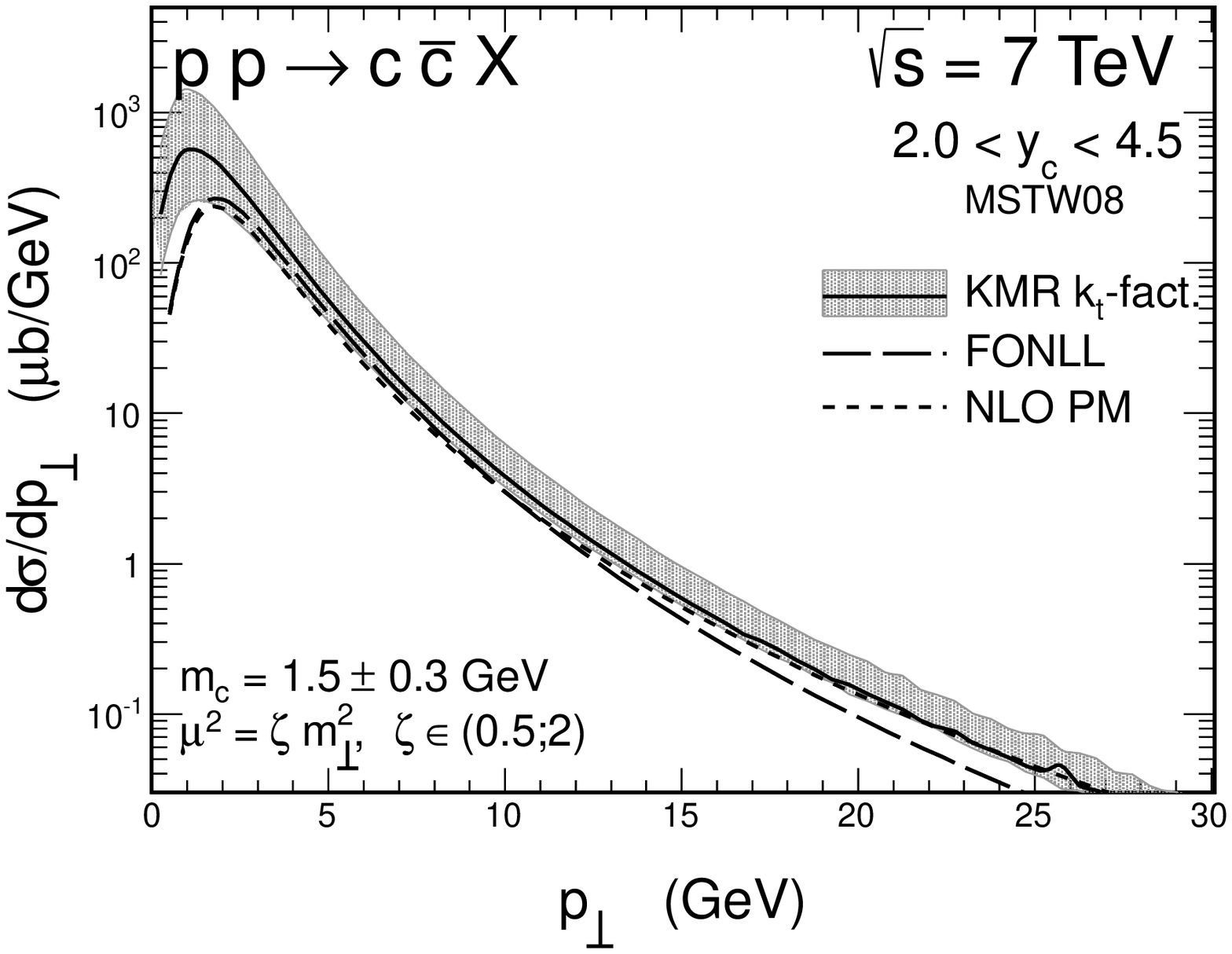}}
\end{minipage}
   \caption{
\small Theoretical uncertainties on transverse momentum distribution
of $c$ or $\bar c$ production due to the choice of factorization/renormalization scale and those related to charm quark mass for the KMR UGDF (solid line with the shaded bands). Left panel shows the
cross section for the whole range of quark/antiquark rapidities while
the left panel for the rapidity range relevant for the LHCb experiment. For comparison the FONLL and NLO PM predictions are also shown.
 }
 \label{fig:pt-charm-1}
\end{figure}

\begin{figure}[!h]
\begin{minipage}{0.47\textwidth}
 \centerline{\includegraphics[width=1.0\textwidth]{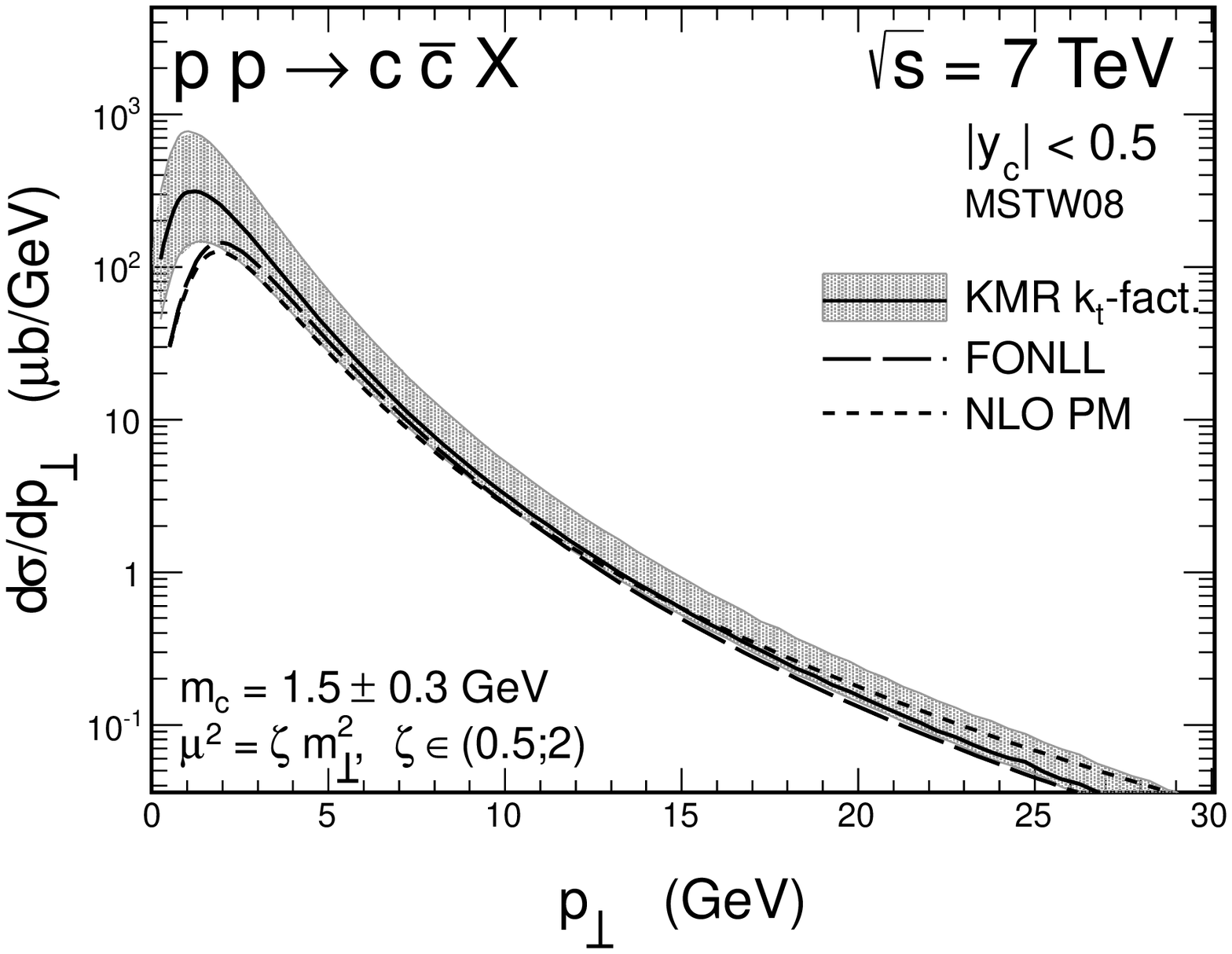}}
\end{minipage}
\hspace{0.5cm}
\begin{minipage}{0.47\textwidth}
 \centerline{\includegraphics[width=1.0\textwidth]{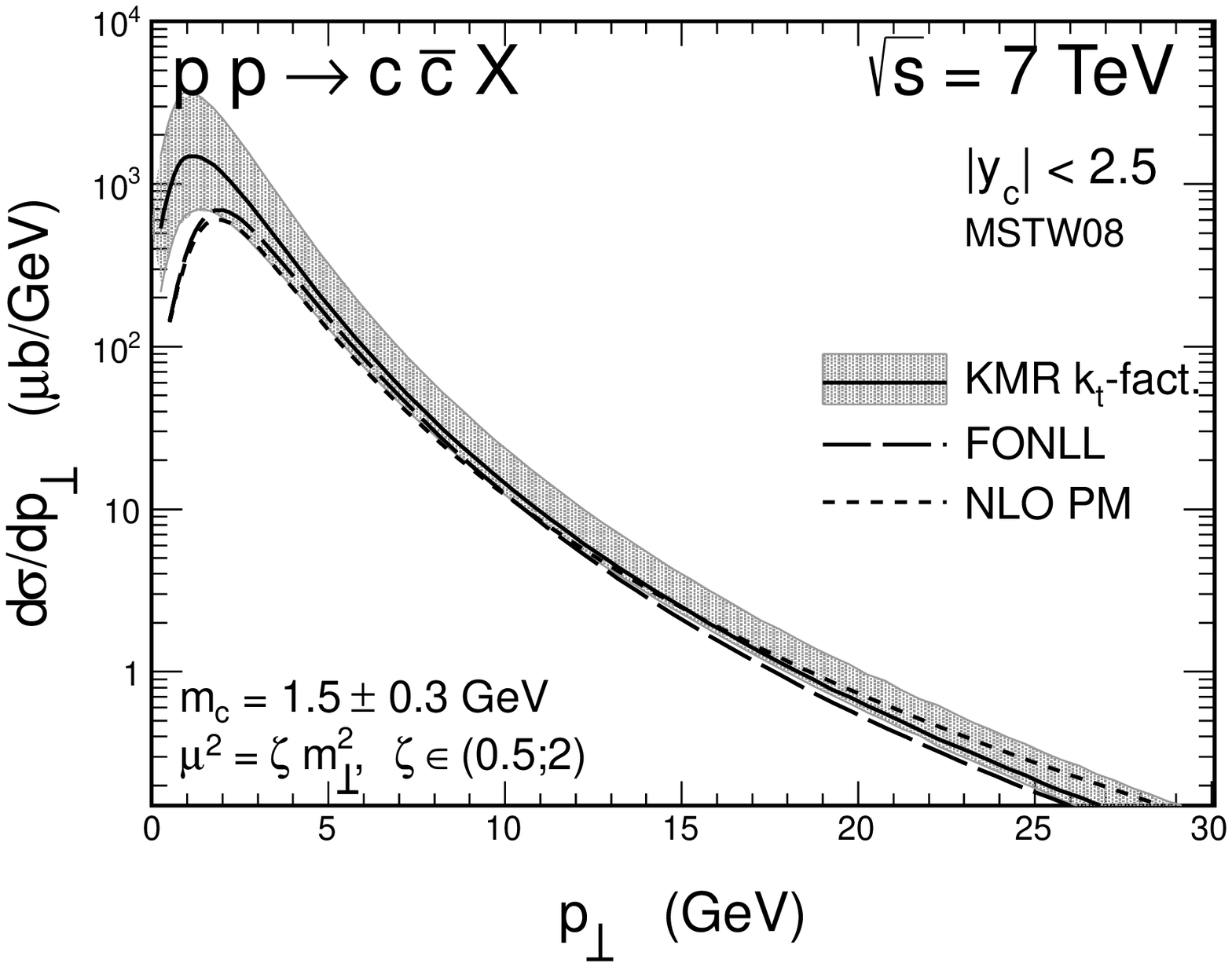}}
\end{minipage}
   \caption{
\small The same as in Fig.\ref{fig:pt-charm-1} but for the ALICE (left
panel) and ATLAS or CMS (right panel) kinematics.}
 \label{fig:pt-charm-2}
\end{figure}

The production of charmed mesons strongly depends on
the choice of UGDF model as will be discussed in the next section.
As will become clear there the KMR UGDF within rather large theoretical uncertainties provides the best description
of the LHC experimental data. The major part of these uncertainties comes from the perturbative part of the calculation.
Therefore in the following we wish
to spend some time to define uncertainties of the corresponding calculations at the quark level.
In Fig.~\ref{fig:pt-charm-1} and Fig.~\ref{fig:pt-charm-2} we present the
uncertainties of our predictions, obtained by changing charm quark mass 
$m_c = 1.5\pm 0.3$ GeV and by varying renormalization and factorization 
scales $\mu^2=\zeta m_{t}^2$, where $\zeta \in (0.5;2)$. The gray shaded
bands represent these both sources of uncertainties summed in quadrature.
The smaller transverse momentum the larger
uncertainty. For comparison we show also results for 
the FONLL \cite{FONLL} and MC@NLO (denoted on figures as NLO PM) \cite{MC@NLO} approaches. Our result of the 
$k_t$-factorization approach is consistent within the uncertainty bands with 
those rather standard NLO collinear calculations. Only at small quark $p_{t}$'s some difference appears. This is the region where transverse momenta of incident gluons play an important role. Particularly, a detailed treatment of the nonperturbative $k_{t}$ region in UGDF may lead to a dumping or an enhancement of the cross section at small $p_{t}$.

\section{Production of charmed mesons}

The hadronization of heavy quarks is usually done
with the help of fragmentation functions. The inclusive distributions of
charmed mesons can be obtained through a convolution of inclusive distributions
of charm quarks/antiquarks and $c \to $ D fragmentation functions:
\begin{equation}
\frac{d \sigma(pp \rightarrow D \bar{D} X)}{d y_D d^2 p_{t,D}} \approx
\int_0^1 \frac{dz}{z^2} D_{c \to D}(z)
\frac{d \sigma(pp \rightarrow c \bar{c} X)}{d y_c d^2 p_{t,c}}
\Bigg\vert_{y_c = y_D \atop p_{t,c} = p_{t,D}/z}
 \; ,
\label{Q_to_h}
\end{equation}
where $p_{t,c} = \frac{p_{t,D}}{z}$ and $z$ is the fraction of
longitudinal momentum of heavy quark carried by meson.
We have made typical approximation assuming that $y_{c}$  is
unchanged in the fragmentation process, i.e. $y_D = y_c$.

As a default set in our calculations we use standard Peterson model of fragmentation function \cite{Peterson} with
the parameter $\varepsilon_{c} = 0.05$. This value was extracted by ZEUS and H1 analyses and seems to be relevant for LO calculations.
Hoowever, in the fragmentation scheme applied in the FONLL framework, rather harder functions (or smaller $\varepsilon_{c}$) are suggested \cite{Cacciari-RHIC}. This issue together with effects of applying other fragmentation functions from the literature \cite{BCFY,Kartvel,CS} will be discussed in more detail when discussing differential distributions.

In Table~\ref{table-inc} we have collected integrated cross sections for
the production of different species of $D$ mesons.
Measured cross sections from different LHC experiments are compared to theoretical predictions obtained with three sets of UGDFs.
The error bars shown for the KMR UGDF reflect uncertainty
due to the choice of factorization/renormalization scale ($\mu$) and related to the
mass of the quark ($m_c$). The fractional uncertainties
due to these both sources for other UGDFs are similar.
Only cross sections obtained with the KMR UGDF are consistent within error bars with
the experimental data.

In the cases of measurements with the full coverage of the meson transverse momentum range, the theoretical cross sections are almost insensitive
to the fragmentation model. Quite different situation is observed when small $p_t$ region is excluded. In the latter case, using the Peterson model with $\varepsilon_{c} = 0.02$
(which gives results closer to the FONLL predictions) we note the enhancement of the integrated cross sections by about $20 \%$.

Let us start presentation of differential distributions for different LHC experiments.

\begin{table}[tb]%
\caption{Integrated cross sections for production of different $D$ mesons at LHC.}
\newcolumntype{Z}{>{\centering\arraybackslash}X}
\label{table-inc}
\centering %
\begin{tabularx}{17.5cm}{ZZZZZZZ}
\toprule[0.1em] %
\\[-2.4ex] 

 \multirow{2}*{Acceptance}     & \multirow{2}*{Mode} & \multirow{2}*{$\sigma_{tot}^{EXP} \;\;[\mu$b]} & \multicolumn{4}{c}{$\sigma_{tot}^{THEORY} \;\;[\mu$b]} \\ [+0.4ex]
                        &              &                         & \multicolumn{2}{c}{KMR $^{+}_{-}(\mu)$ $^{+}_{-}(m_{c})$} & Jung setA$0+$ & KMS \\[+0.4ex] 
 
\toprule[0.1em]\\[-2.4ex] 

 ALICE                                     & $(D^{0} + \bar{D^{0}})/2$  & $516\pm41^{+69}_{-175}$ & \multicolumn{2}{c}{$514$ $^{+169}_{-130}$ $^{+384}_{-198}$}   & 317  & 313 \\ [+0.4ex]
 $|y| < 0.5$                               & $(D^{+} + D^{-})/2$        & $248\pm30^{+52}_{-92}$  & \multicolumn{2}{c}{$206$ $^{+68}_{-52}$ $^{+154}_{-79}$}      & 127  & 125 \\ [+0.4ex]
                                           & $(D^{*+} + D^{*-})/2$      & $247\pm27^{+36}_{-81}$  & \multicolumn{2}{c}{$208$ $^{+69}_{-53}$ $^{+156}_{-80}$}      & 129  & 127 \\  [+0.4ex]
    ALICE                                  &    &  & \multicolumn{2}{c}{}      &  & \\  [+0.4ex]
 $|y| < 0.5$                               & $(D^{+}_{S} + D^{-}_{S})/2$     & $53\pm12^{+13}_{-15}$  & \multicolumn{2}{c}{$20$ $^{+5}_{-4}$ $^{+7}_{-5}$ $^{(+20\%)}$}      & $13$ $^{(+20\%)}$  & $13$ $^{(+20\%)}$ \\  [+0.4ex]
  \multirow{1}*{$2<p_{\perp}<12$ GeV}      &    &  & \multicolumn{2}{c}{}      &  &  \\  [+0.4ex]
 \hline \\[-2.4ex]
 LHCb                                      & $D^{0} + \bar{D^{0}}$      & $1488\pm182$            & \multicolumn{2}{c}{$1744$ $^{+565}_{-418}$ $^{+1435}_{-700}$} & 1162 & 872 \\[+0.4ex]
 $2 < y < 4.5$                             & $D^{+} + D^{-}$            & $717\pm109$             & \multicolumn{2}{c}{$697$ $^{+226}_{-167}$ $^{+574}_{-280}$}   & 465  & 349 \\[+0.4ex]
 \multirow{1}*{$0 < p_{\perp} < 8$ GeV}    & $D^{*+} + D^{*-}$          & $676\pm137$             & \multicolumn{2}{c}{$705$ $^{+229}_{-169}$ $^{+582}_{-284}$}   & 471  & 354 \\[+0.4ex]
                                           & $D^{+}_{S} + D^{-}_{S}$    & $194\pm38$              & \multicolumn{2}{c}{$246$ $^{+80}_{-59}$ $^{+203}_{-99}$}      & 164  & 123 \\[+0.4ex]
\hline \\[-2.4ex]

 ATLAS                                    & $D^{+} + D^{-}$             & $238\pm13^{+35}_{-23}$ & \multicolumn{2}{c}{$137$ $^{+31}_{-20}$ $^{+30}_{-24}$ $^{(+20\%)}$}       & $103$ $^{(+20\%)}$  & $93$ $^{(+20\%)}$ \\ [+0.4ex]
 $|\eta| < 2.1$                           & $D^{+}_{S} + D^{-}_{S}$     & $168\pm34^{+27}_{-25}$  & \multicolumn{2}{c}{$48$ $^{+12}_{-7}$ $^{+11}_{-8}$ $^{(+20\%)}$}      & $36$ $^{(+20\%)}$  & $33$ $^{(+20\%)}$ \\ [+0.4ex]
 $p_{\perp} > 3.5$ GeV                    & $D^{*+} + D^{*-}$           & $285\pm16^{+32}_{-27}$  & \multicolumn{2}{c}{$155$ $^{+37}_{-22}$ $^{+37}_{-28}$ $^{(+20\%)}$}     & $115$ $^{(+20\%)}$  & $104$ $^{(+20\%)}$ \\  [+0.4ex]

\bottomrule[0.1em]

\end{tabularx}

\end{table}

\subsection{ALICE}

\begin{figure}[!h]
\begin{minipage}{0.47\textwidth}
 \centerline{\includegraphics[width=1.0\textwidth]{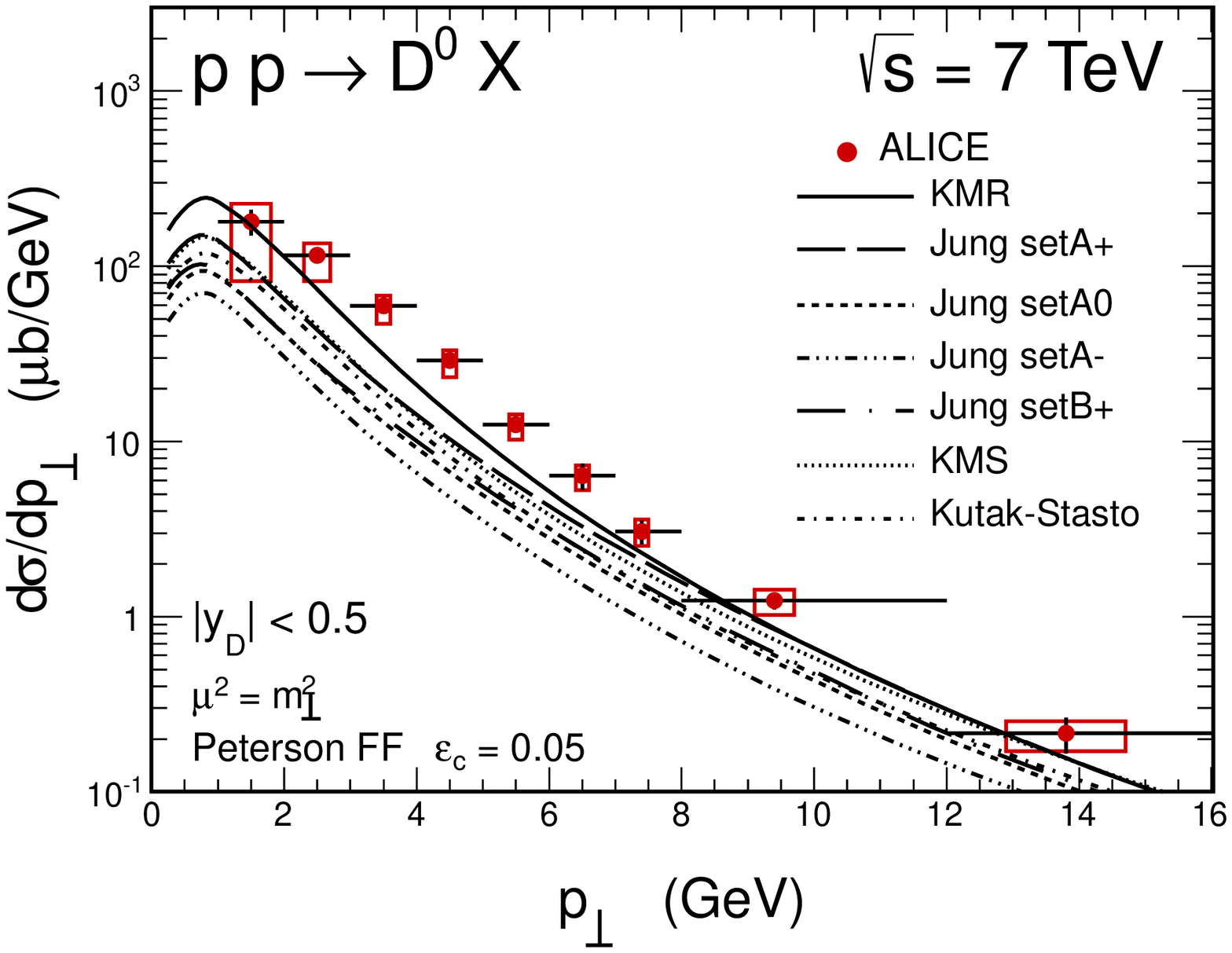}}
\end{minipage}
\hspace{0.5cm}
\begin{minipage}{0.47\textwidth}
 \centerline{\includegraphics[width=1.0\textwidth]{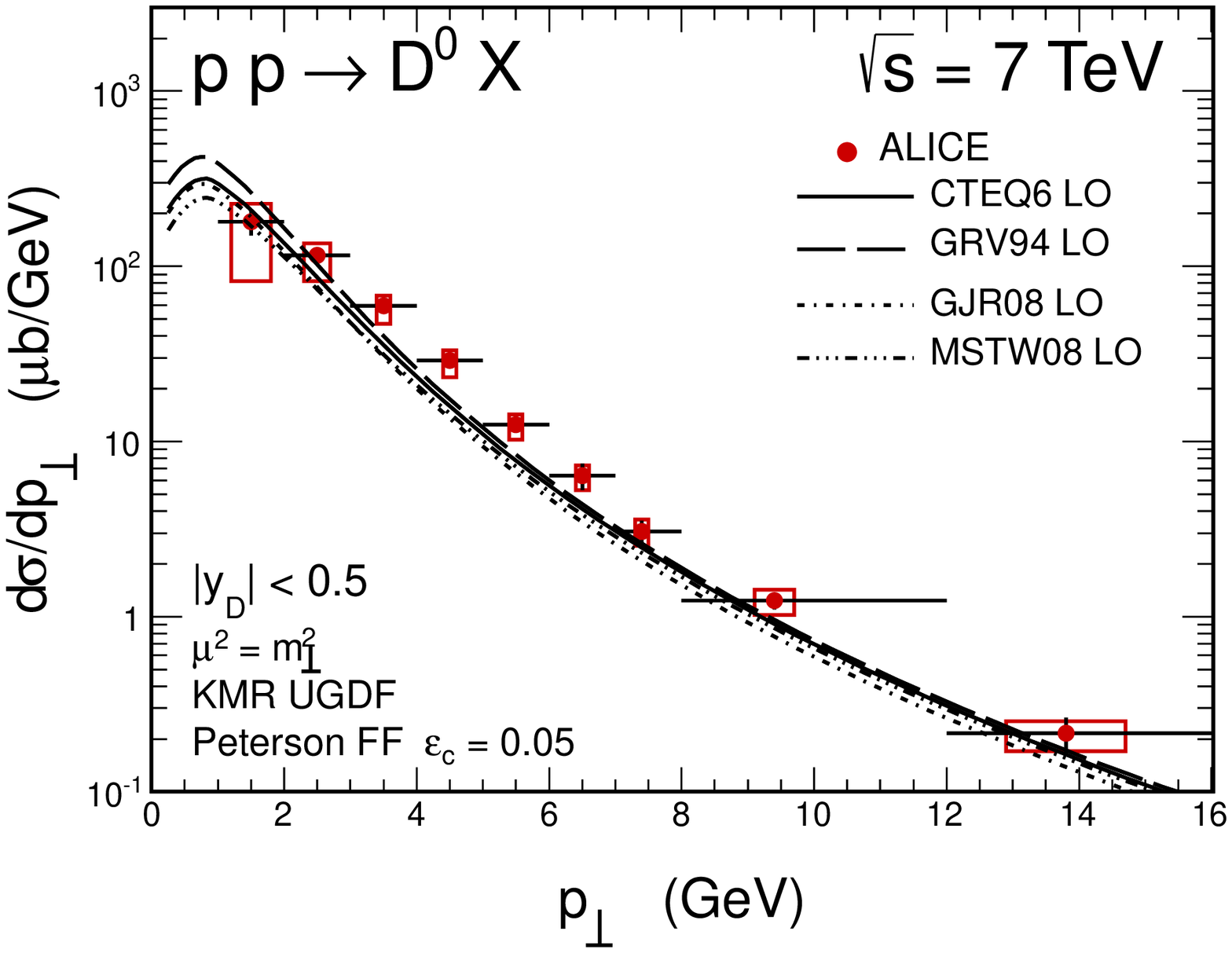}}
\end{minipage}
   \caption{
\small Transverse momentum distribution of $D^0$ mesons for 
the ALICE measurement. The left panel shows results for different UGDFs
while the right panel shows uncertainties due to the choice of collinear
gluon distributions in calculation of the KMR UGDF. }
 \label{fig:pt-alice-D-1}
\end{figure}

\begin{figure}[!h]
\begin{minipage}{0.47\textwidth}
 \centerline{\includegraphics[width=1.0\textwidth]{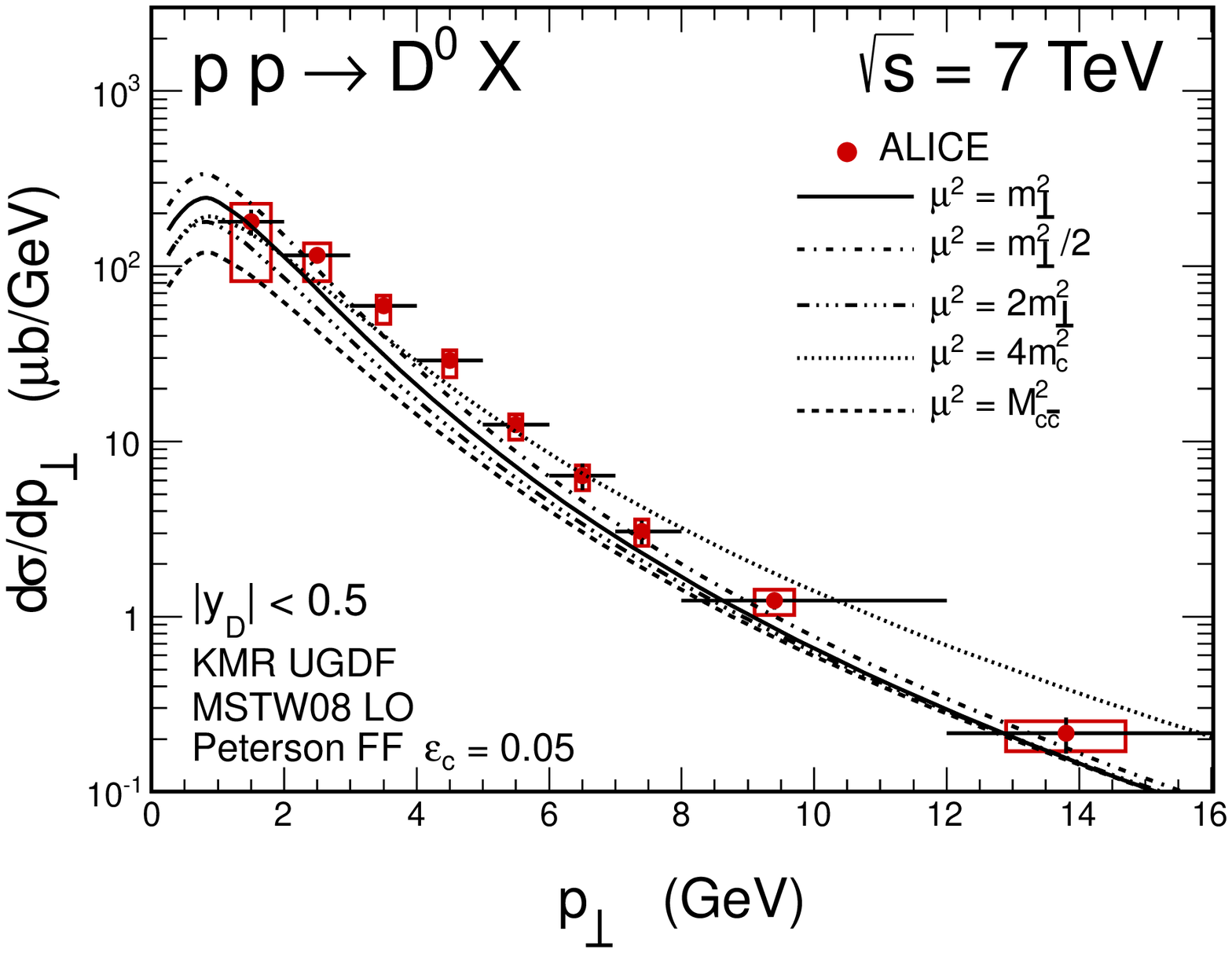}}
\end{minipage}
\hspace{0.5cm}
\begin{minipage}{0.47\textwidth}
 \centerline{\includegraphics[width=1.0\textwidth]{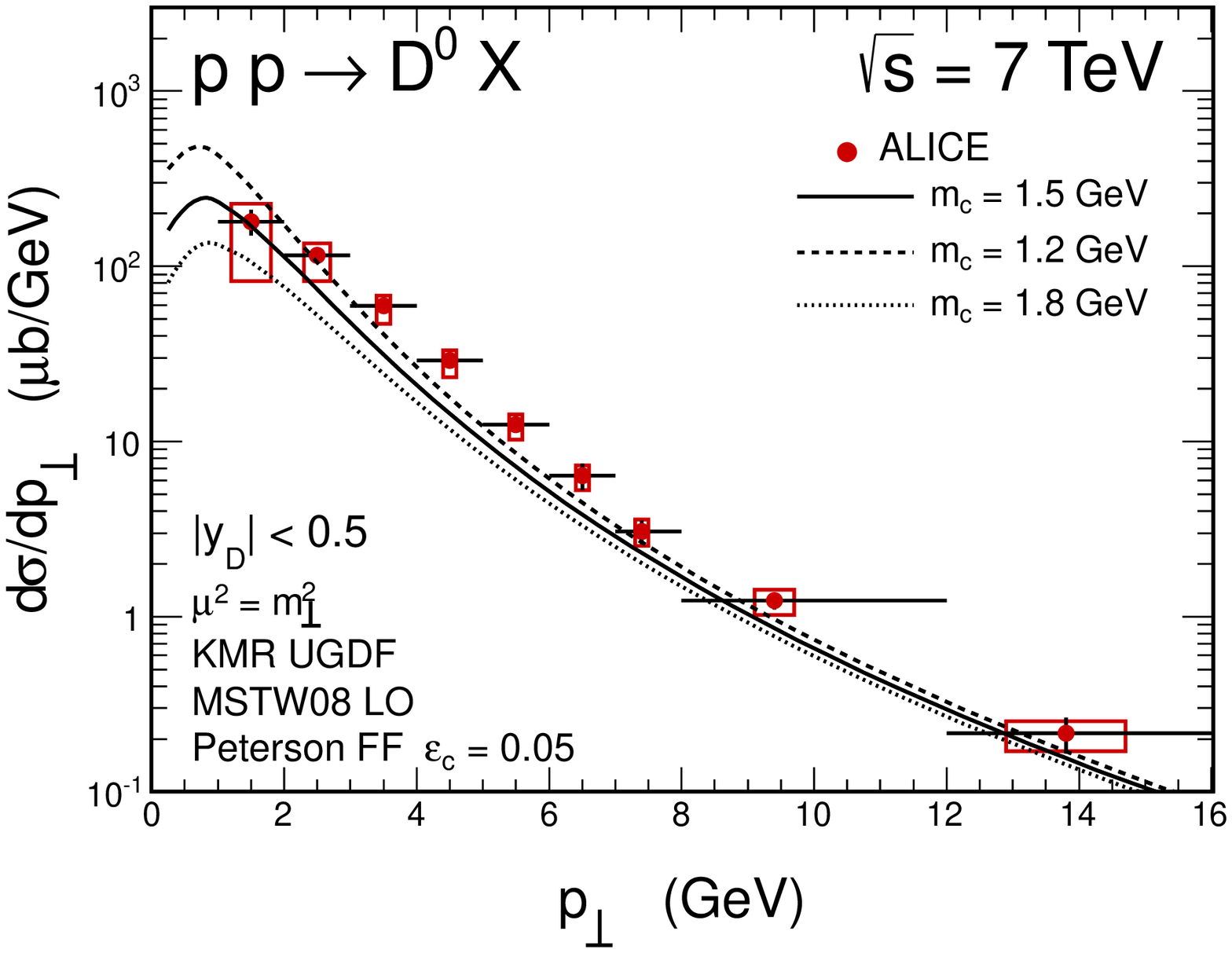}}
\end{minipage}
   \caption{
\small Uncertainties of the theoretical cross section for the $D^0$ meson
production within the ALICE acceptance due to
the choice of the scale (left) and due to the quark mass (right). }
\label{fig:pt-alice-D-2}
\end{figure}

Let us focus first on the production of charmed mesons at midrapidities. The
ALICE collaboration has performed a measurement of transverse
momentum distribution of $D^0$, $D^+$, $D^{*+}$, $D_s^+$ \cite{ALICEincD,ALICEincDs}. 
In the very limited range of (pseudo)rapidity one tests unintegrated gluon distributions in a
pretty narrow region of longitudinal momentum fractions 
(see Fig.~\ref{fig:ylogx-kmr}).
In Fig.~\ref{fig:pt-alice-D-1} we show transverse momentum distribution
of $D^0$ mesons. In the left panel we present results for different UGDF
known from the literature. Most of the existing distributions fail
to describe the ALICE data. The KMR UGDF provides the best description
of the measured distributions. Therefore in the following we shall concentrate on
the results obtained with the KMR UGDF.
In the right panel we show uncertainties due to the
choice of usual integrated collinear gluon distributions (PDFs) used for calculating
the KMR UGDF.
In the latter case the biggest uncertainty can be observed at small
transverse momenta, i.e. in the region of small gluon longitudinal
momentum fraction. We use rather up-to-date MSTW08 \cite{MSTW08}, CTEQ6 \cite{CTEQ6} and GJR08 \cite{GJR08}
parametrizations as well as GRV94 \cite{GRV94} which is fairly older but was very often used in last years
in similar analyzes.
For more detailed discussion of the PDFs aspects in charm production we refer the reader to 
Ref.~\cite{LMS-2011-sub}. In Fig.~\ref{fig:pt-alice-D-2} we show separately uncertainties due to the choice of
factorization/renormalization scale (left panel) and those due to the choice
of the quark mass (right panel). The uncertainties due to the choice of scales
is rather large. The uncertainties due to quark mass are significant only at small transverse momenta. They are calculated
by varying the quark mass $m_c = 1.5 \pm 0.3$ \GeV and are representative for all other UGDFs.

\begin{figure}[!h]
\begin{minipage}{0.47\textwidth}
 \centerline{\includegraphics[width=1.0\textwidth]{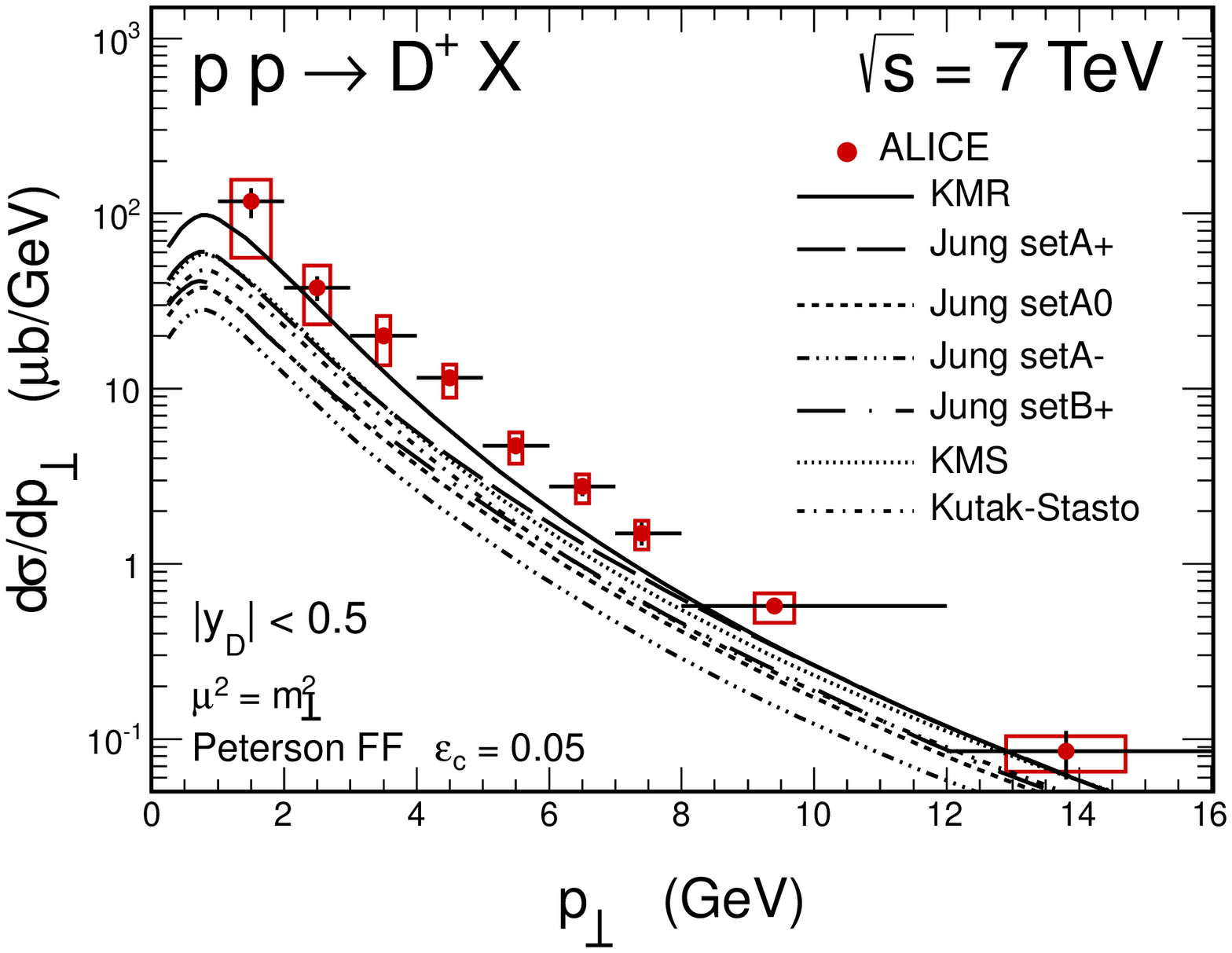}}
\end{minipage}
\hspace{0.5cm}
\begin{minipage}{0.47\textwidth}
 \centerline{\includegraphics[width=1.0\textwidth]{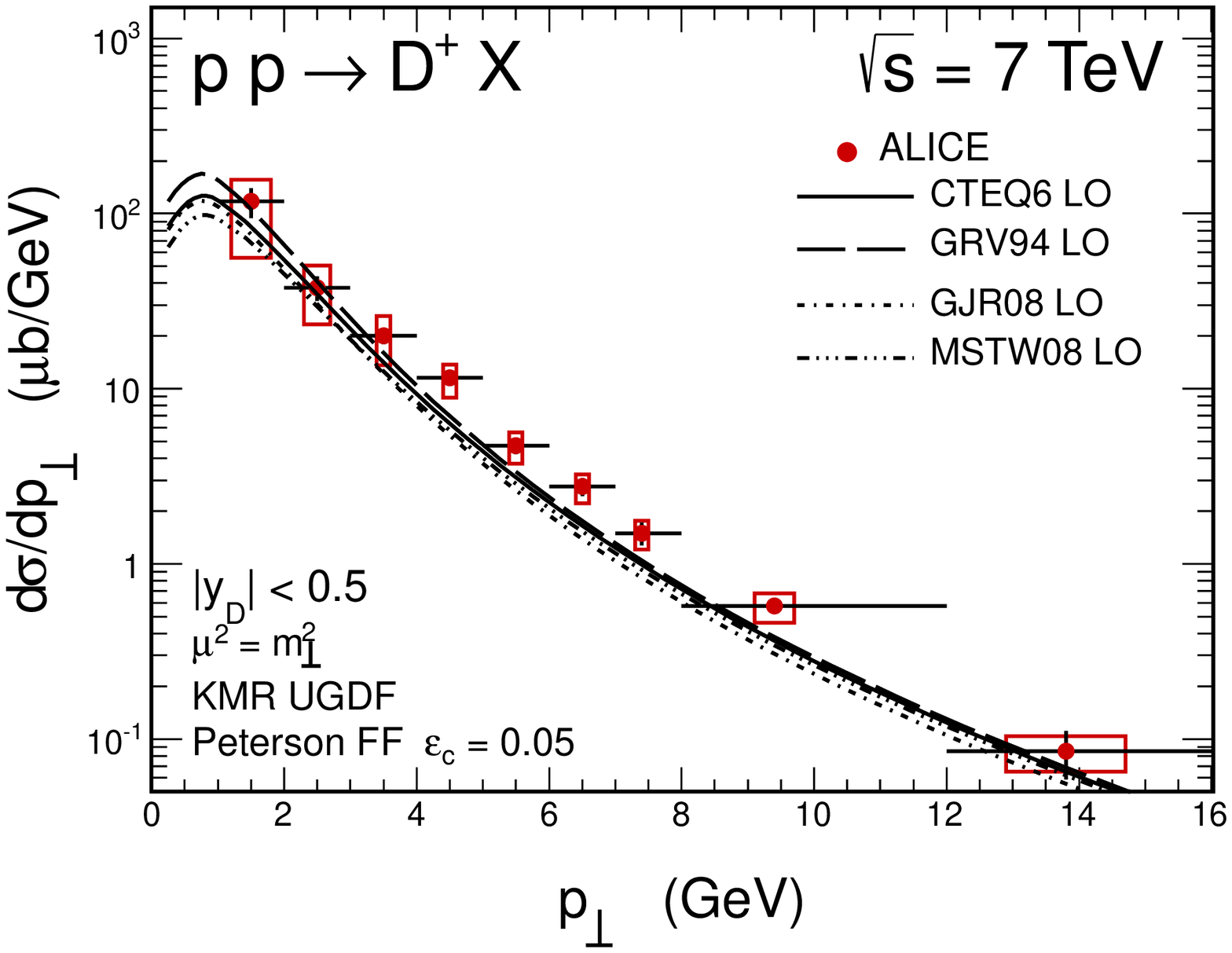}}
\end{minipage}
   \caption{
\small The same as in Fig.~\ref{fig:pt-alice-D-1} but for the production
of $D^+$ mesons. }
 \label{fig:pt-alice-D-3}
\end{figure}

\begin{figure}[!h]
\begin{minipage}{0.47\textwidth}
 \centerline{\includegraphics[width=1.0\textwidth]{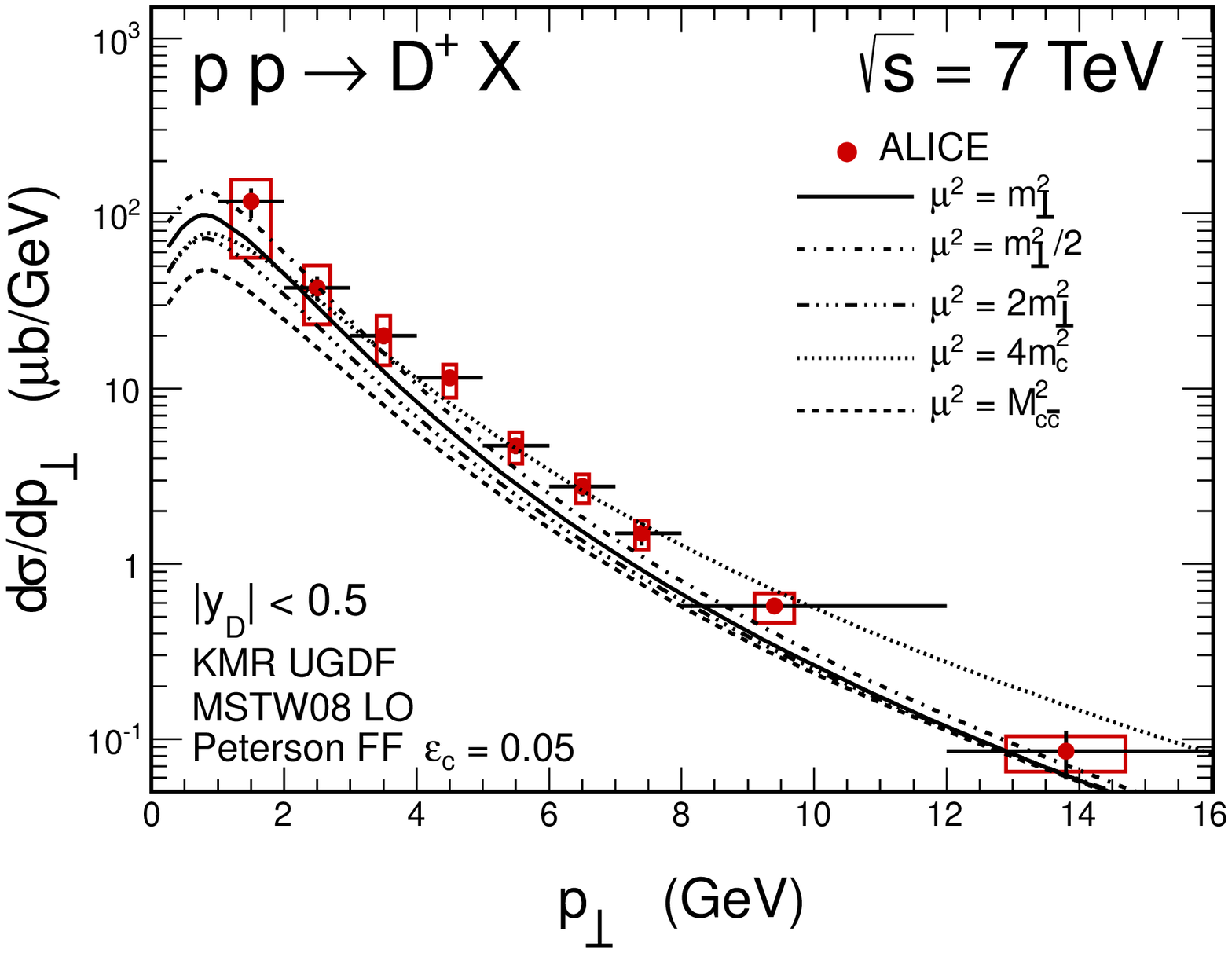}}
\end{minipage}
\hspace{0.5cm}
\begin{minipage}{0.47\textwidth}
 \centerline{\includegraphics[width=1.0\textwidth]{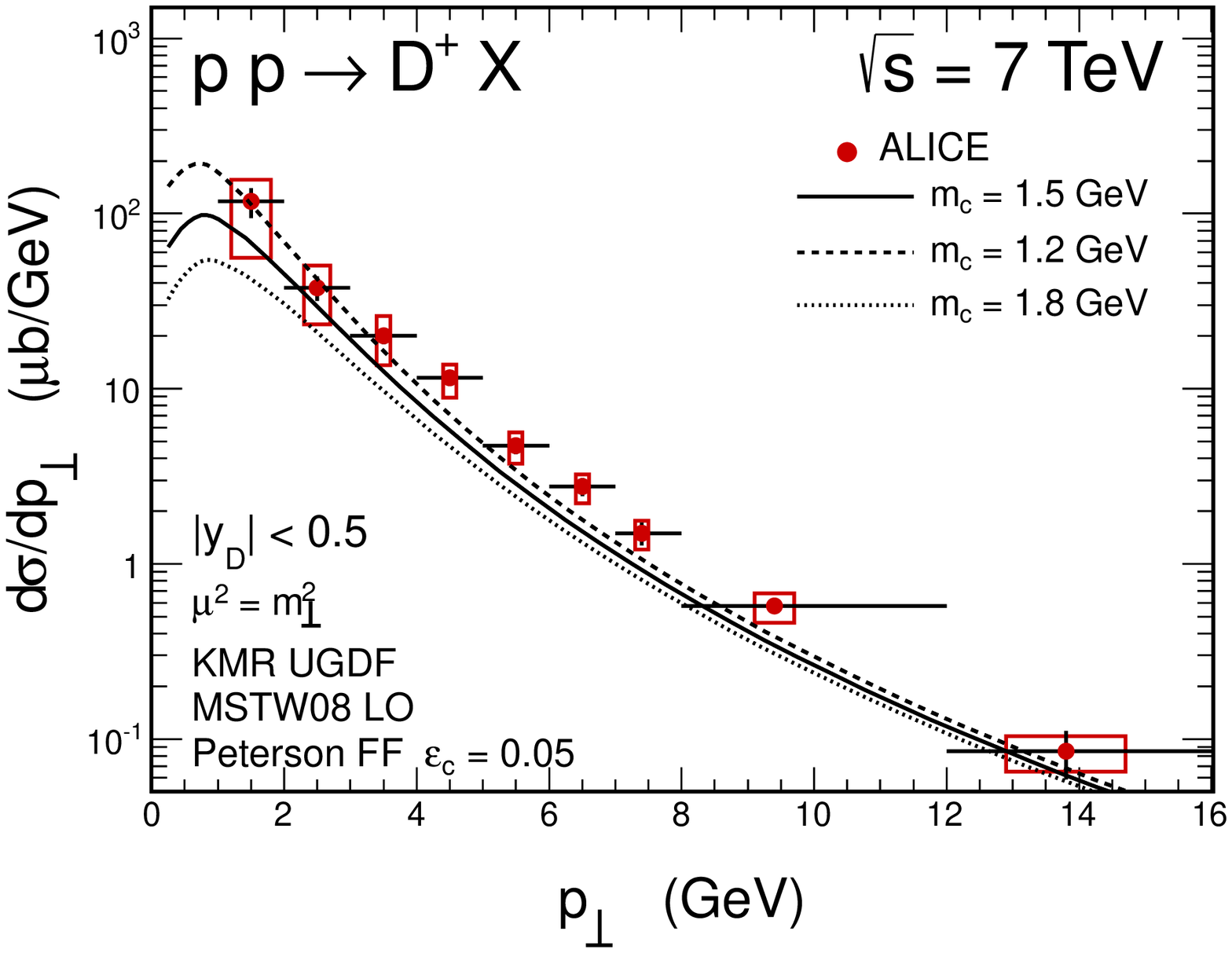}}
\end{minipage}
   \caption{
\small The same as in Fig.~\ref{fig:pt-alice-D-2} but for the $D^+$ meson. }
 \label{fig:pt-alice-D-4}
\end{figure}

In Fig.~\ref{fig:pt-alice-D-3} and Fig.~\ref{fig:pt-alice-D-4}
we present corresponding plots for $D^+$ mesons. The situation here is very similar
to the case of $D^0$ mesons.

\begin{figure}[!h]
\begin{minipage}{0.47\textwidth}
 \centerline{\includegraphics[width=1.0\textwidth]{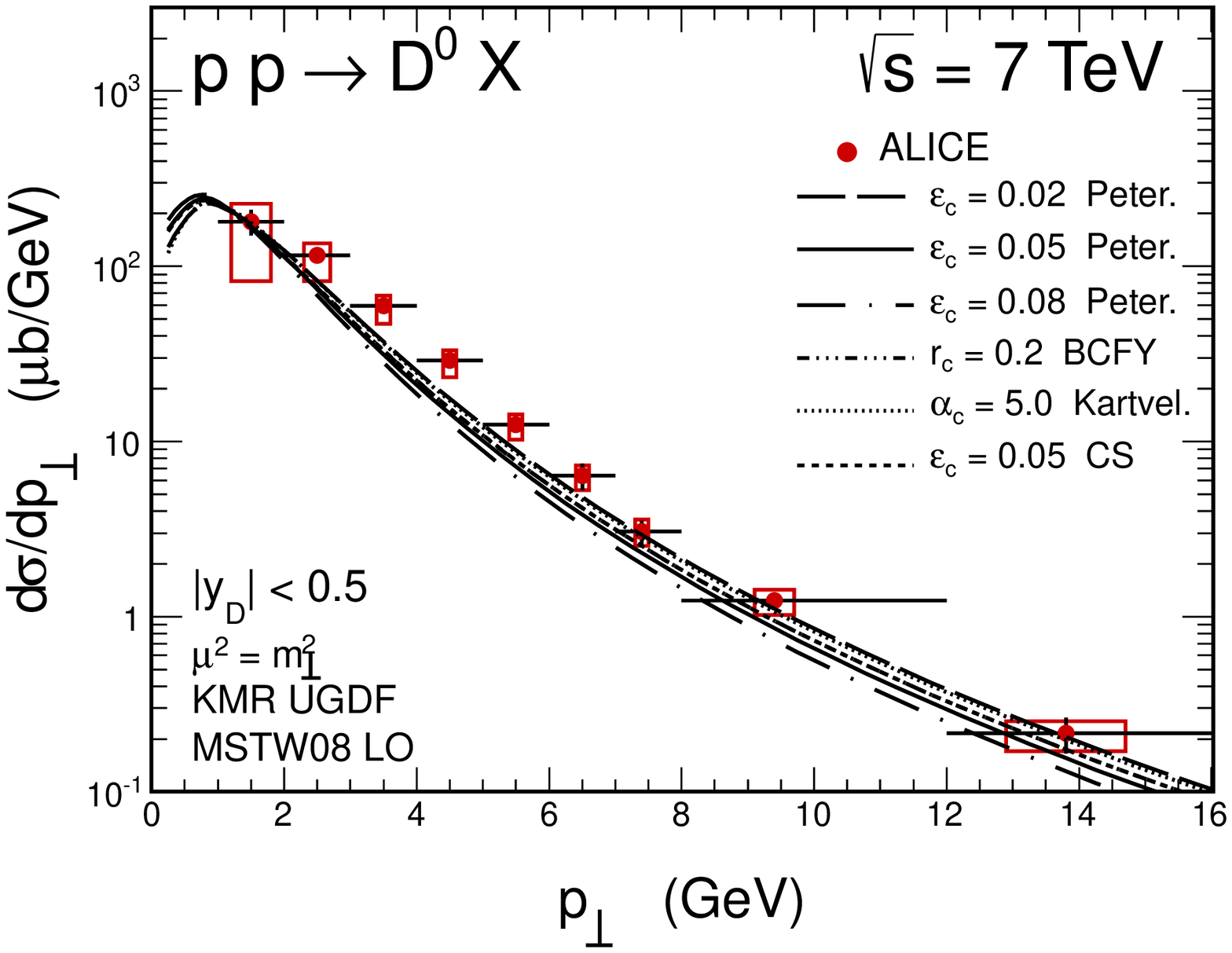}}
\end{minipage}
\hspace{0.5cm}
\begin{minipage}{0.47\textwidth}
 \centerline{\includegraphics[width=1.0\textwidth]{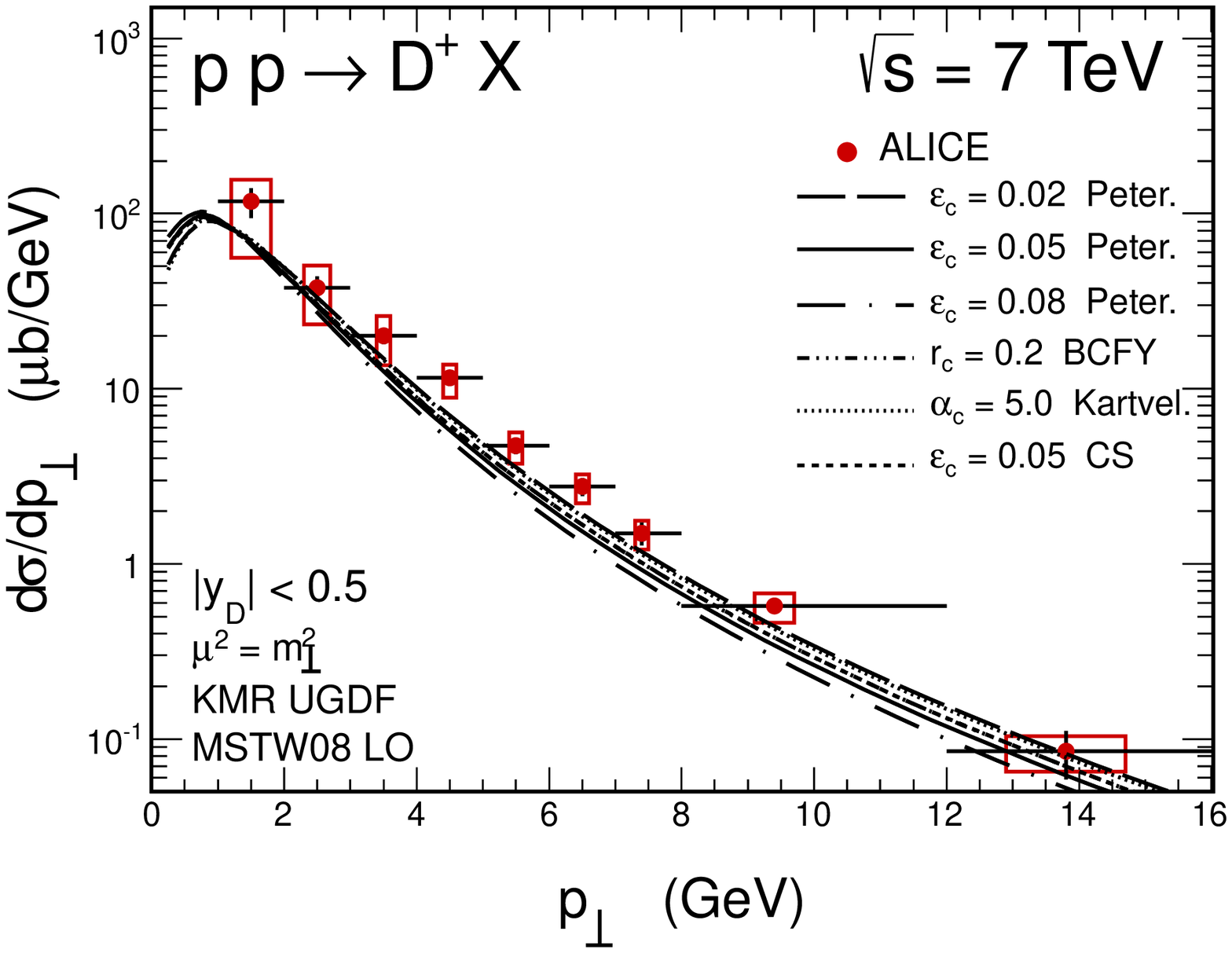}}
\end{minipage}
   \caption{
\small Uncertainties due to fragmentation parameter $\varepsilon_c$ in the
Peterson fragmentation function and related to the choice of other fragmentation models
for the KMR UGDFs.}
 \label{fig:pt-alice-D-6}
\end{figure}

Let us quantify now uncertainties due to the fragmentation process.
Fig.~\ref{fig:pt-alice-D-6} shows results for $D^0$ (left panel) and $D^+$ (right panel) mesons
for different fragmentation functions from the literature.
We use here the Peterson model with three different sets of $\varepsilon_c$ parameter, as well as
Braaten et al. \cite{BCFY}, Kartvelishvili et al. \cite{Kartvel} and Collins-Spiller \cite{CS} parametrizations.
All of the applied functions give similar results. The effects related to the fragmentation process
seem to be important only at larger meson $p_t$'s, starting from $p_t = 3$ \GeV.

\begin{figure}[!h]
\begin{minipage}{0.47\textwidth}
 \centerline{\includegraphics[width=1.0\textwidth]{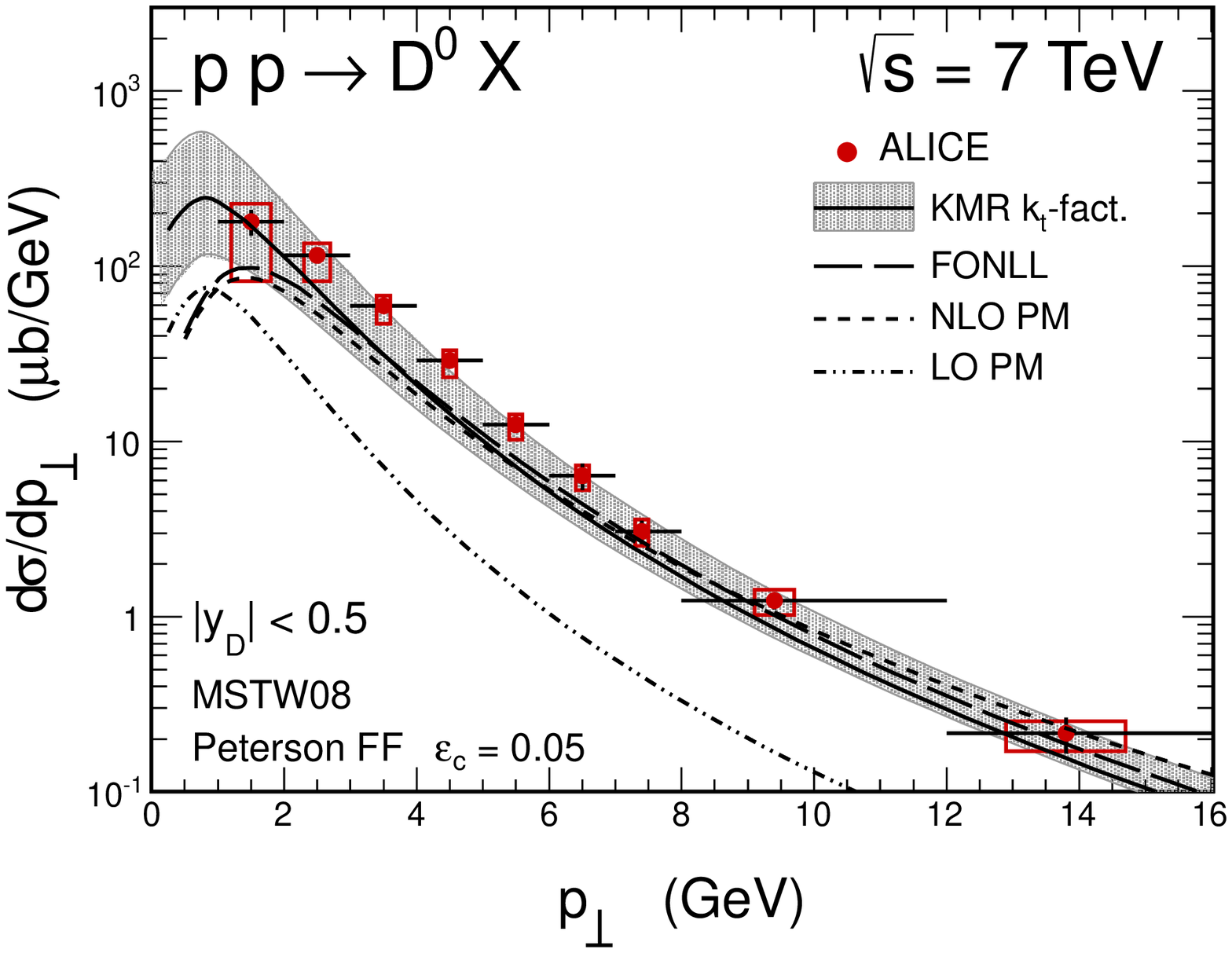}}
\end{minipage}
\hspace{0.5cm}
\begin{minipage}{0.47\textwidth}
 \centerline{\includegraphics[width=1.0\textwidth]{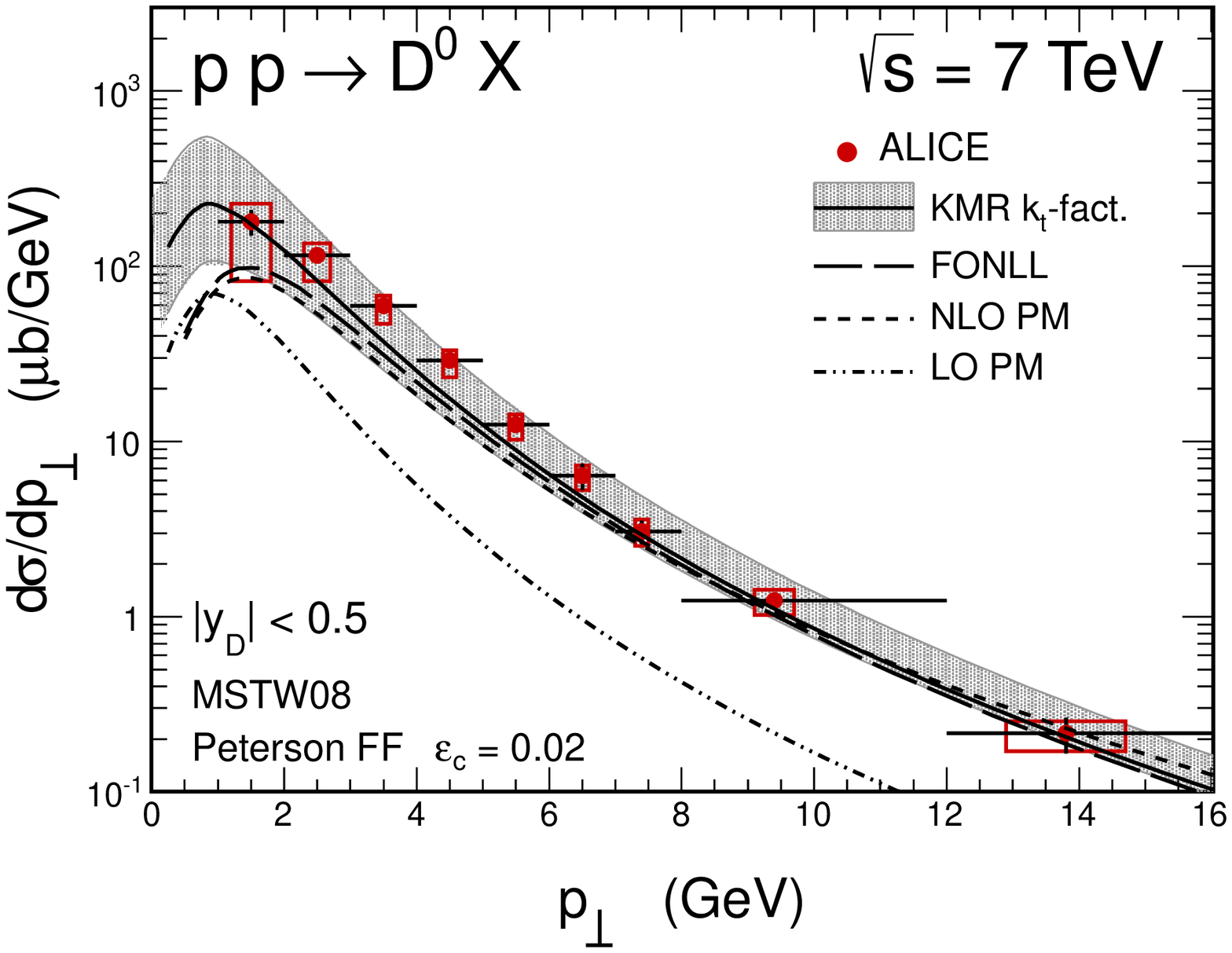}}
\end{minipage}
   \caption{
\small Transverse momentum distribution of $D^0$ mesons for the ALICE
kinematical region for different values of the Peterson $\varepsilon_c$
parameter. Together with our predictions for the KMR UGDF (solid line with shaded band) results of different popular approaches
are also shown.}
 \label{fig:pt-alice-D-5a}
\end{figure}

\begin{figure}[!h]
\begin{minipage}{0.47\textwidth}
 \centerline{\includegraphics[width=1.0\textwidth]{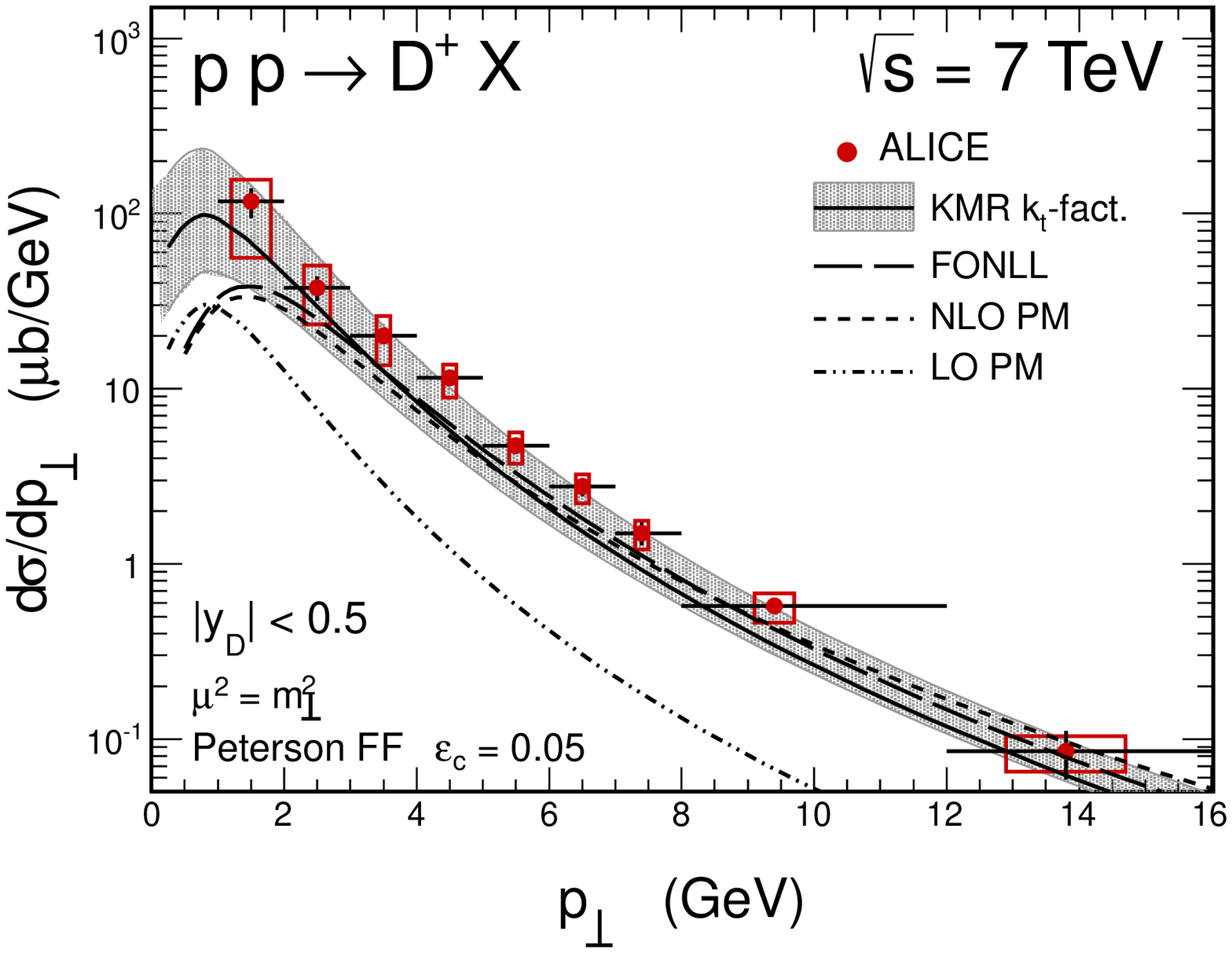}}
\end{minipage}
\hspace{0.5cm}
\begin{minipage}{0.47\textwidth}
 \centerline{\includegraphics[width=1.0\textwidth]{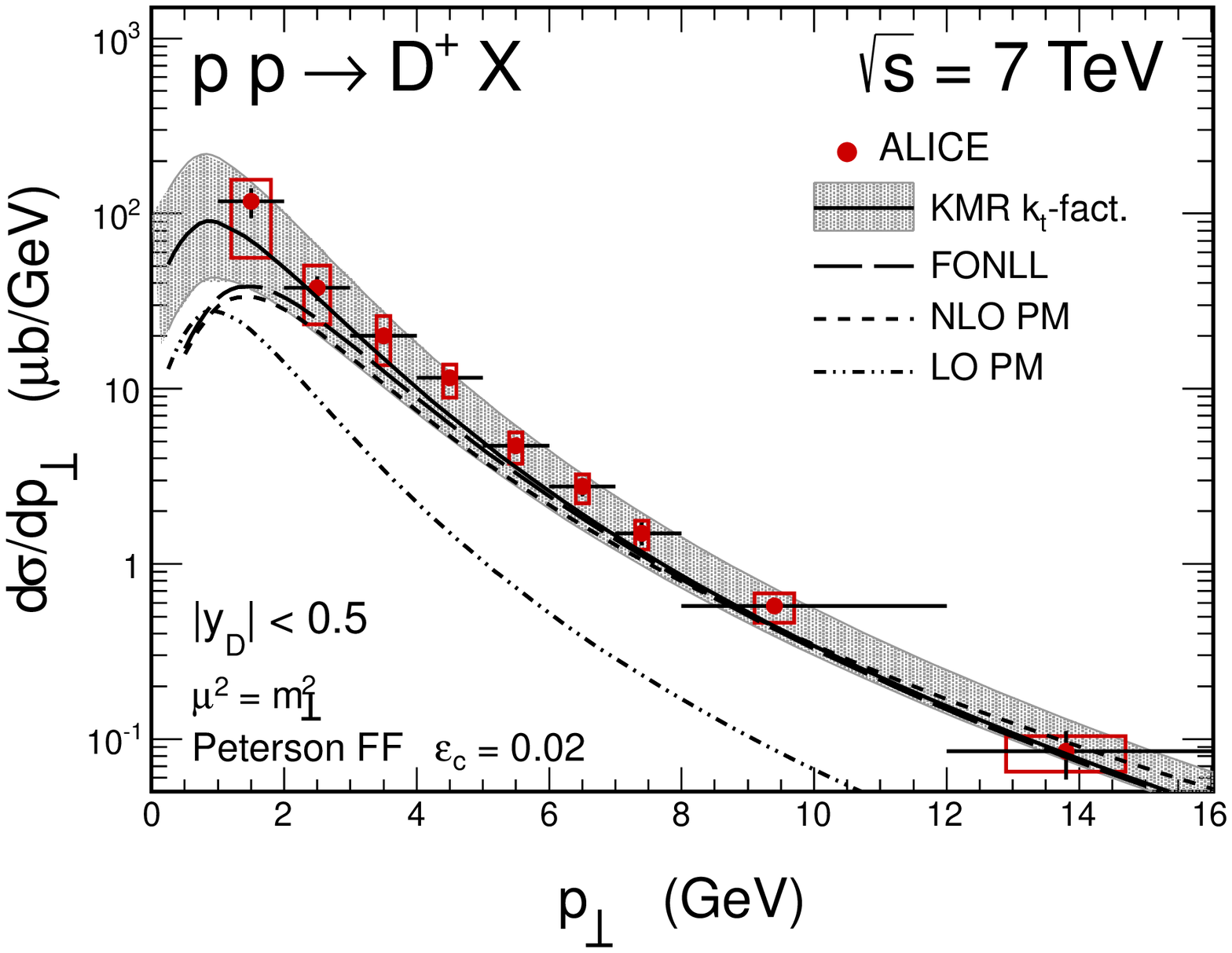}}
\end{minipage}
   \caption{
\small The same as in Fig.~\ref{fig:pt-alice-D-5a} but for the $D^+$ meson.}
 \label{fig:pt-alice-D-5b}
\end{figure}

Let us compare now results of our approach to the results of some other
popular approaches used in the literature.
In Fig.~\ref{fig:pt-alice-D-5a} and Fig.~\ref{fig:pt-alice-D-5b} we present
such a comparison. Our results obtained within the $k_t$-factorization
approach with the KMR UGDF are very similar to those obtained within
NLO PM and FONLL models. The cross sections obtained within
leading-order collinear approximation (LO PM) are much smaller, in particular
for larger transverse momenta. Comparing left and right panels of these figures
one can observe an improvement of the large $p_t$ data description
when $\varepsilon_c = 0.02$ in the Peterson function is taken. This choice corresponds
to the upper limit of our uncertainties in the hadronization. It makes our 
results closer to those from FONLL. In the FONLL approach as a default fragmentation scheme
for charm quarks the BCFY model with $r_c = 0.1$ is used. However, the Peterson parametrization
with $\varepsilon_c = 0.02$ gives in general very similar characteristics.
Since our $k_t$-factorization calculation is very similar to the FONLL predictions at the quark level (see Fig.~\ref{fig:pt-charm-2}) application of the
harder fragmentation functions may be justified.

\begin{figure}[!h]
\begin{minipage}{0.47\textwidth}
 \centerline{\includegraphics[width=1.0\textwidth]{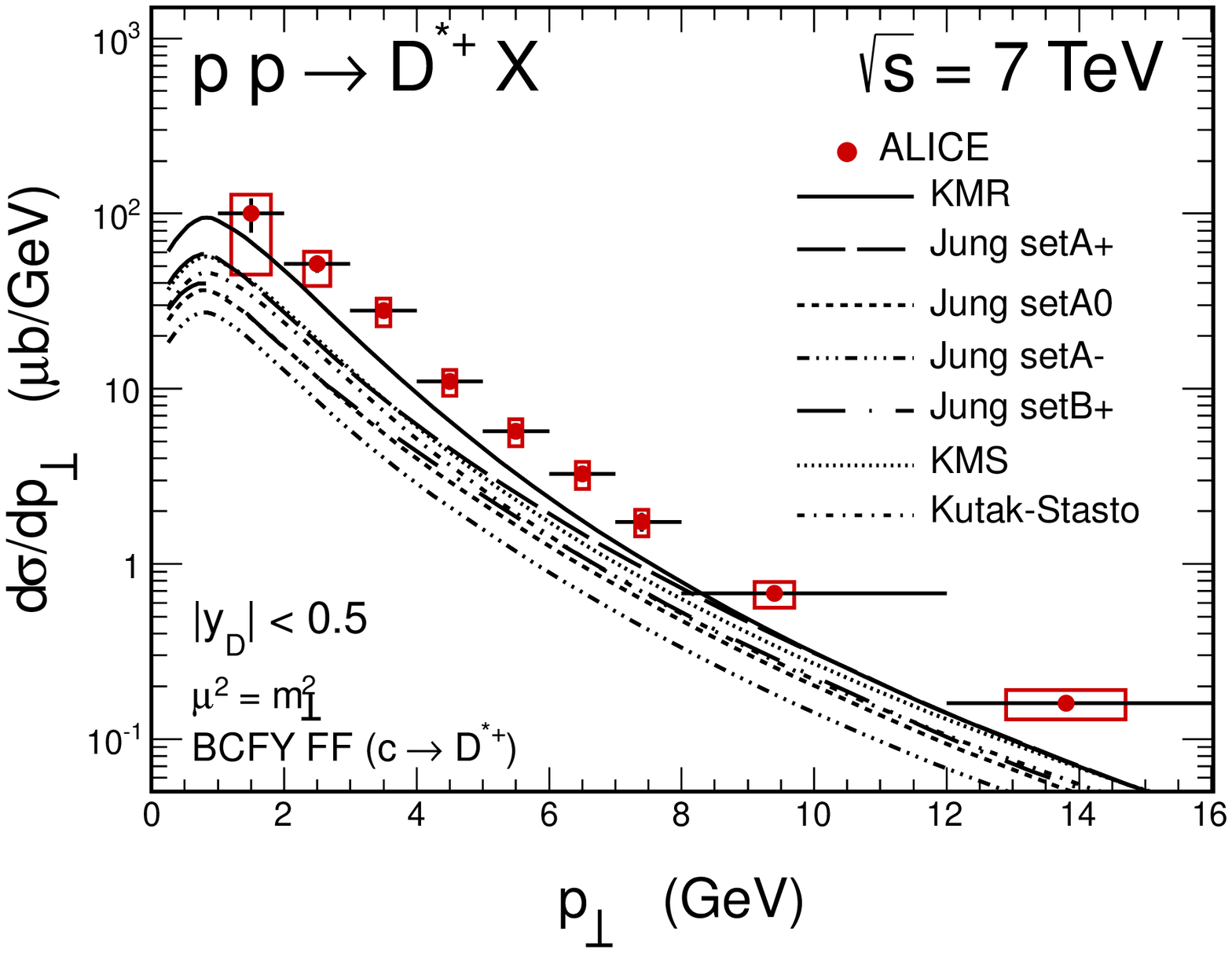}}
\end{minipage}
\hspace{0.5cm}
\begin{minipage}{0.47\textwidth}
 \centerline{\includegraphics[width=1.0\textwidth]{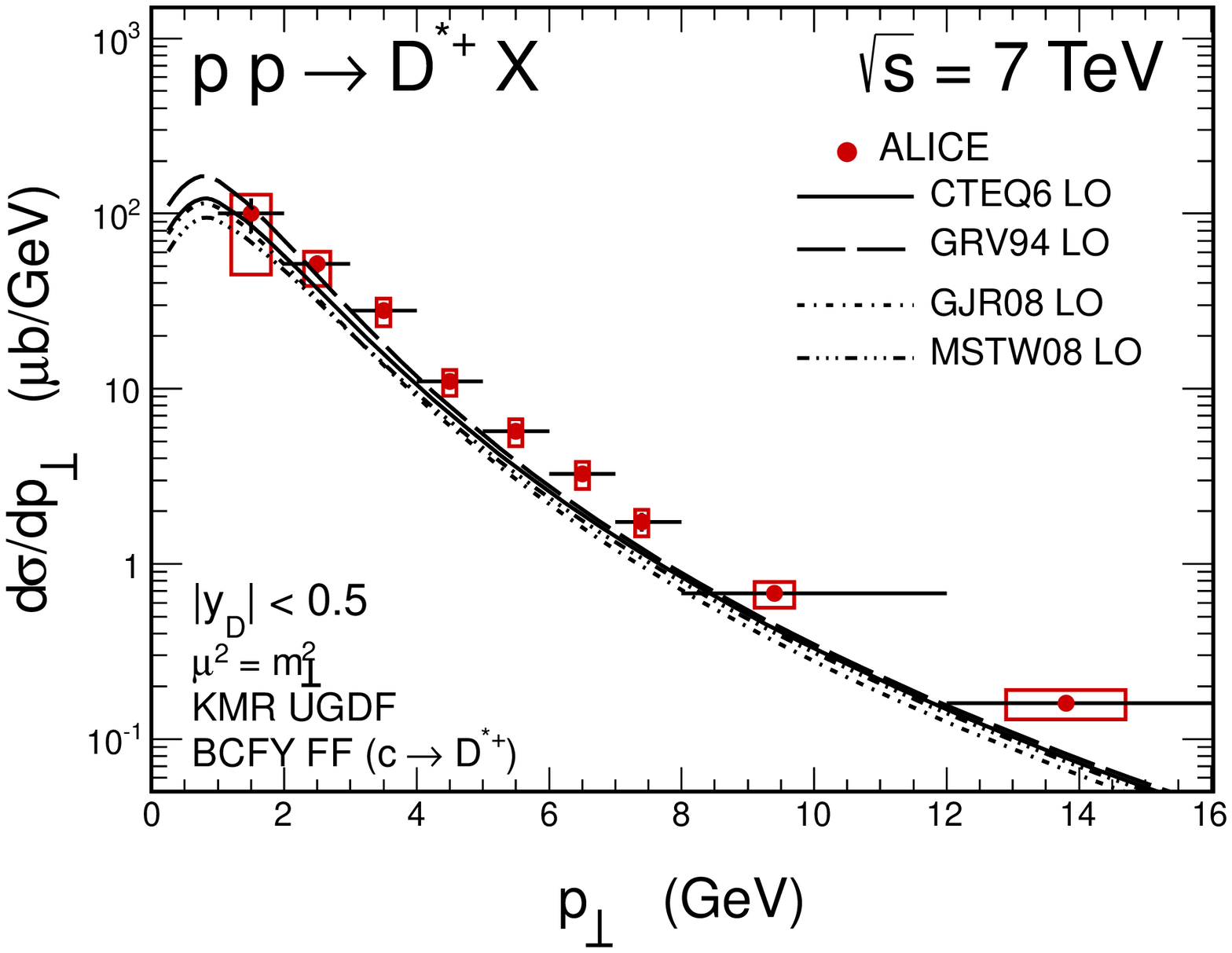}}
\end{minipage}
   \caption{
\small Transverse momentum distribution of $D^{*+}$ mesons for different
UGDFs (left) and the uncertainties due to the choice of collinear
PDFs used in calculating the KMR UGDF (right). }
 \label{fig:pt-alice-D-7}
\end{figure}

\begin{figure}[!h]
\begin{minipage}{0.47\textwidth}
 \centerline{\includegraphics[width=1.0\textwidth]{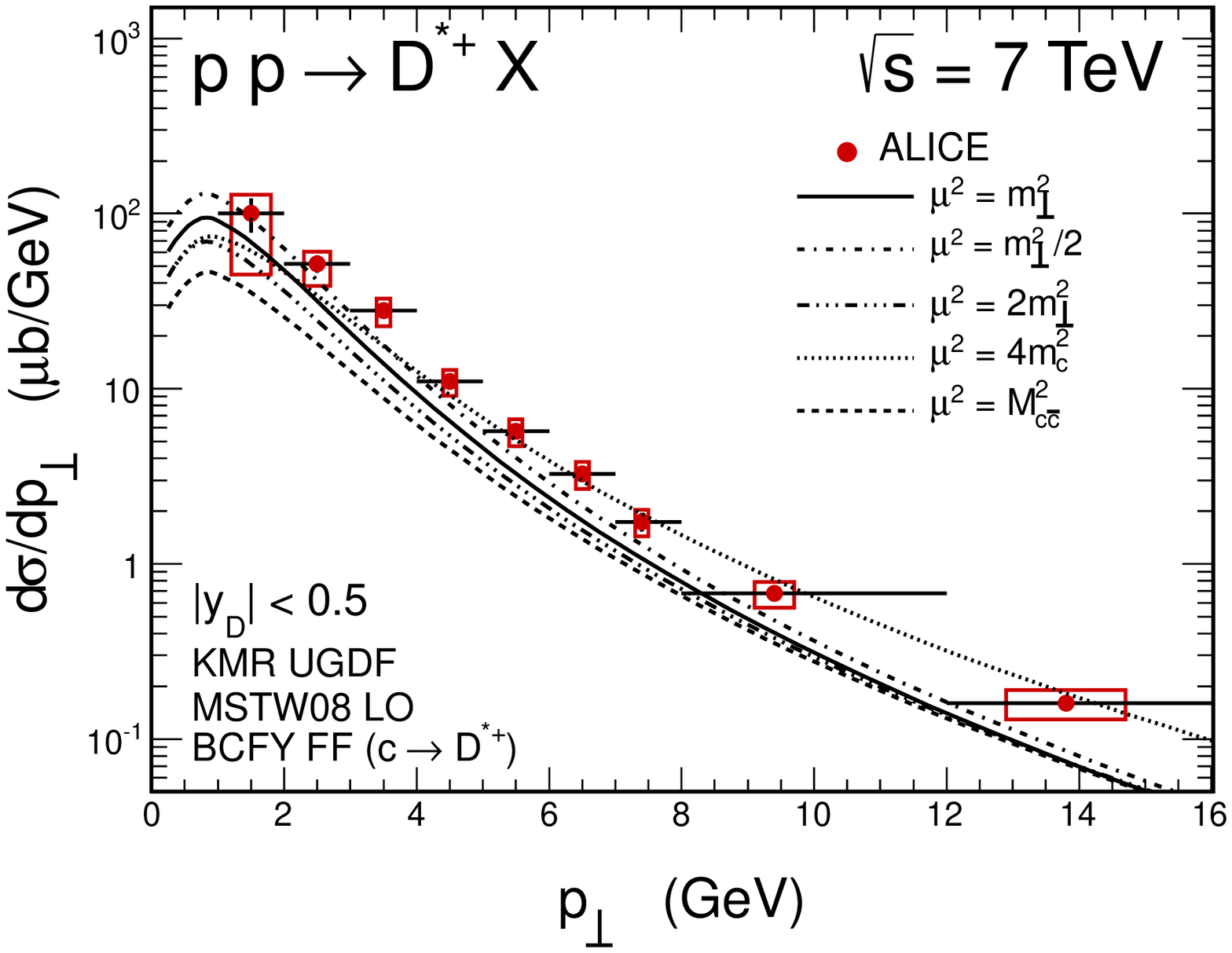}}
\end{minipage}
\hspace{0.5cm}
\begin{minipage}{0.47\textwidth}
 \centerline{\includegraphics[width=1.0\textwidth]{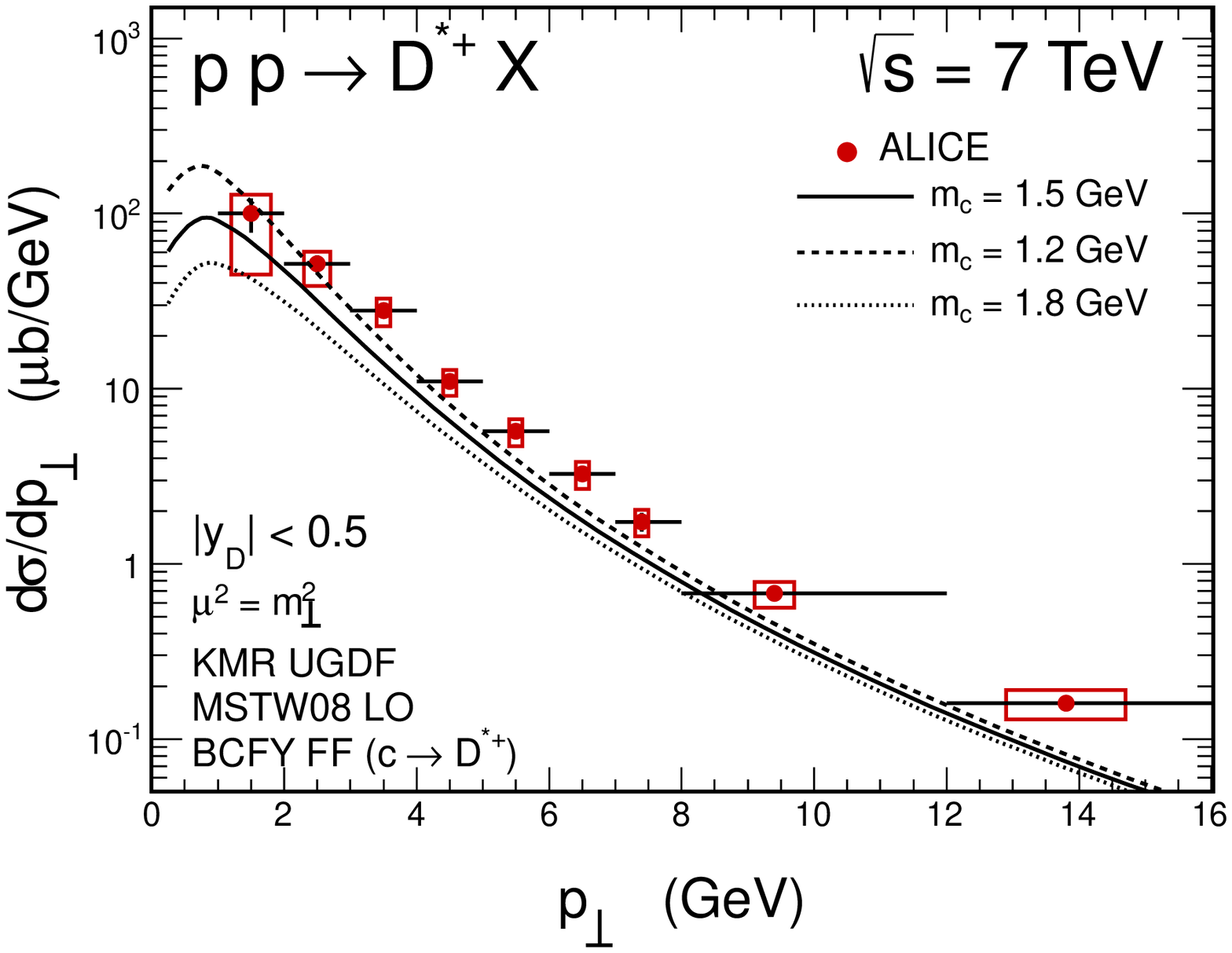}}
\end{minipage}
   \caption{
\small Transverse momentum distribution of $D^{*+}$ mesons. Shown are
uncertainties due to the choice of scales in the KMR UGDF (left panel) and due to 
the quark masses (right panel).}
 \label{fig:pt-alice-D-8}
\end{figure}

\begin{figure}[!h]
\begin{minipage}{0.47\textwidth}
 \centerline{\includegraphics[width=1.0\textwidth]{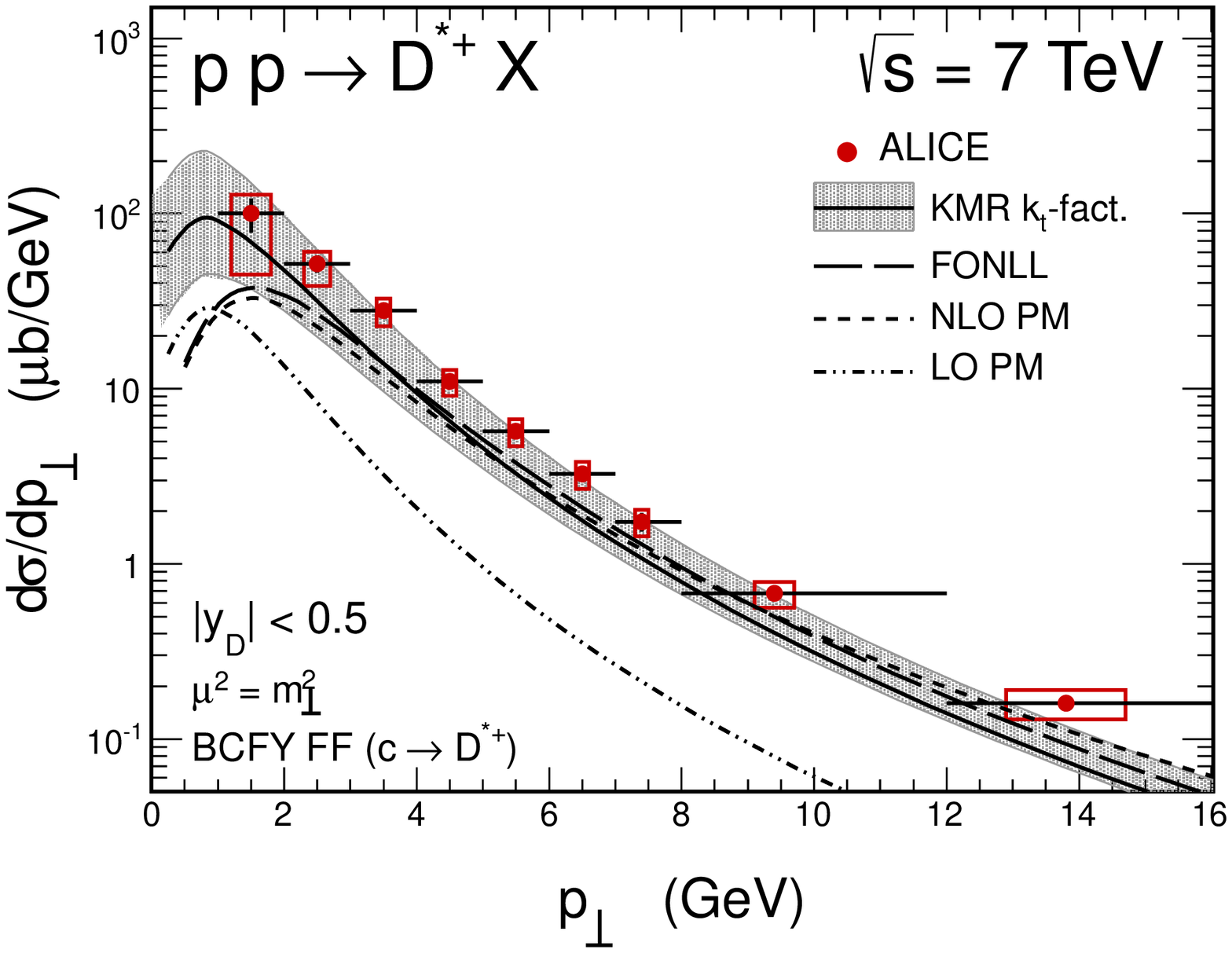}}
\end{minipage}
   \caption{
\small Transverse momentum distribution of $D^{*+}$ mesons for
different approaches known from the literature. The grey band represents overall 
uncertainties of the $k_t$-factorization approach with the KMR UGDF.}
 \label{fig:pt-alice-D-9}
\end{figure}

Now let us consider for the moment distributions for vector mesons $D^{*+}$.
In Fig.~\ref{fig:pt-alice-D-7} we show transverse momentum distributions
of $D^{*+}$ for different UGDFs and uncertainties in calculating 
distributions with the KMR UGDF due to the choice of collinear gluon distributions.
As in the previous cases, the choice of collinear PDFs has some importance
only at small transverse momenta. The same conlcusions as in the cases of pseudoscalar mesons come from
Fig.~\ref{fig:pt-alice-D-8}, where uncertainties due to the scales (left) and related to the quark mass (right)
are presented. In Fig.~\ref{fig:pt-alice-D-9} both of these sources are taken together and our predictions with the KMR UGDF are confronted
once again with LO and NLO PM collinear calculations. Here we use the Braaten et al. model for fragmentation which has, the only one on the market,
parametrization of fragmentation function for the transition of heavy quark into vector meson state.

\begin{figure}[!h]
\begin{minipage}{0.47\textwidth}
 \centerline{\includegraphics[width=1.0\textwidth]{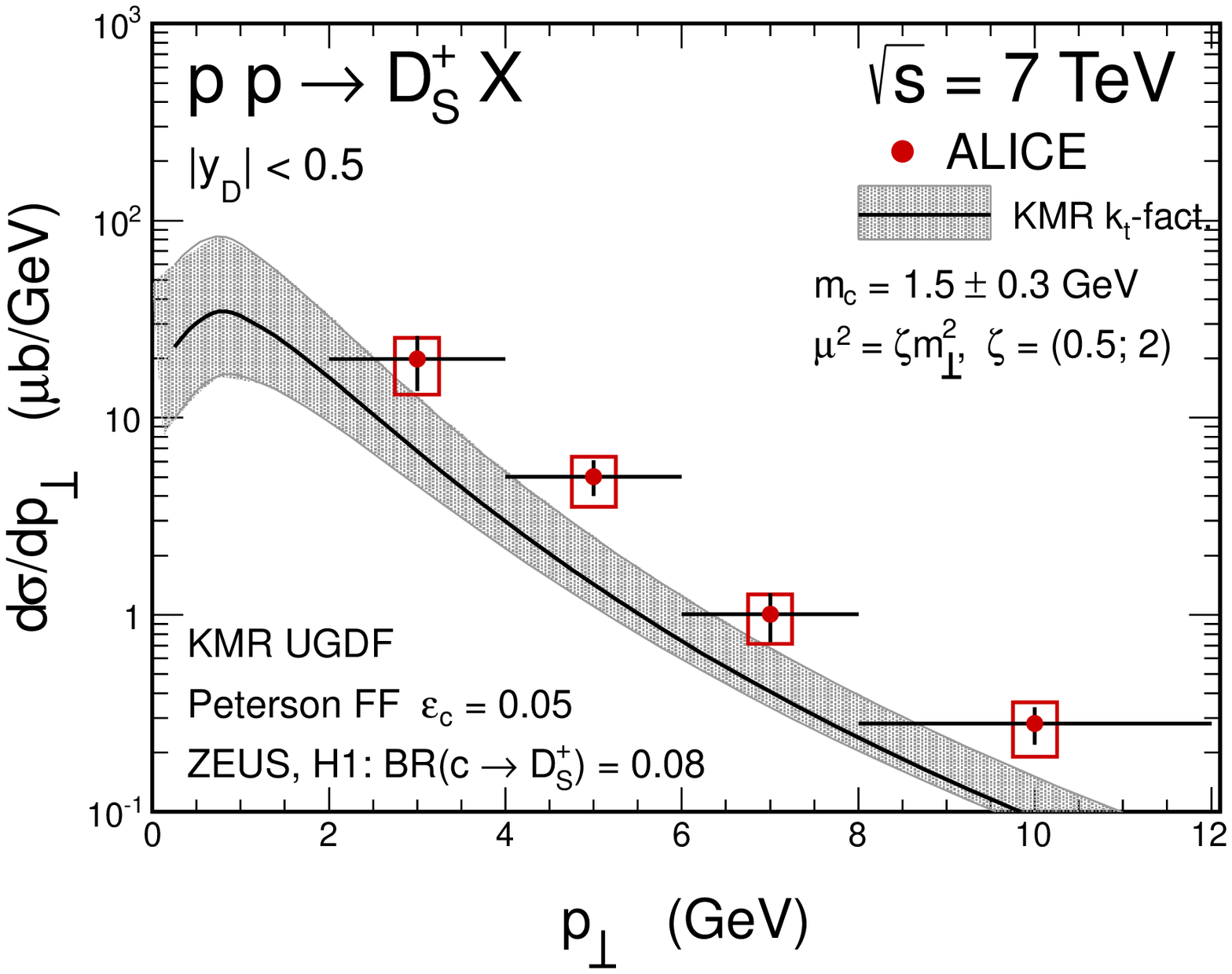}}
\end{minipage}
\hspace{0.5cm}
\begin{minipage}{0.47\textwidth}
 \centerline{\includegraphics[width=1.0\textwidth]{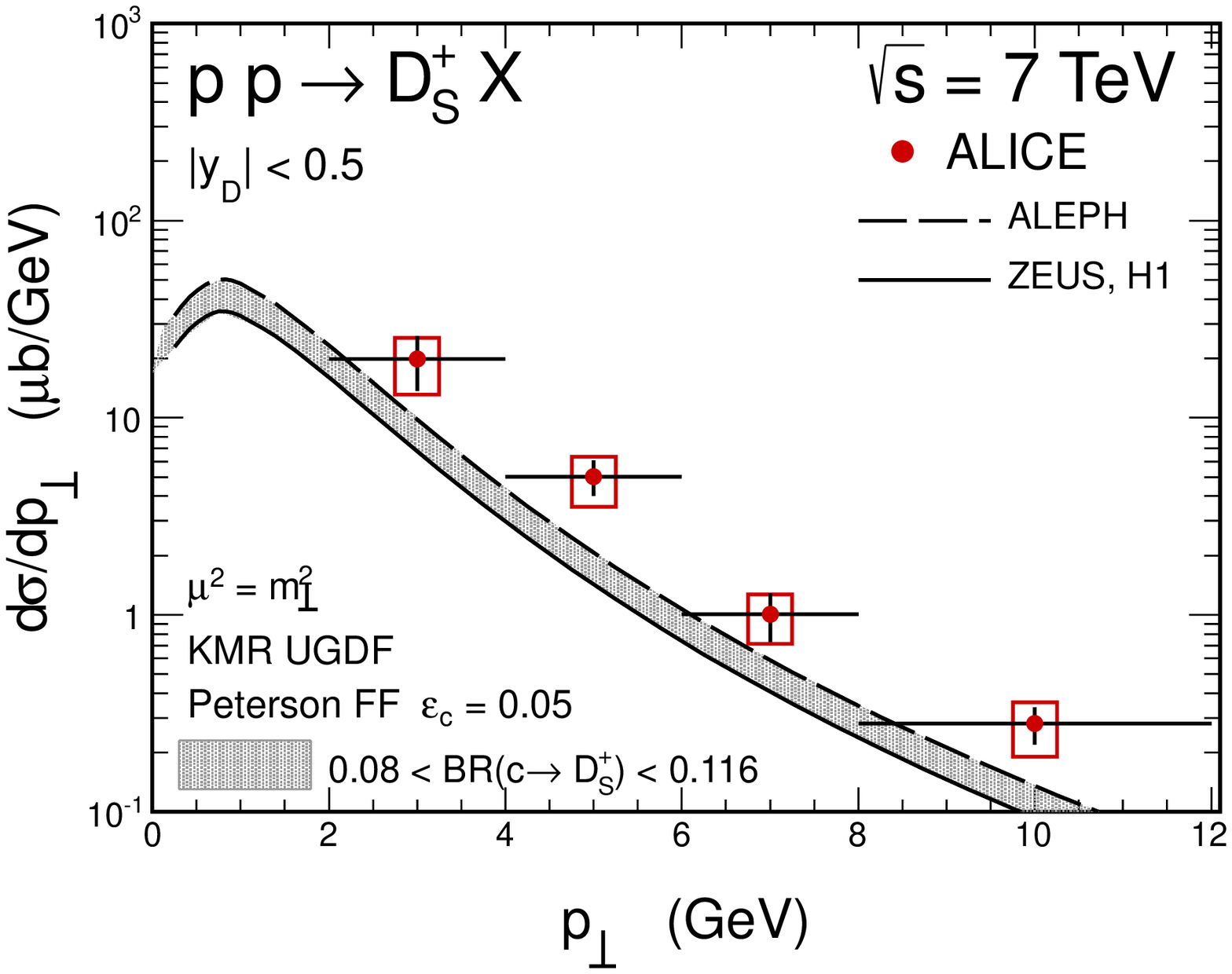}}
\end{minipage}
   \caption{
\small Transverse momentum distribution of $D_s^+$ mesons. We show uncertainties due to the choice of scales and masses (left panel) and uncertainties related to poorly known fragmentation branching fraction BR($c \rightarrow D^{+}_{S}$) (right panel).}
 \label{fig:pt-alice-D-10}
\end{figure}

Finally in Fig.~\ref{fig:pt-alice-D-10} we show distributions for $D_s^+$
mesons, i.e. mesons built of charm and strange quarks/antiquarks. 
The corresponding cross section is considerably smaller than for 
the charm mesons containing light (up or down) quarks/antiquarks.
The general situation is, however, very similar. The KMR UGDF provides
the best agreement with the ALICE data. Results for other UGDFs are
much below the experimental data which means, in our opinion, that they do not pass
the powerfull test. In the case of $D^{+}_{S}$ meson a different, quite important
source of uncertainties appears, namely the poorly known fragmentation fractions 
BR($c \rightarrow D^{+}_{S}$). Changing the value from $0.08$ (ZEUS, H1) to
$0.116$ (ALEPH) a significant enhancement of the theoretical predictions is achieved. 

\subsection{ATLAS}

The ATLAS experiment covers much broader range of pseudorapidities than ALICE.
As a consequence one tests broader region of longitudinal momentum fractions 10$^{-4} < x_1,x_2 <$ 10$^{-2}$.
The gluon distributions in this range of $x_1$ and $x_2$ values carried by gluons 
are rather well known. Also application of the known UGDFs from the literature
should be reliable.

\begin{figure}[!h]
\begin{minipage}{0.47\textwidth}
 \centerline{\includegraphics[width=1.0\textwidth]{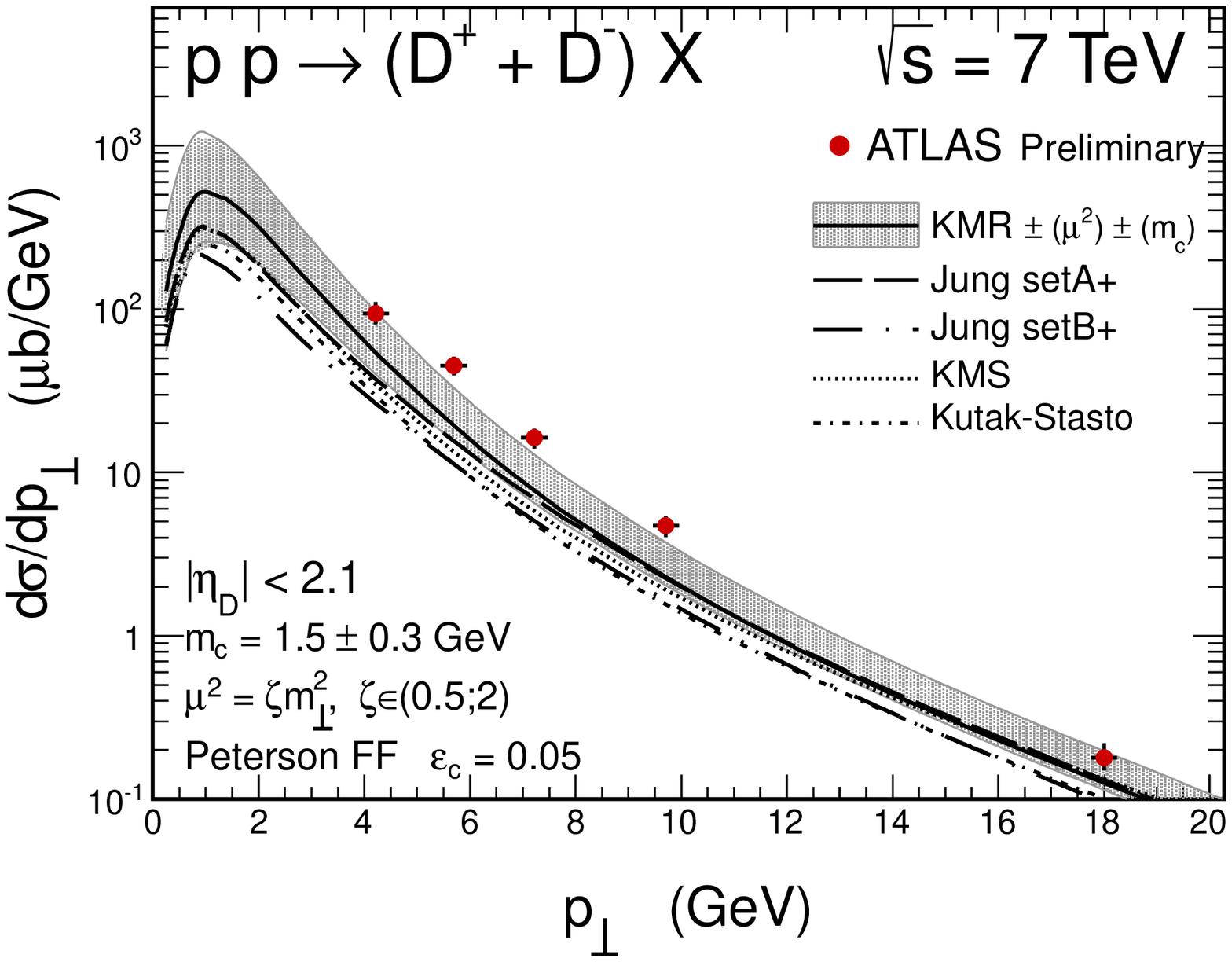}}
\end{minipage}
\hspace{0.5cm}
\begin{minipage}{0.47\textwidth}
 \centerline{\includegraphics[width=1.0\textwidth]{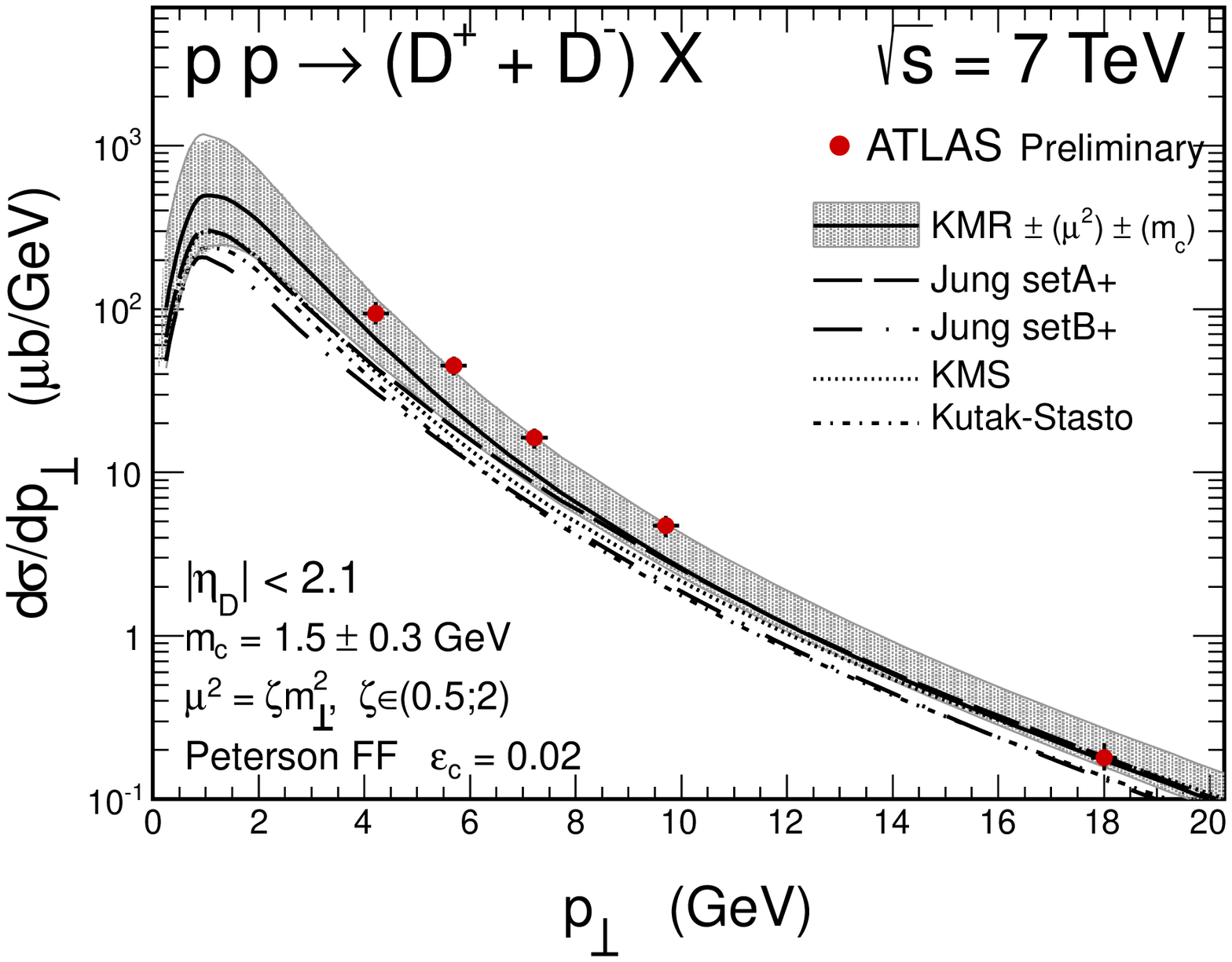}}
\end{minipage}
   \caption{
\small Transverse momentum distribution of $D^{\pm}$ mesons for different
UGDFs from the literature compared with the preliminary ATLAS
experimental data for $\varepsilon_c = 0.05$ (left) and $\varepsilon_c = 0.02$ (right).
}
 \label{fig:pt-atlas-D-1a}
\end{figure}

\begin{figure}[!h]
\begin{minipage}{0.47\textwidth}
 \centerline{\includegraphics[width=1.0\textwidth]{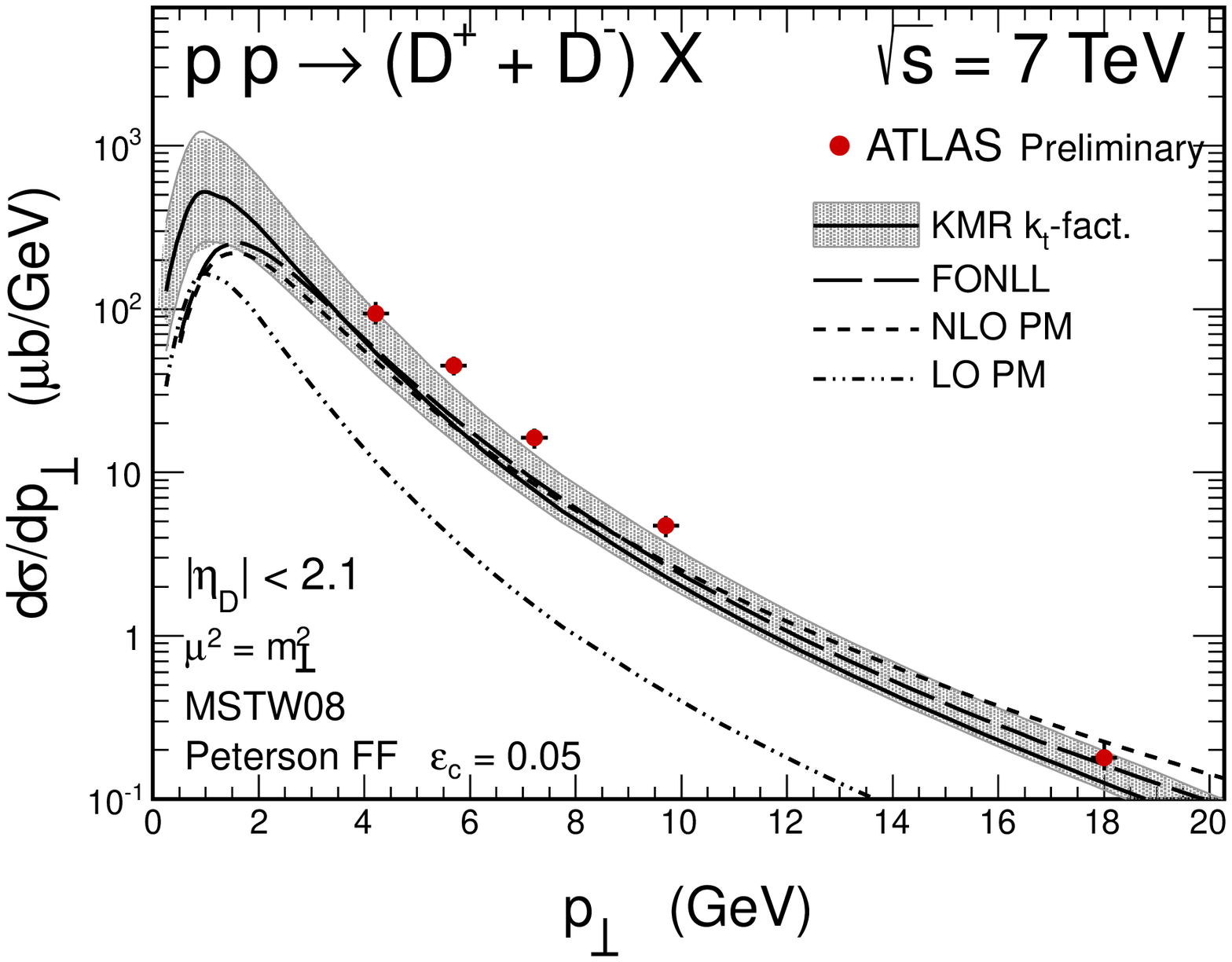}}
\end{minipage}
\hspace{0.5cm}
\begin{minipage}{0.47\textwidth}
 \centerline{\includegraphics[width=1.0\textwidth]{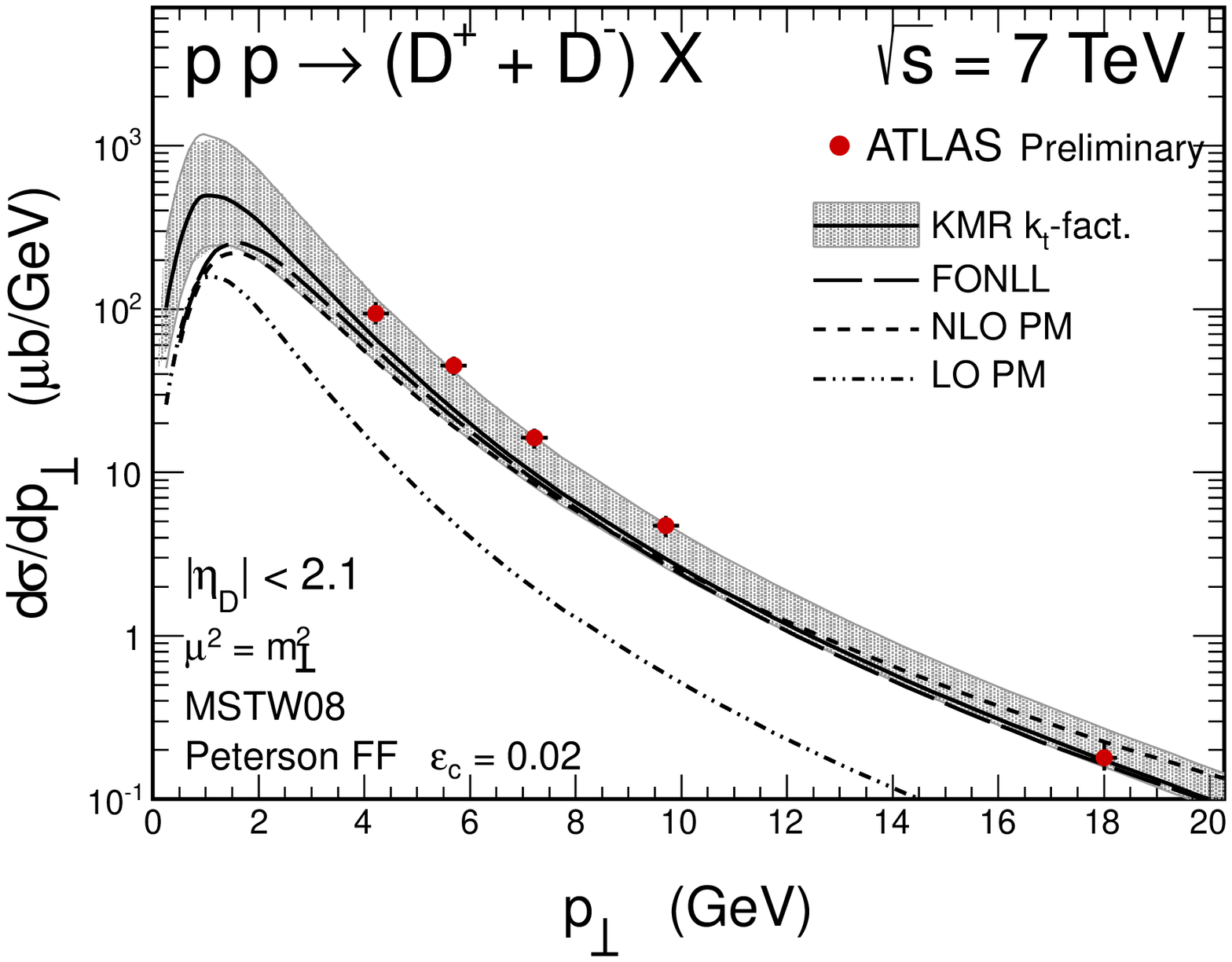}}
\end{minipage}
   \caption{
\small Transverse momentum distribution of $D^{\pm}$ mesons for different
standard approaches compared with the ATLAS experimental data \cite{ATLASincD} for $\varepsilon_c = 0.05$ (left) and $\varepsilon_c = 0.02$ (right).
The details are specified in the figures.}
 \label{fig:pt-atlas-D-2a}
\end{figure}

Fig.~\ref{fig:pt-atlas-D-1a} shows transverse momentum distributions of charged pseudoscalar $D^{\pm}$ mesons for different
models of unintegrated distributions, for $\varepsilon_c = 0.05$ (left) and $\varepsilon_c = 0.02$ (right).
General situation is very similar as for the ALICE experiment although the agreement is somewhat worse. Only the upper limit of the KMR result is compatible
with the ATLAS experimental data. This may be caused by much broader
range of pseudorapidities in the case of the ATLAS detector. Potentially, this can be related to double-parton scattering effects to be discussed
elsewhere \cite{MS2013}. Also the other standard approaches give results below the ATLAS data as can be seen in Fig.~\ref{fig:pt-atlas-D-2a}. 

\begin{figure}[!h]
\begin{minipage}{0.47\textwidth}
 \centerline{\includegraphics[width=1.0\textwidth]{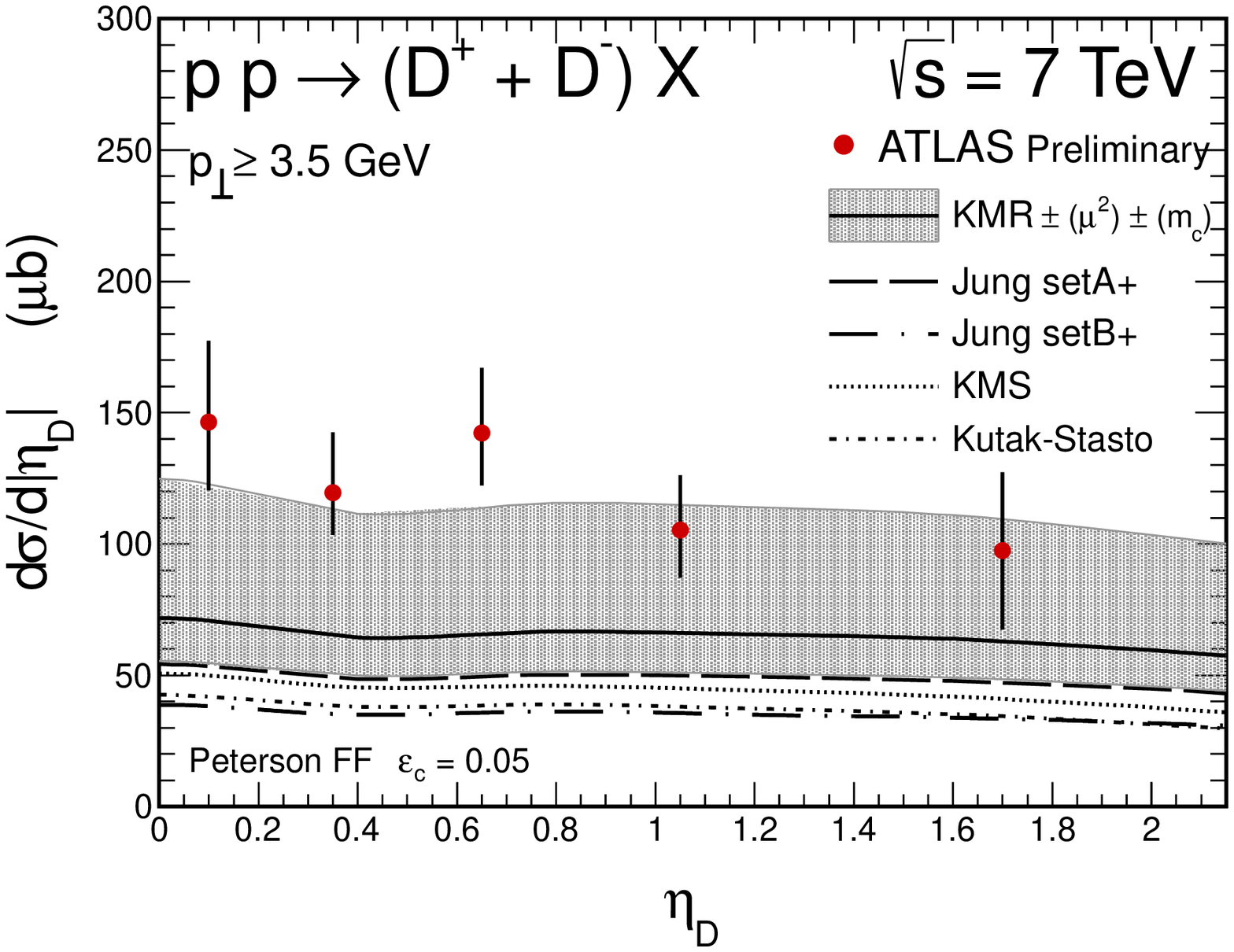}}
\end{minipage}
\hspace{0.5cm}
\begin{minipage}{0.47\textwidth}
 \centerline{\includegraphics[width=1.0\textwidth]{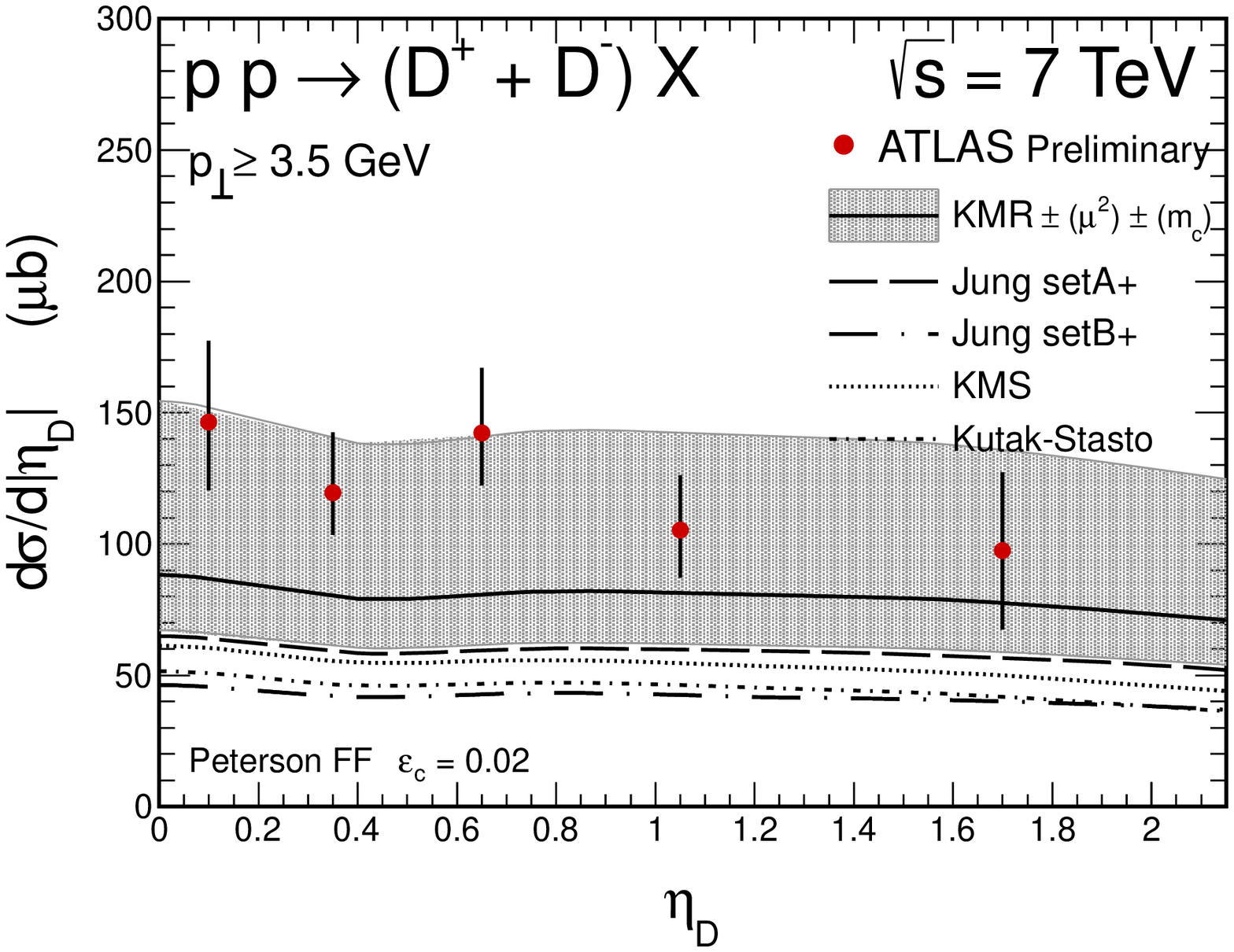}}
\end{minipage}
   \caption{
\small Distribution in $D^{\pm}$ meson pseudorapidity. The results for
different UGDFs are compared with the ATLAS preliminary data \cite{ATLASincD}
for different values of the parameter
$\varepsilon_c$ of the Peterson fragmentation function: $\varepsilon_c =$ 0.05
(left), $\varepsilon_c = $ 0.02 (right).
}
 \label{fig:eta-atlas-D-1b}
\end{figure}

\begin{figure}[!h]
\begin{minipage}{0.47\textwidth}
 \centerline{\includegraphics[width=1.0\textwidth]{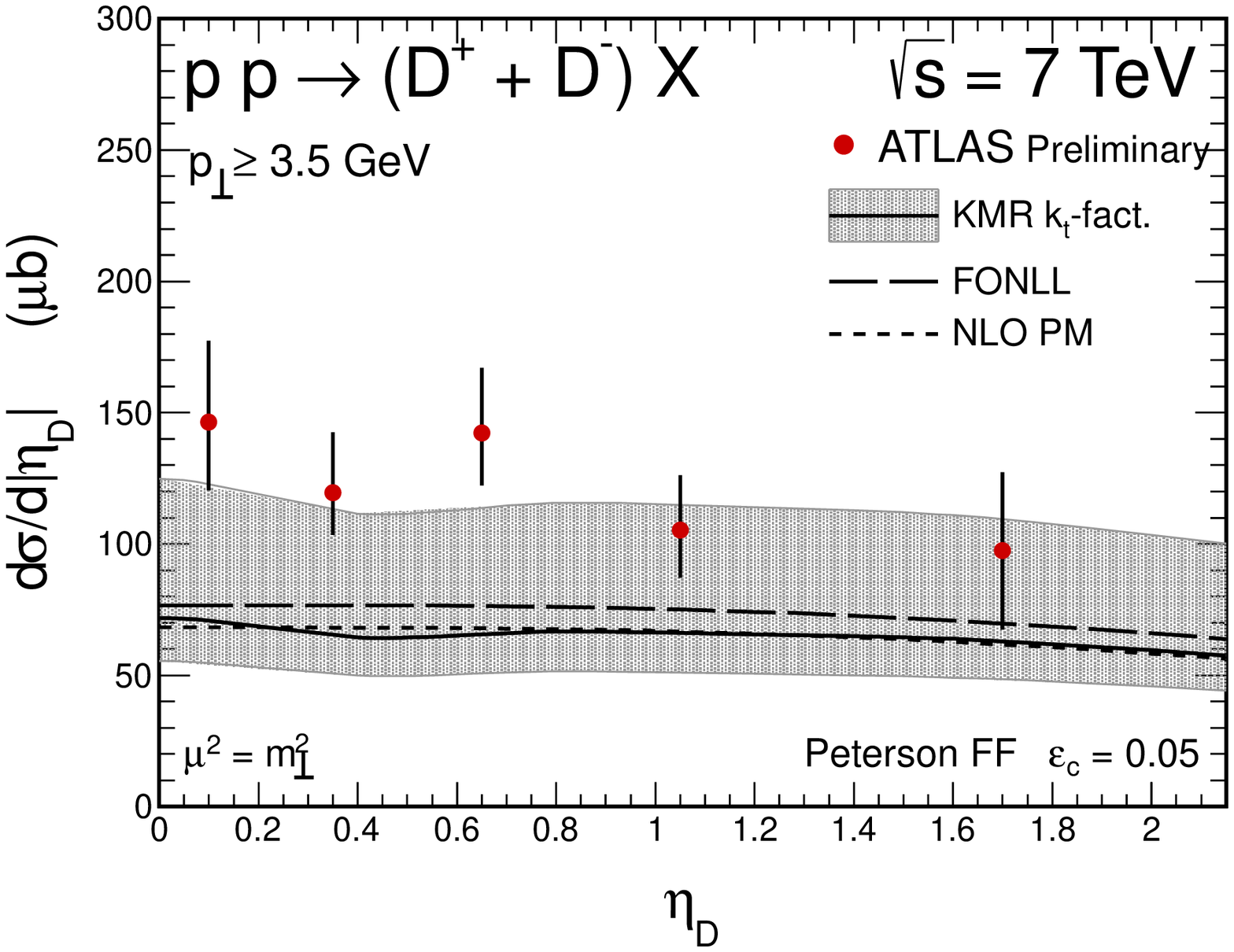}}
\end{minipage}
\hspace{0.5cm}
\begin{minipage}{0.47\textwidth}
 \centerline{\includegraphics[width=1.0\textwidth]{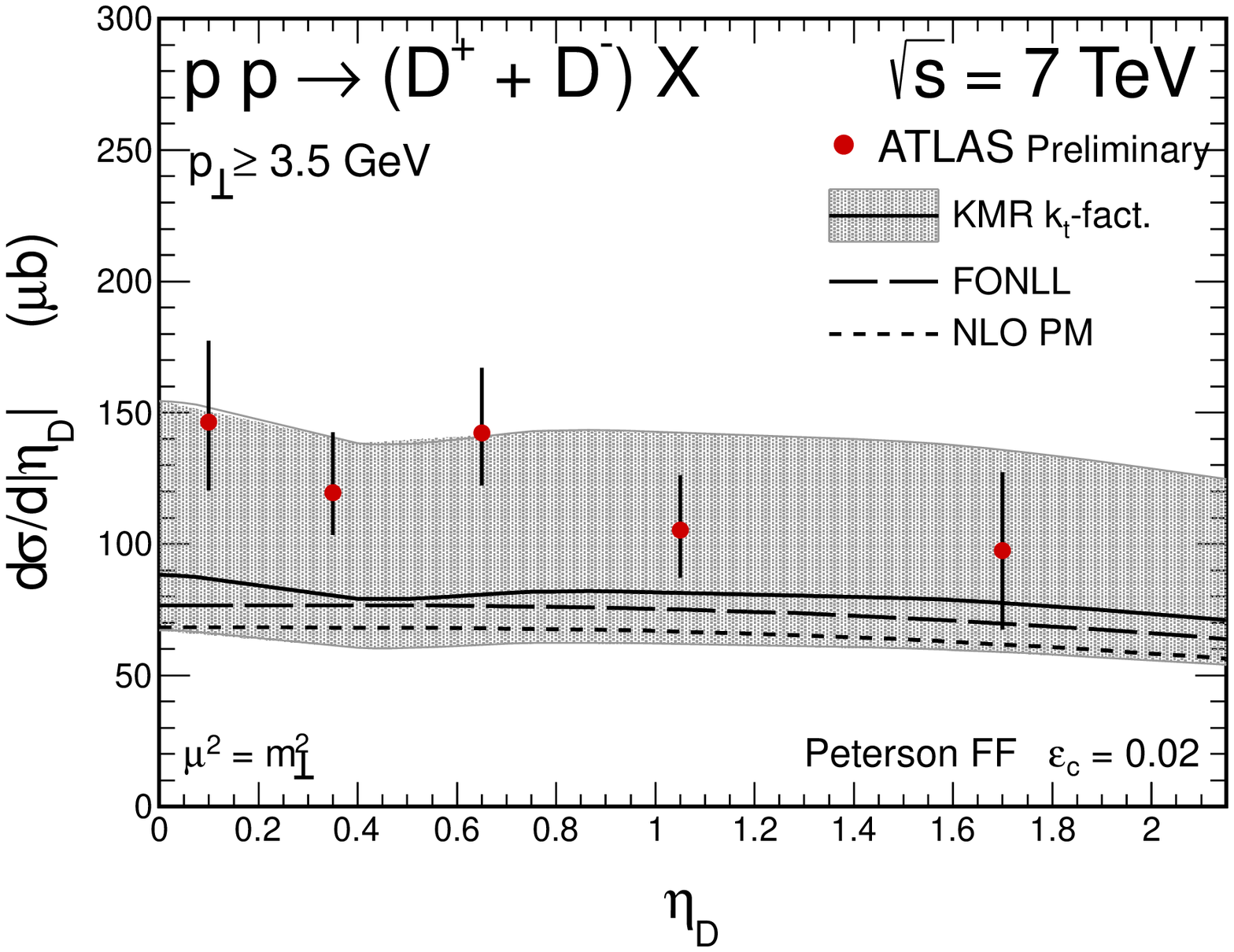}}
\end{minipage}
   \caption{
\small Distribution in $D^{\pm}$ meson pseudorapidity. The results of our 
calculation are compared with those for other calculations and with the 
ATLAS preliminary data \cite{ATLASincD} for different values of the parameter
$\varepsilon_c$ of the Peterson fragmentation function: $\varepsilon_c =$ 0.05
(left),
$\varepsilon_c = $ 0.02 (right).
}
 \label{fig:eta-atlas-D-2b}
\end{figure}

Due to fairly large span of pseudorapidities the ATLAS collaboration can extract
also pseudorapidity distributions.
We wish to show now also results for charm meson pseudorapidity
distributions. In Fig.~\ref{fig:eta-atlas-D-1b} and
Fig.~\ref{fig:eta-atlas-D-2b} we show pseudorapidity distributions for charged $D^{\pm}$
meson. These distributions are rather flat. The results
are also compared to recent (preliminary) ATLAS data. Only the upper limits of large error bars of the theoretical results obtained with the KMR distributions are consistent
with the ATLAS data. The results with other UGDFs clearly
underpredict the experimental data.

\begin{figure}[!h]
\begin{minipage}{0.47\textwidth}
 \centerline{\includegraphics[width=1.0\textwidth]{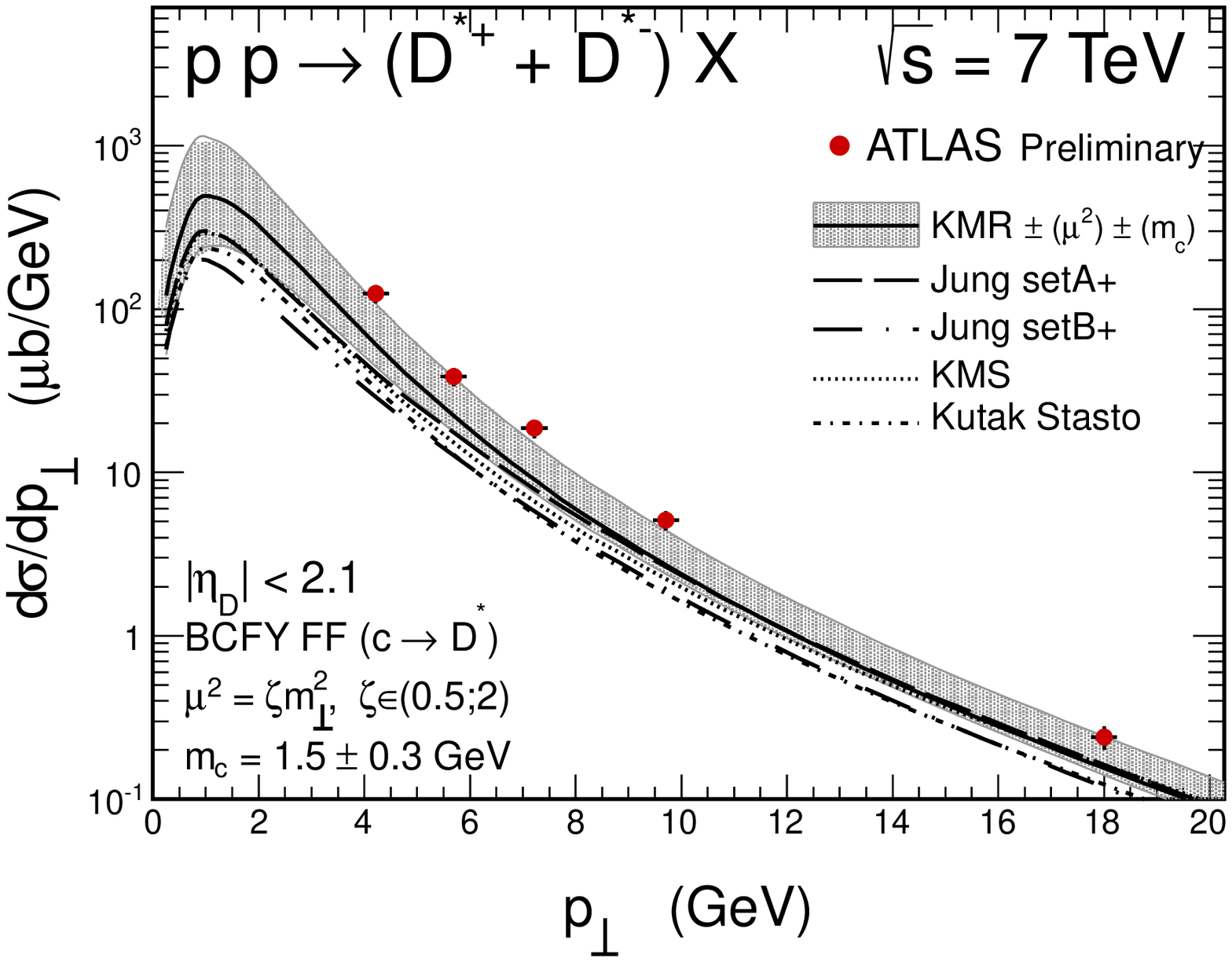}}
\end{minipage}
\hspace{0.5cm}
\begin{minipage}{0.47\textwidth}
 \centerline{\includegraphics[width=1.0\textwidth]{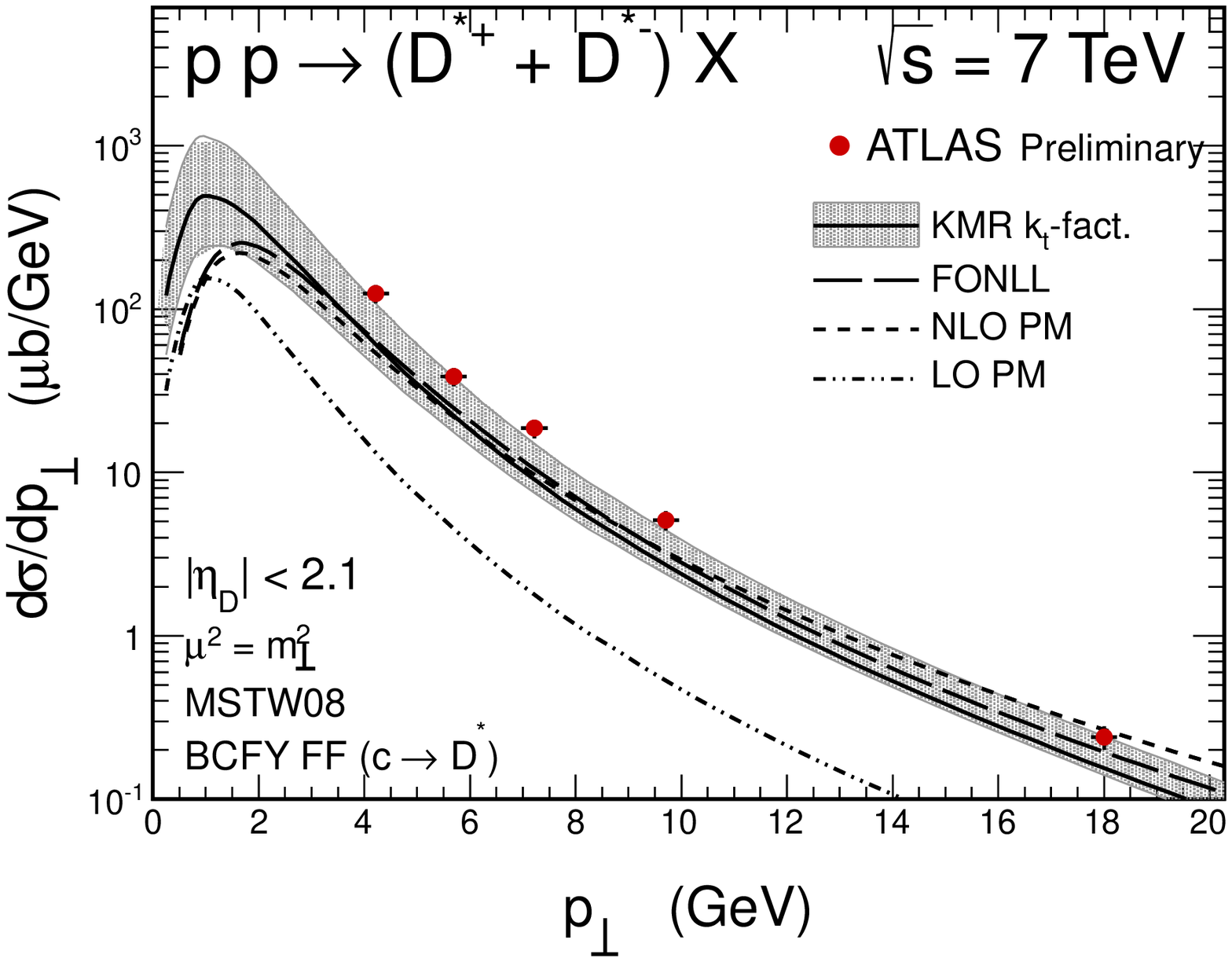}}
\end{minipage}
   \caption{
\small  Distribution in $D^{*+}$ meson transverse momentum. The results of our 
calculation are compared with the ATLAS preliminary data \cite{ATLASincD}. In the left panel we show results for 
different UGDFs and in the right panel our results are compared with other approaches. }
 \label{fig:pt-atlas-D-3}
\end{figure}

As for pseudoscalar mesons above, in Fig.~\ref{fig:pt-atlas-D-3} we show transverse momentum distributions
for charged vector mesons. The situation is pretty much the same as for pseudoscalar charged mesons discussed previously in Fis.~\ref{fig:pt-atlas-D-1a} and Fig.~\ref{fig:pt-atlas-D-2a}.

In Fig.~\ref{fig:eta-atlas-D-2b} we compare results obtained within the
$k_t$-factorization approach (grey band) with results obtained within
other approaches. The central value of the $k_t$-factorization approach with the KMR UGDF is consistent with the FONLL and NLO PM predictions.

\subsection{LHCb}

Finally let us focus on the measurements in the forward rapidity region $2 < y < 4.5$.
Recently the LHCb collaboration presented first results for
the production of $D^0$, $D^+$, $D^{*+}$ and $D_s^+$ mesons \cite{LHCbincD}.
In this region of phase space one tests asymmetric gluon longitudinal
momentum fractions: $x_1 \sim$ 10$^{-5}$ and $x_2 >$ 10$^{-2}$ 
(see Fig.~\ref{fig:ylogx-kmr}). This is
certainly more difficult region for reliable calculation and 
interpretation of experimental data. First of all gluon distributions
were never tested at such small $x_1$ values. Secondly many UGDFs from
the literature may be not good enough for $x_2 >$ 10$^{-2}$.
Therefore some care in interpreting the results is required.

\begin{figure}[!h]
\begin{minipage}{0.47\textwidth}
 \centerline{\includegraphics[width=1.0\textwidth]{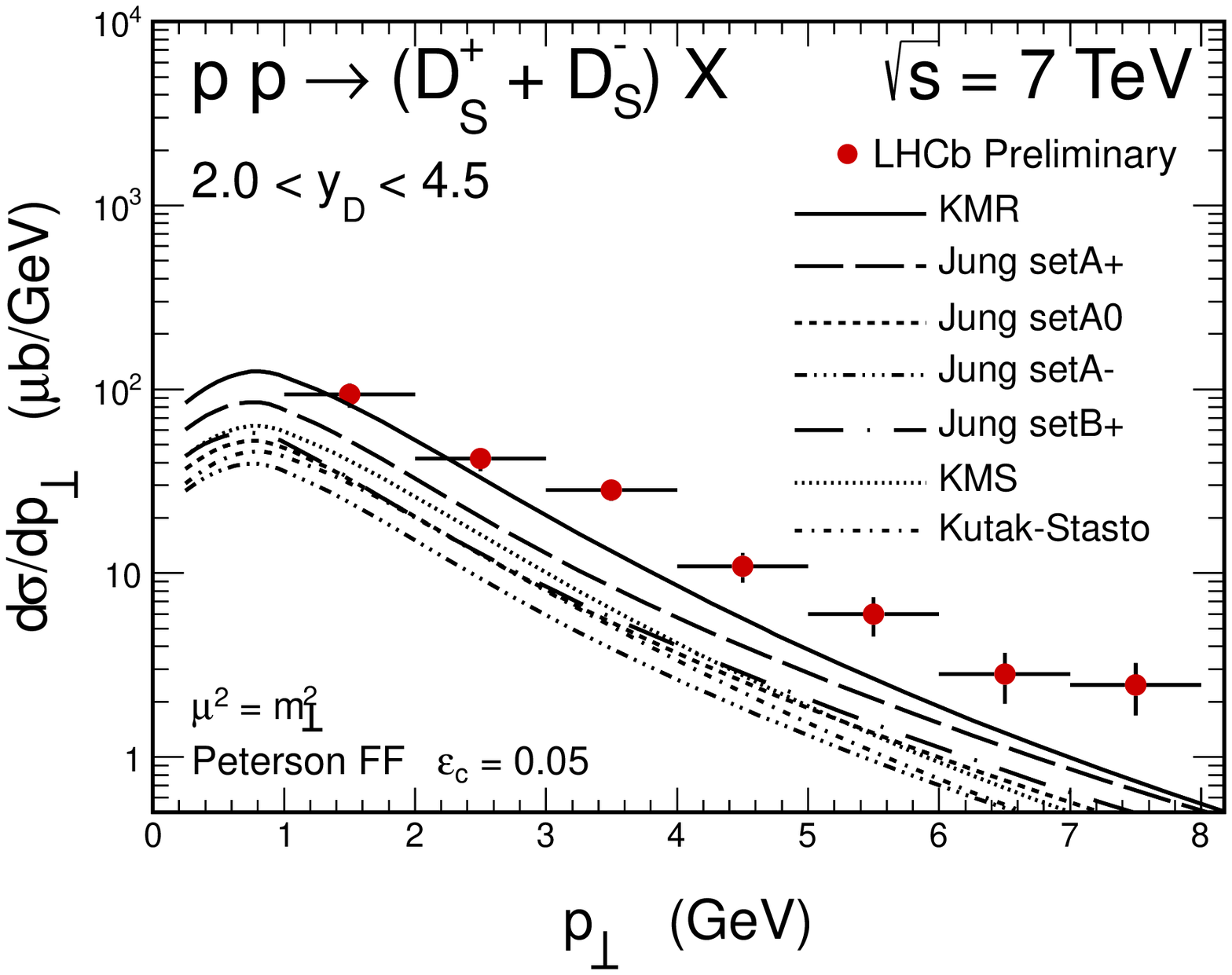}}
\end{minipage}
\hspace{0.5cm}
\begin{minipage}{0.47\textwidth}
 \centerline{\includegraphics[width=1.0\textwidth]{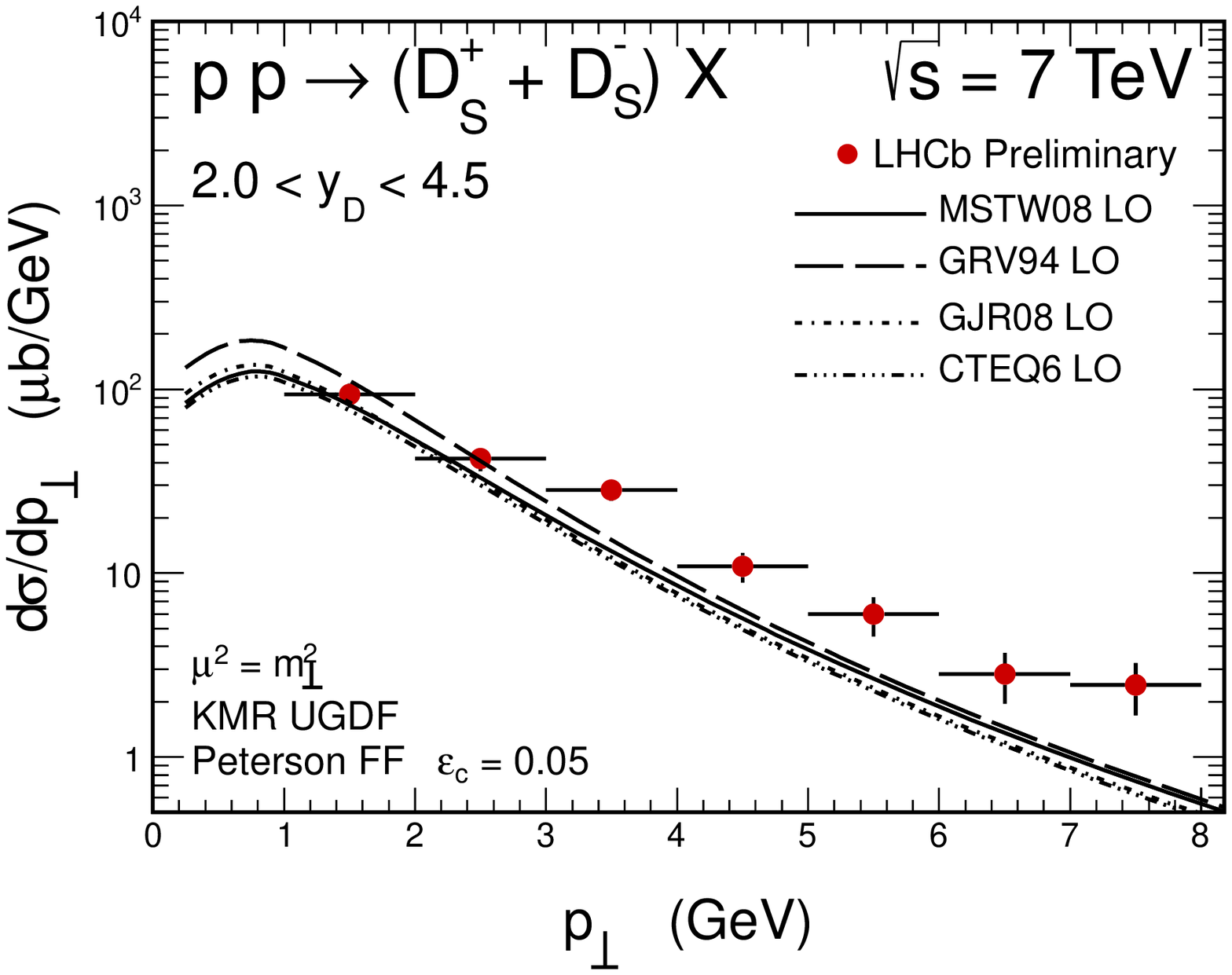}}
\end{minipage}
   \caption{
\small Transverse momentum distribution of $D_s^{\pm}$ mesons for
different UGDFs from the literature (left) and the dependence on the
choice of collinear gluon distribution functions for the KMR UGDF
(right). The results of calculation are compared with the LHCb
collaboration data.
}
 \label{fig:pt-lhcb-D-1}
\end{figure}

\begin{figure}[!h]
\begin{minipage}{0.47\textwidth}
 \centerline{\includegraphics[width=1.0\textwidth]{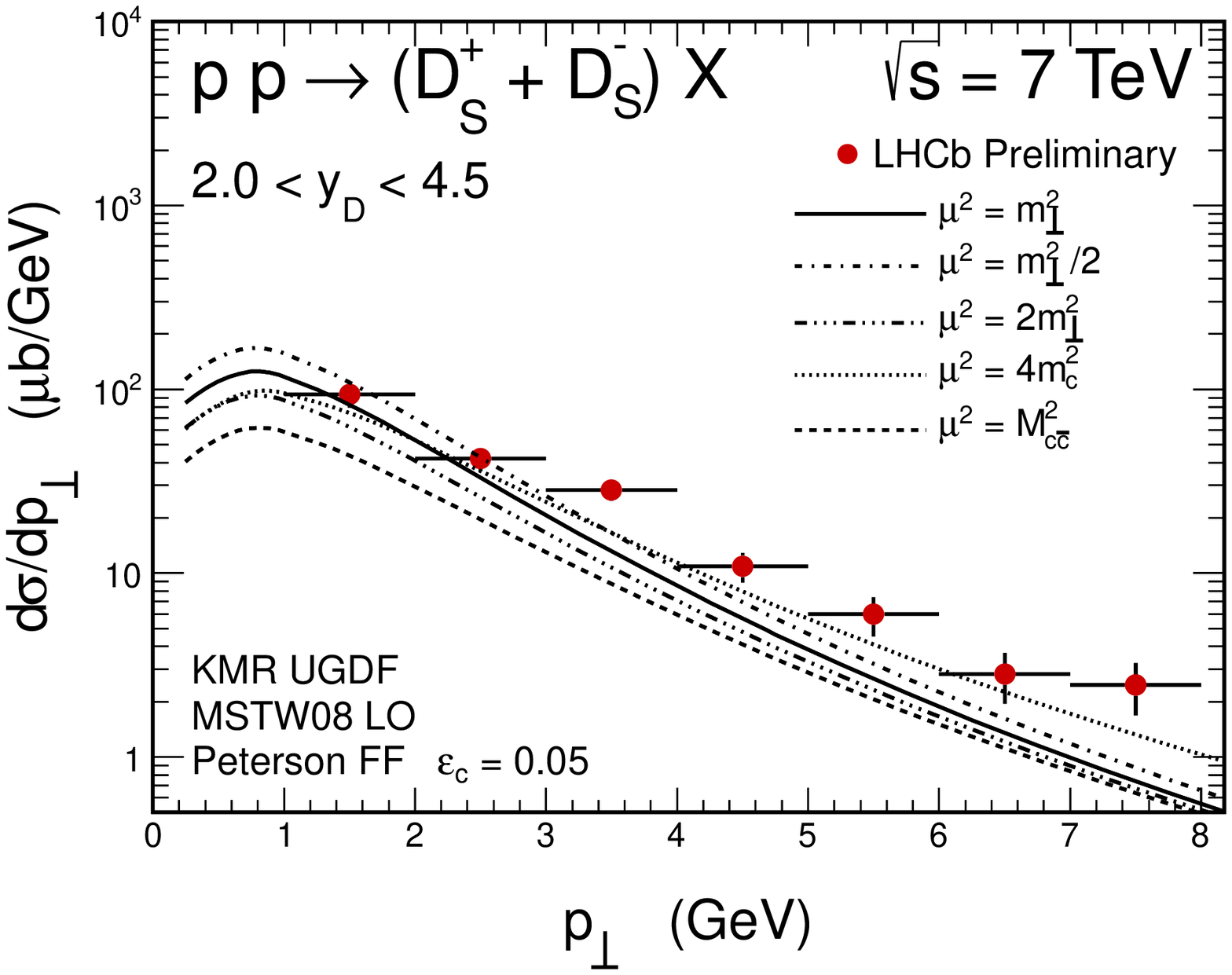}}
\end{minipage}
\hspace{0.5cm}
\begin{minipage}{0.47\textwidth}
 \centerline{\includegraphics[width=1.0\textwidth]{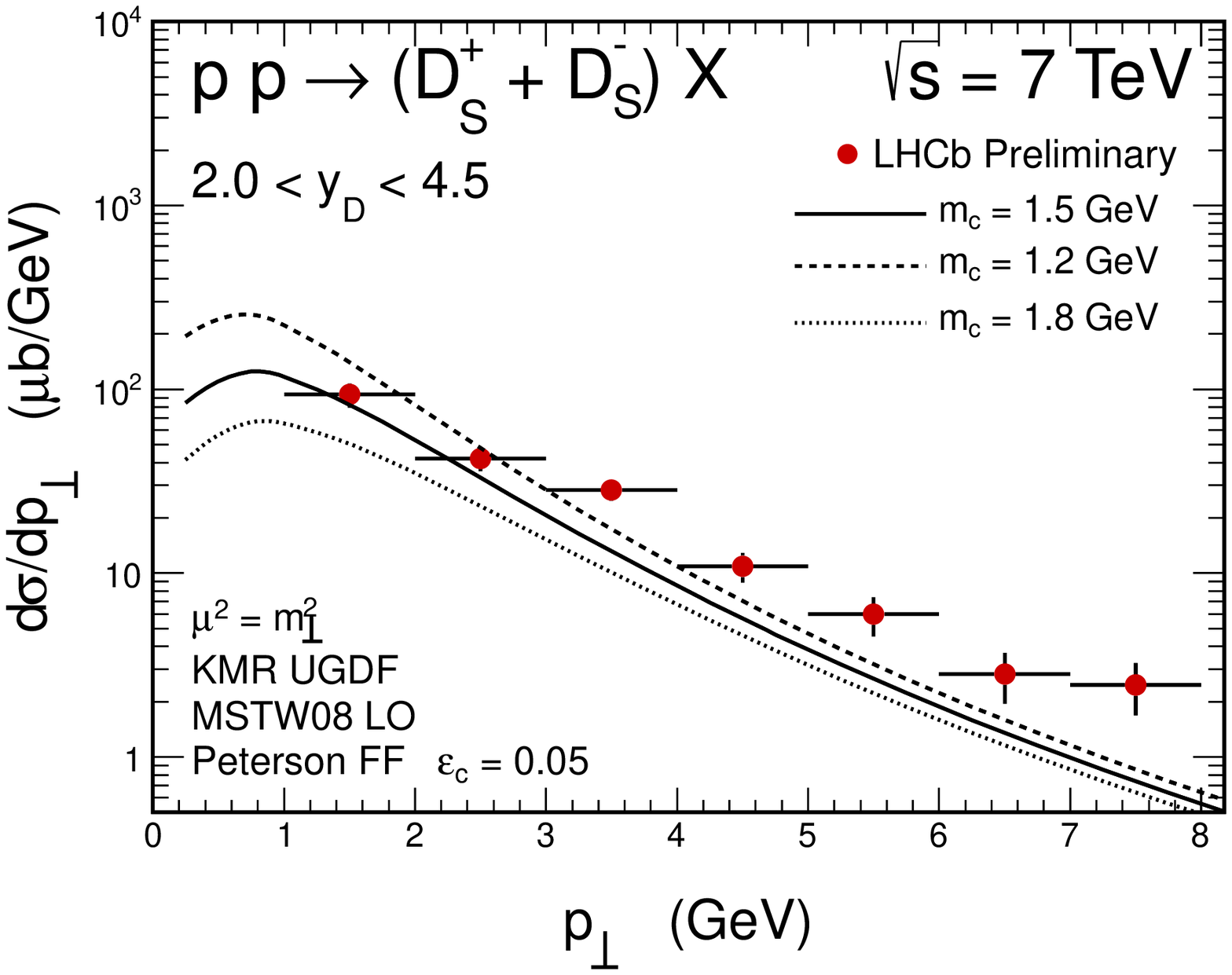}}
\end{minipage}
   \caption{
\small Uncertainties of the theoretical predictions due to the choice of scales for the KMR UGDF (left) and
due to charm quark mass (right). }
 \label{fig:pt-lhcb-D-2}
\end{figure}

\begin{figure}[!h]
\begin{minipage}{0.47\textwidth}
 \centerline{\includegraphics[width=1.0\textwidth]{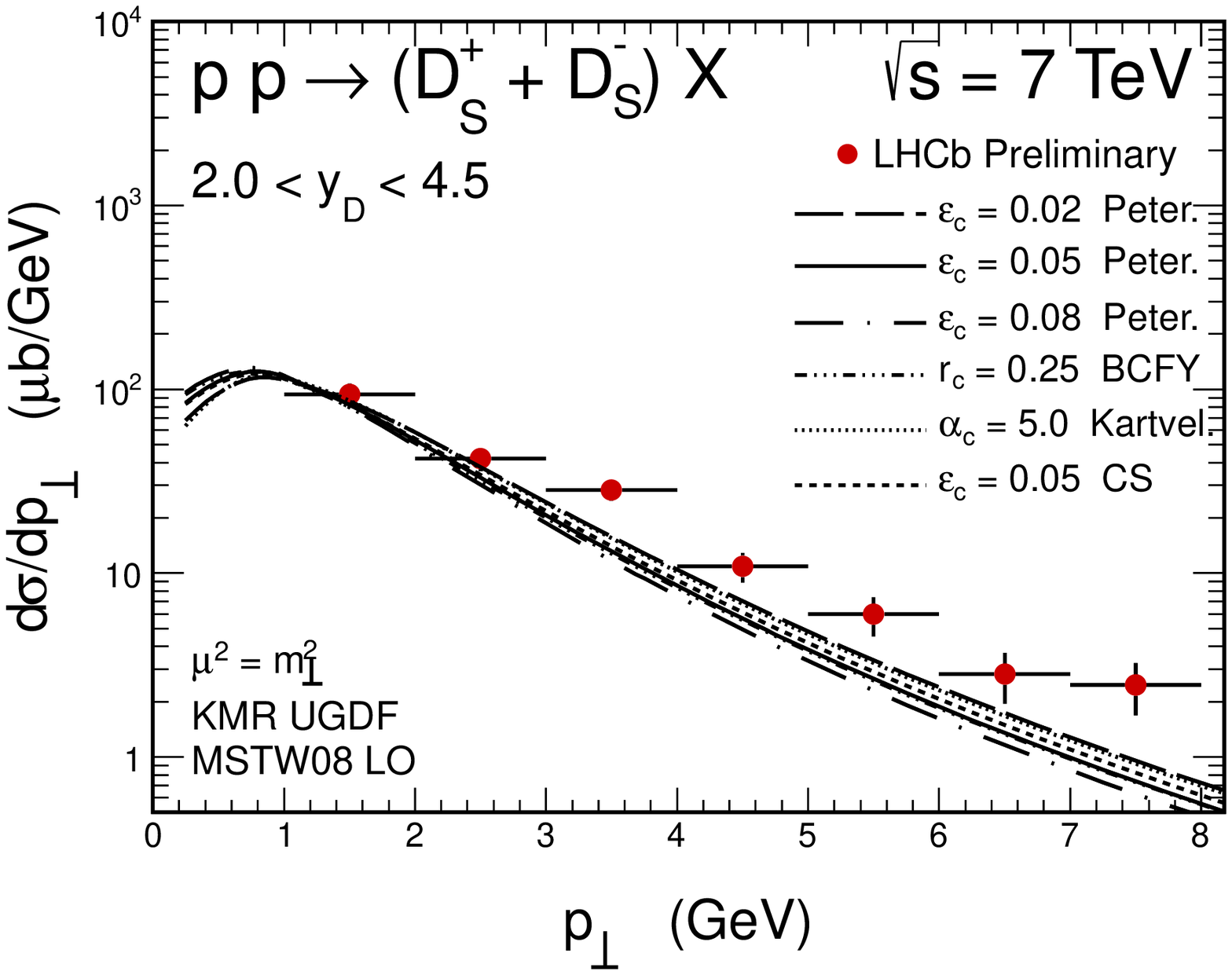}}
\end{minipage}
   \caption{
\small Uncertainties in the fragmentation of $c \to D_s^+$. We show
results obtained with different fragmentation functions from the literature.
}
 \label{fig:pt-lhcb-D-3}
\end{figure}

The LHCb, similar as ALICE, has measured also distributions of rather
rarely produced $D_s^+$ mesons. 
We start from transverse momentum distributions for $D_s^{\pm}$ mesons.
In Fig.~\ref{fig:pt-lhcb-D-1} we present
distributions for different UGDF from the literature and uncertainties
for the KMR UGDF related to the choice of standard PDFs.
In Fig.~\ref{fig:pt-lhcb-D-2} we show uncertainties related to
the choice of factorization/renormalization scale and due to the choice of quark
masses and in Fig.~\ref{fig:pt-lhcb-D-3} uncertainties related to
fragmentation functions.
All these uncertainties are very similar as for the ALICE and ATLAS
kinematics.

\begin{figure}[!h]
\begin{minipage}{0.47\textwidth}
 \centerline{\includegraphics[width=1.0\textwidth]{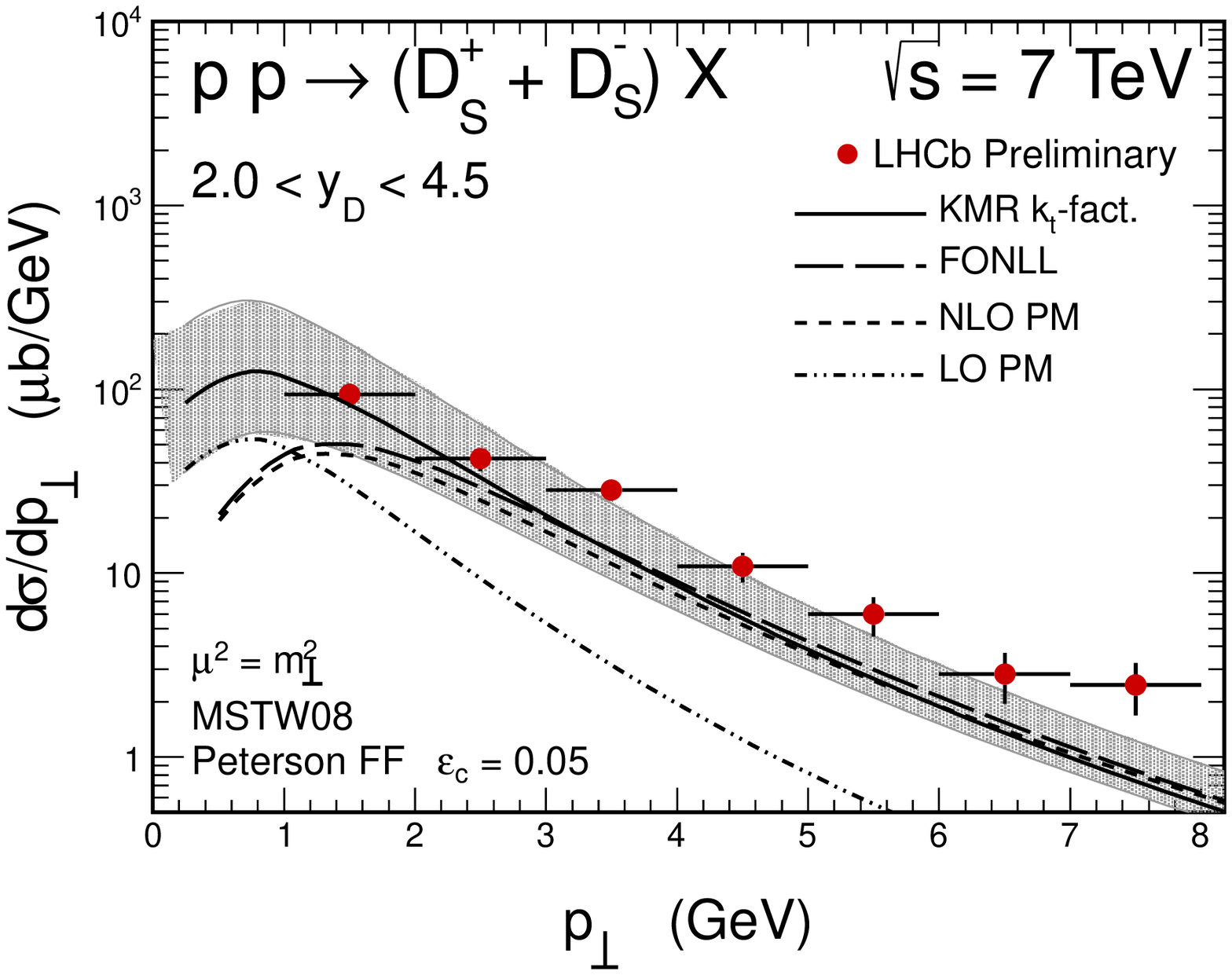}}
\end{minipage}
\hspace{0.5cm}
\begin{minipage}{0.47\textwidth}
 \centerline{\includegraphics[width=1.0\textwidth]{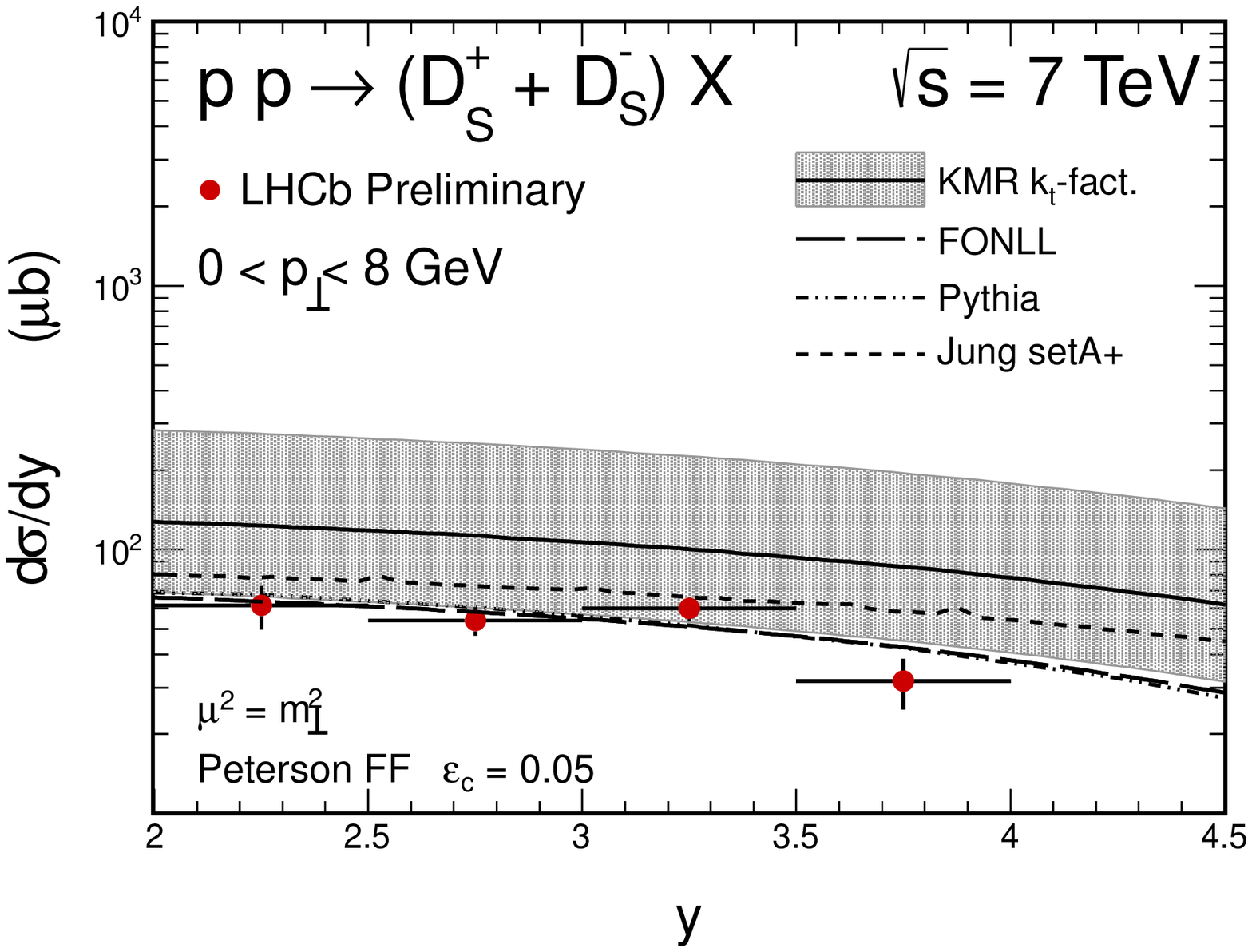}}
\end{minipage}
   \caption{
\small Results with overall uncertainties for transverse momentum (left) and rapidity distributions (right)
of $D_s^{\pm}$ for the $k_t$-factorization approach with the KMR UGDF.
For comparison we show predictions of other popular approaches.
}
 \label{fig:pt-lhcb-D-4}
\end{figure}

In Fig.~\ref{fig:pt-lhcb-D-4} we compare
our predictions, together with predictions of other popular approaches.
Main conclusions are the same again as for the ALICE and ATLAS.

\begin{figure}[!h]
\begin{minipage}{0.47\textwidth}
 \centerline{\includegraphics[width=1.0\textwidth]{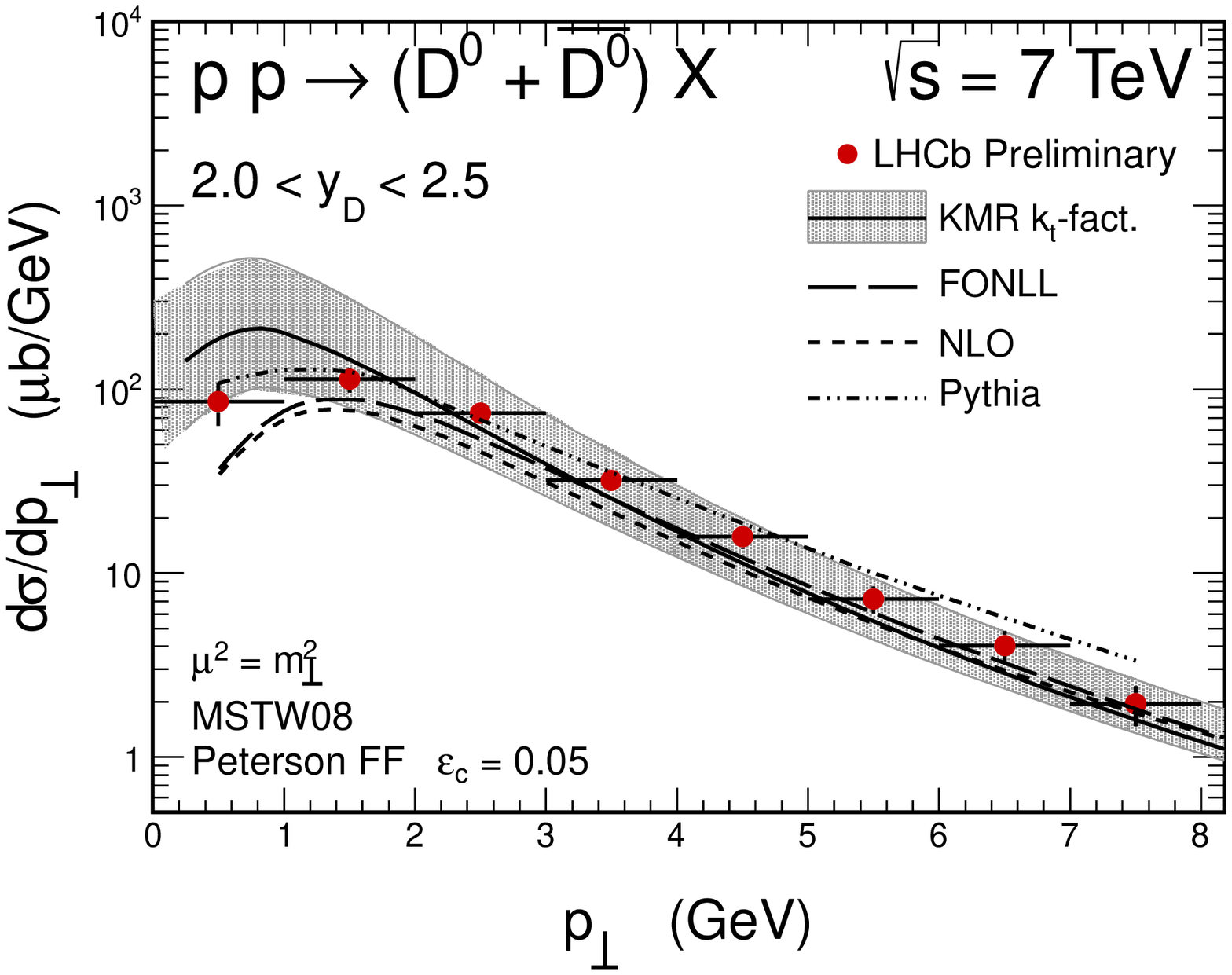}}
\end{minipage}
\hspace{0.5cm}
\begin{minipage}{0.47\textwidth}
 \centerline{\includegraphics[width=1.0\textwidth]{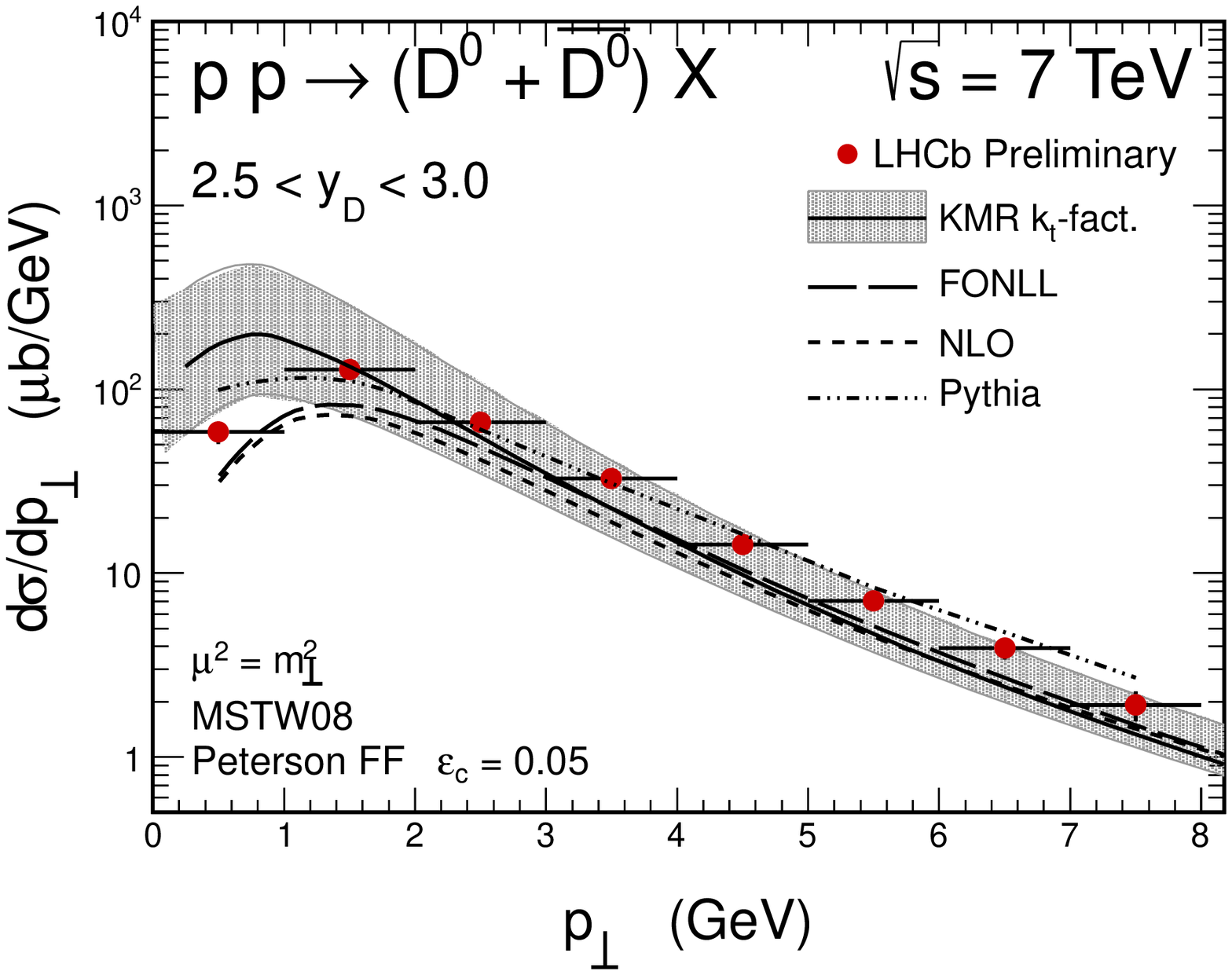}}
\end{minipage}
\begin{minipage}{0.47\textwidth}
 \centerline{\includegraphics[width=1.0\textwidth]{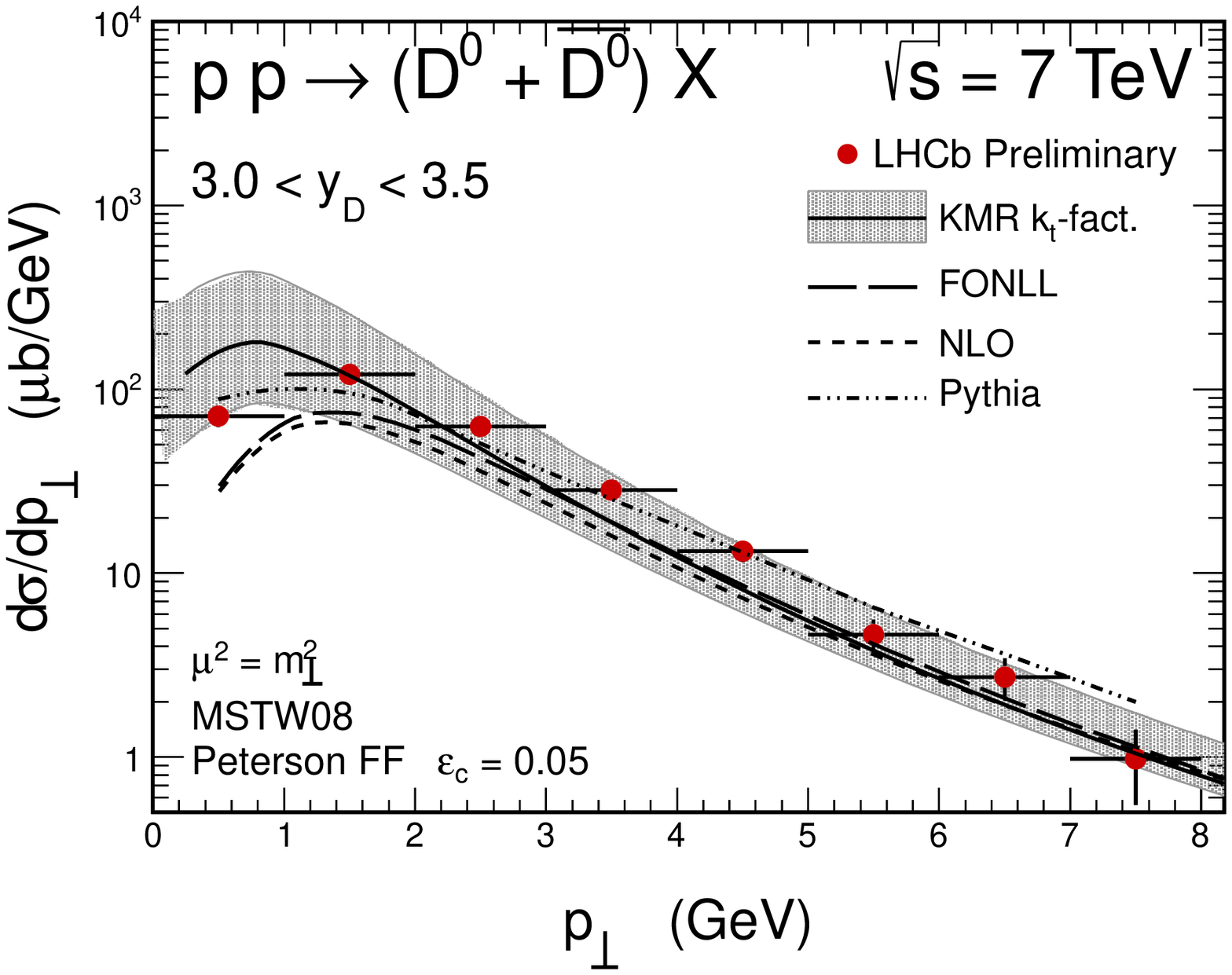}}
\end{minipage}
\hspace{0.5cm}
\begin{minipage}{0.47\textwidth}
 \centerline{\includegraphics[width=1.0\textwidth]{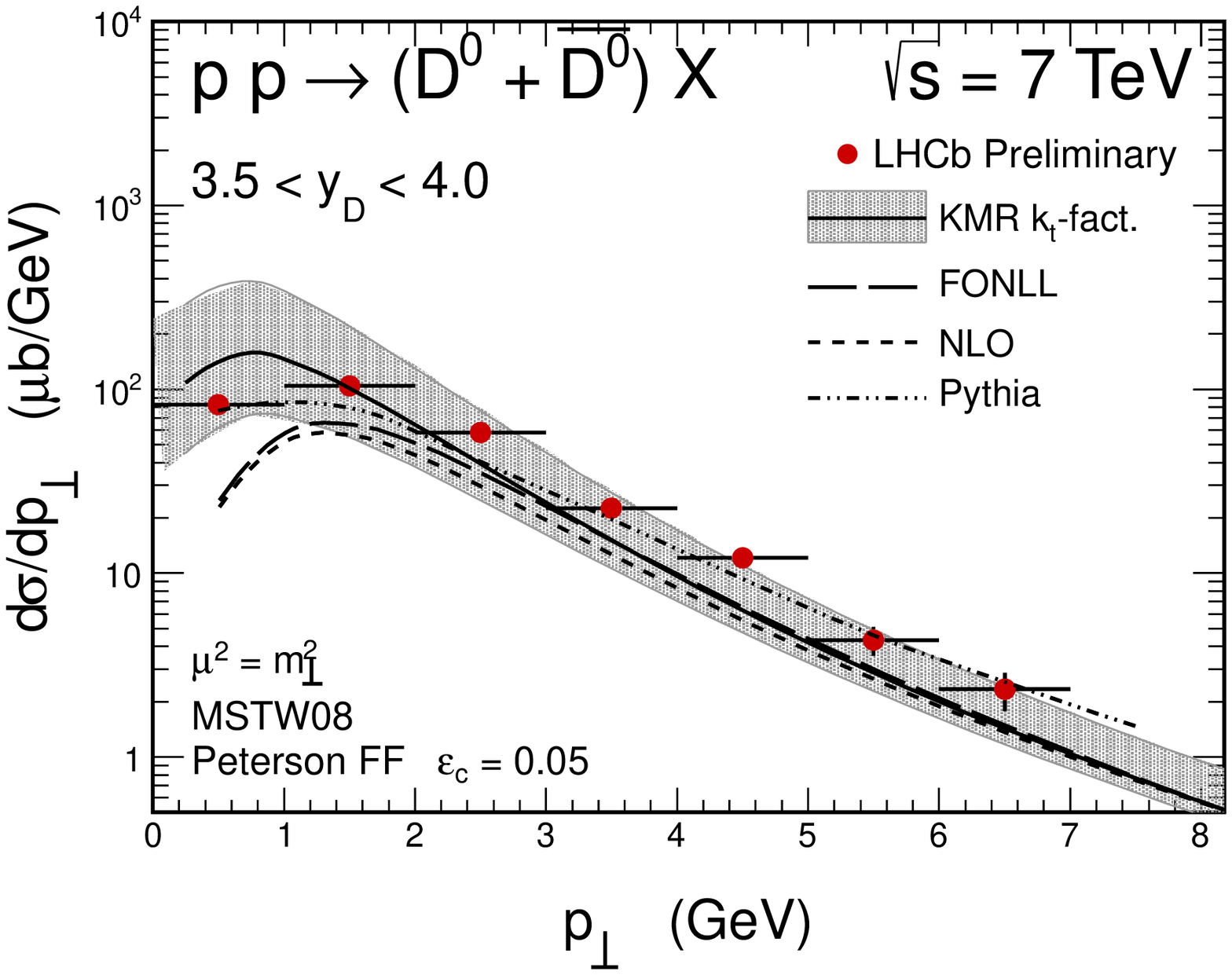}}
\end{minipage}
\begin{minipage}{0.47\textwidth}
 \centerline{\includegraphics[width=1.0\textwidth]{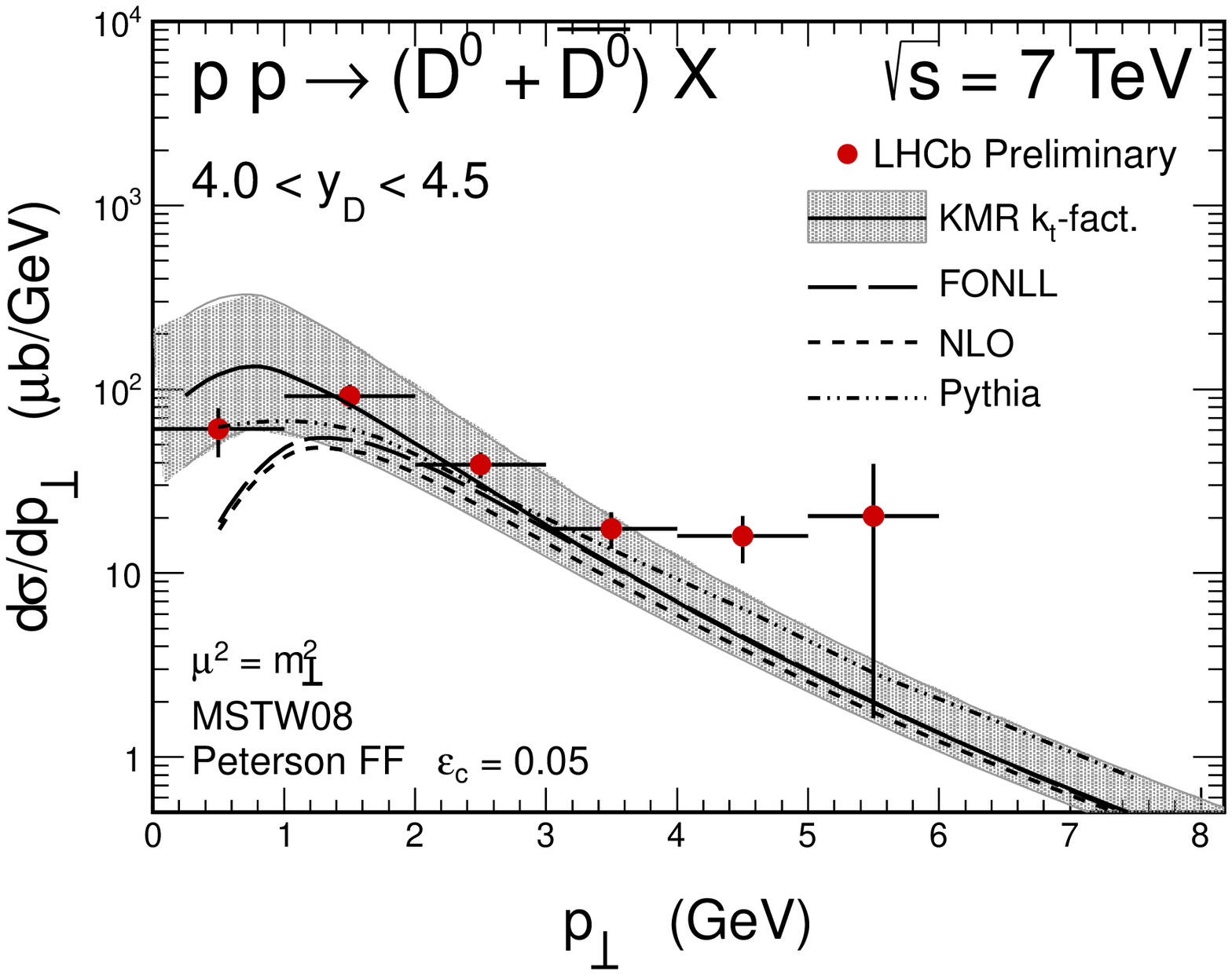}}
\end{minipage}
\hspace{8.2cm}
\begin{minipage}{0.47\textwidth}
\end{minipage}

   \caption{
\small Transverse momentum distribution of neutral $D^0$ mesons for different
ranges of rapidities specified in the figures. We compare results of 
the $k_t$-factorization approach with the KMR UGDF and those obtained 
within other approaches known from the literature.}
 \label{fig:pt-lhcb-D-5}
\end{figure}

\begin{figure}[!h]
\begin{minipage}{0.47\textwidth}
 \centerline{\includegraphics[width=1.0\textwidth]{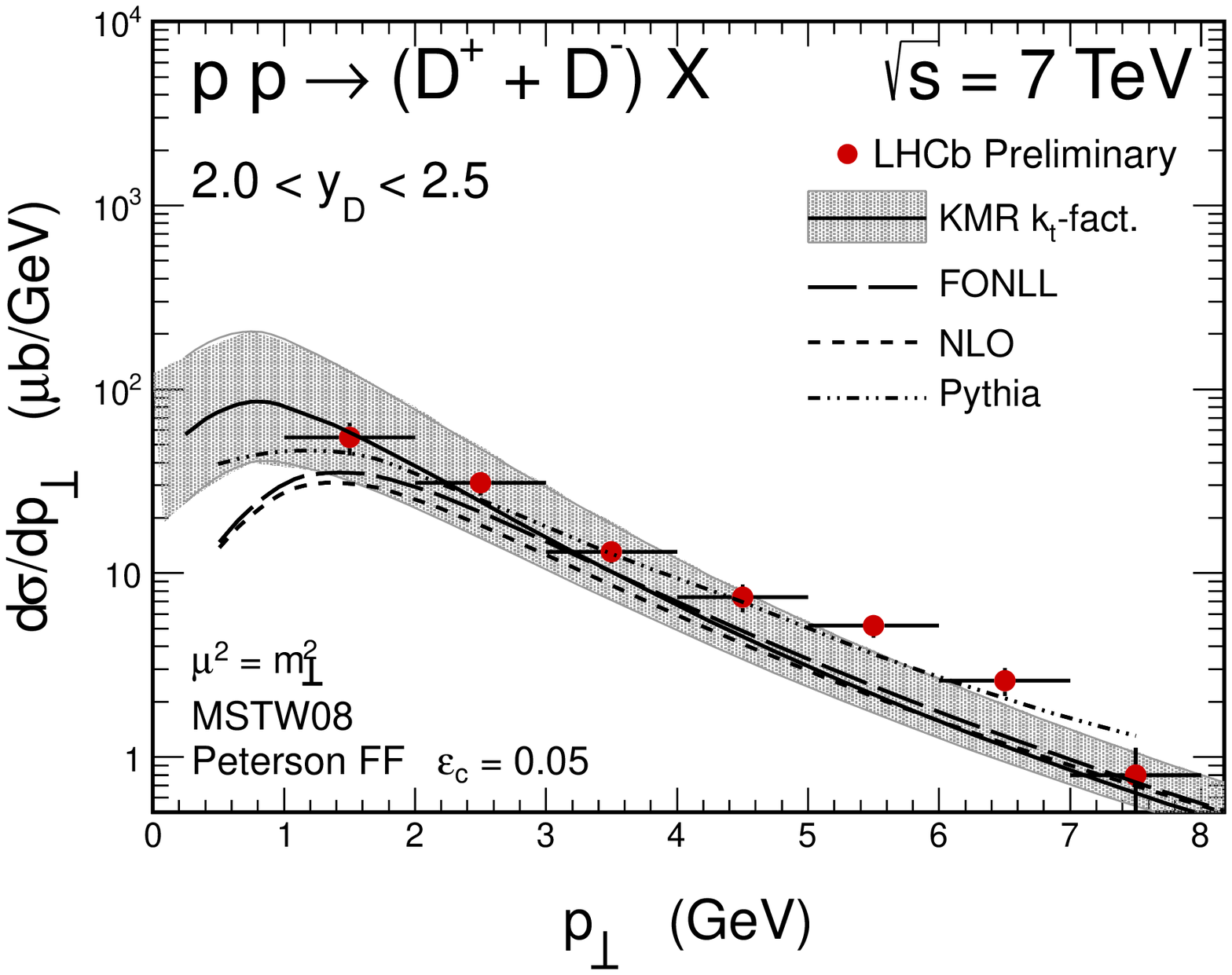}}
\end{minipage}
\hspace{0.5cm}
\begin{minipage}{0.47\textwidth}
 \centerline{\includegraphics[width=1.0\textwidth]{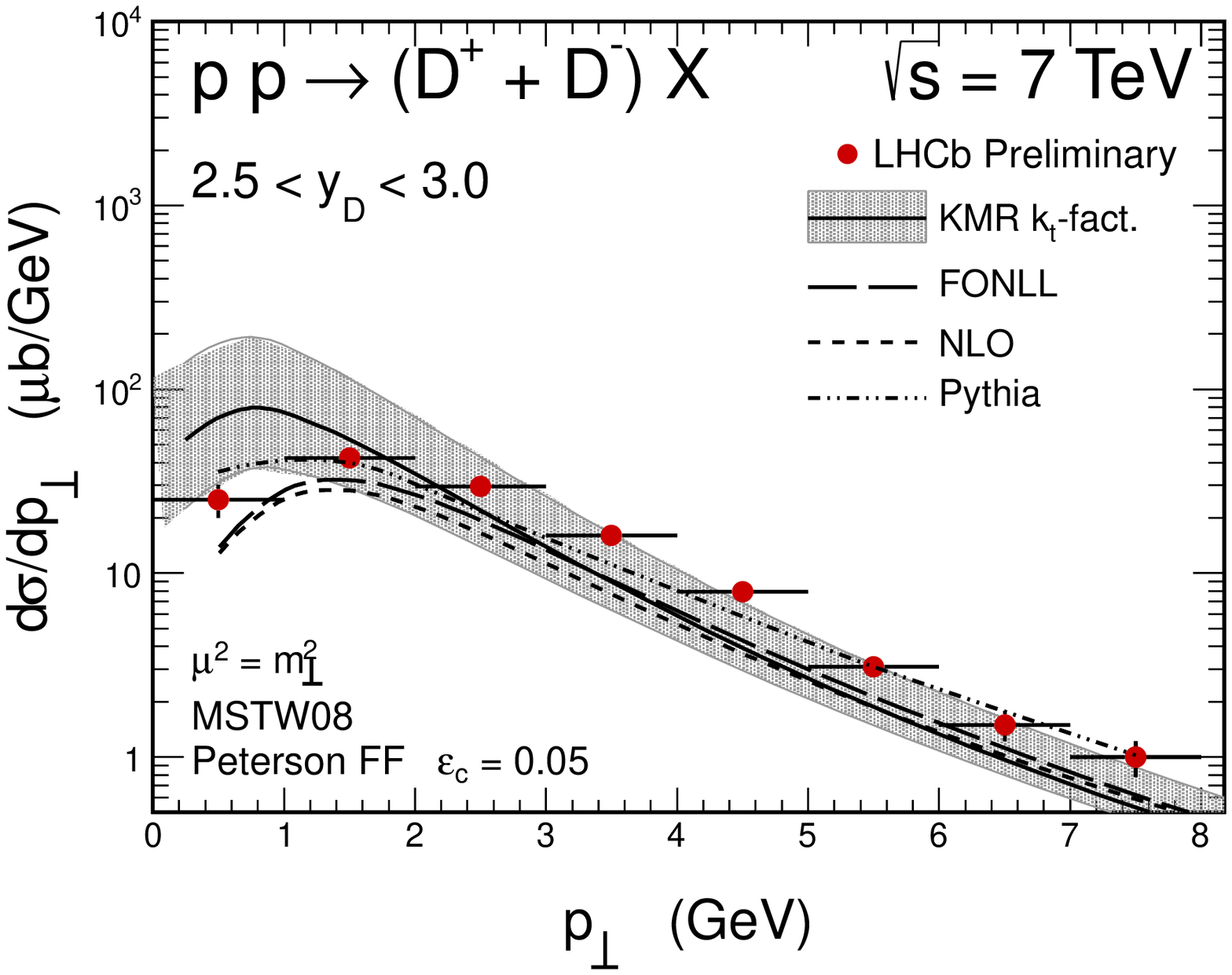}}
\end{minipage}
\begin{minipage}{0.47\textwidth}
 \centerline{\includegraphics[width=1.0\textwidth]{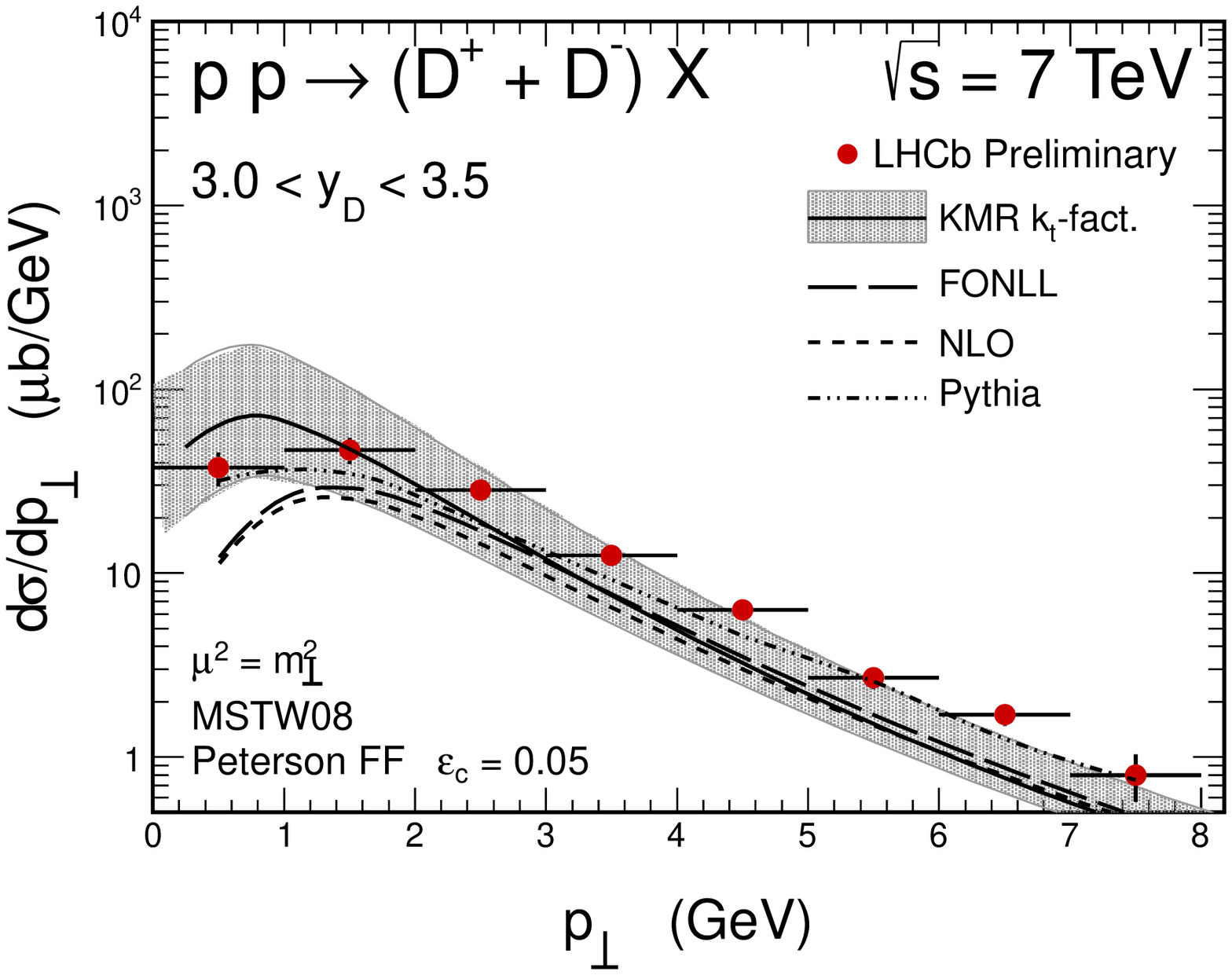}}
\end{minipage}
\hspace{0.5cm}
\begin{minipage}{0.47\textwidth}
 \centerline{\includegraphics[width=1.0\textwidth]{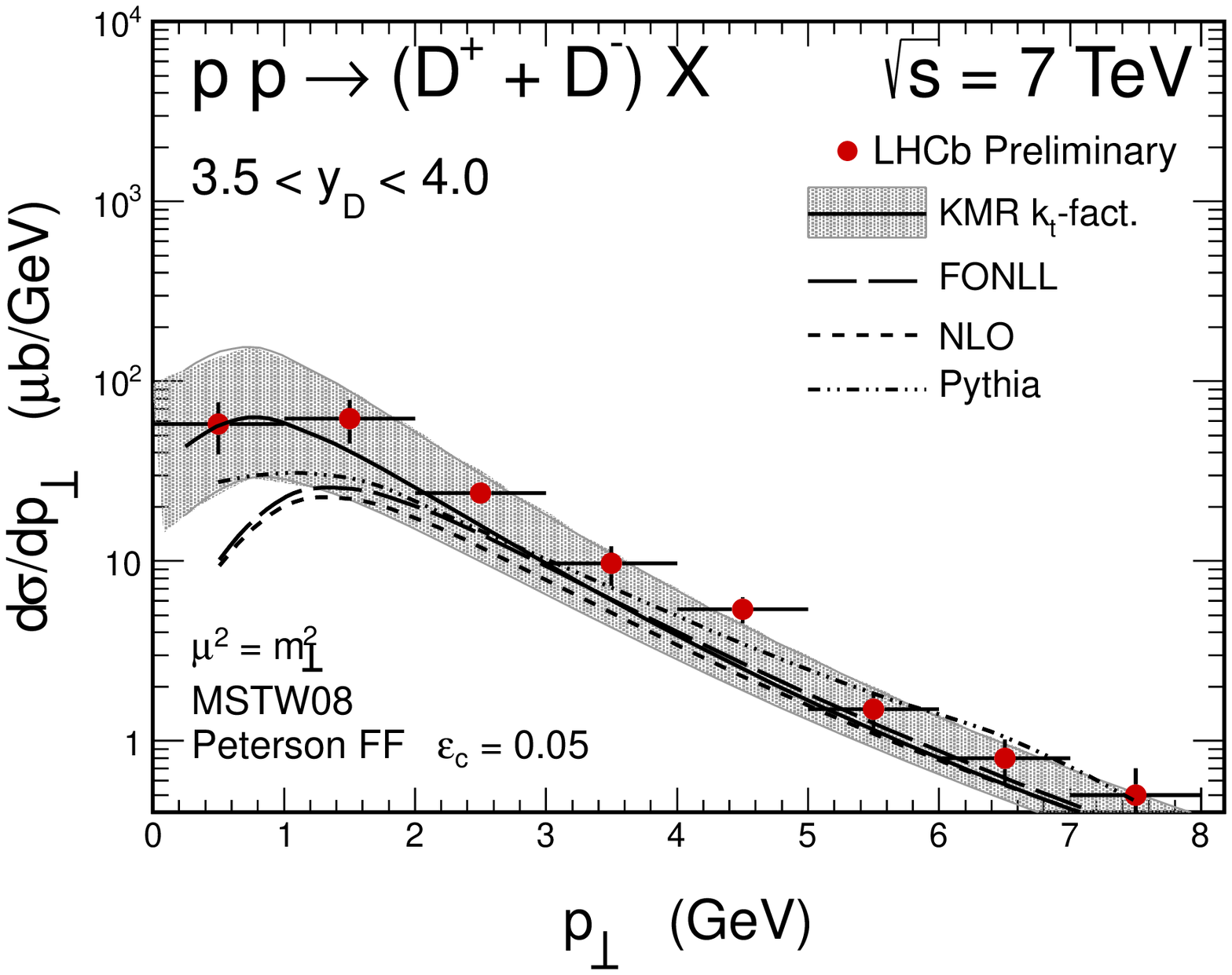}}
\end{minipage}
\begin{minipage}{0.47\textwidth}
 \centerline{\includegraphics[width=1.0\textwidth]{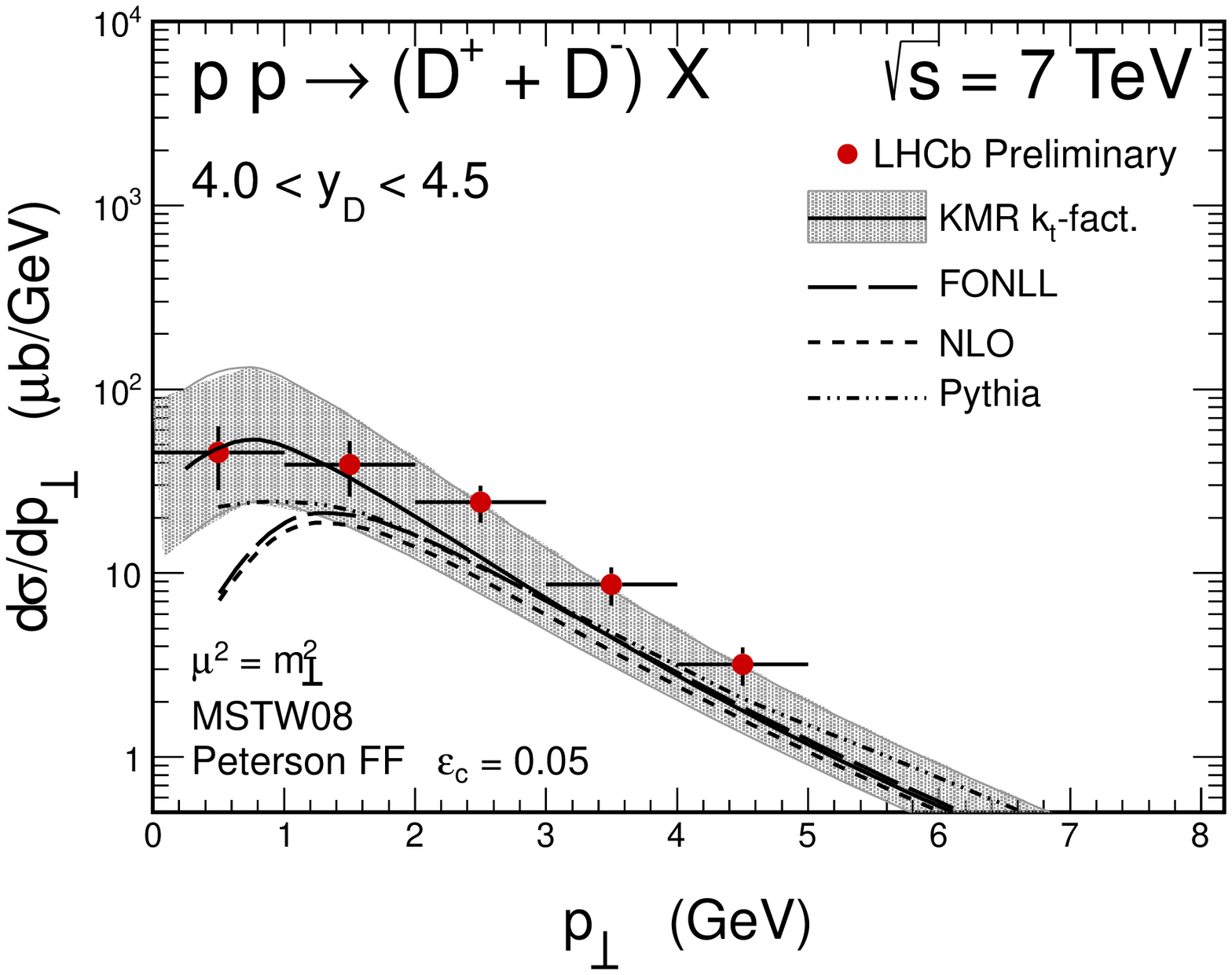}}
\end{minipage}
\hspace{8.2cm}
\begin{minipage}{0.47\textwidth}
\end{minipage}

   \caption{
\small Transverse momentum distribution of charged $D^+$ mesons for different
ranges of rapidities specified in the figures. We compare results of 
the $k_t$-factorization approach with the KMR UGDF and those obtained 
within other approaches known from the literature.
} 
\label{fig:pt-lhcb-D-6}
\end{figure}

\begin{figure}[!h]
\begin{minipage}{0.47\textwidth}
 \centerline{\includegraphics[width=1.0\textwidth]{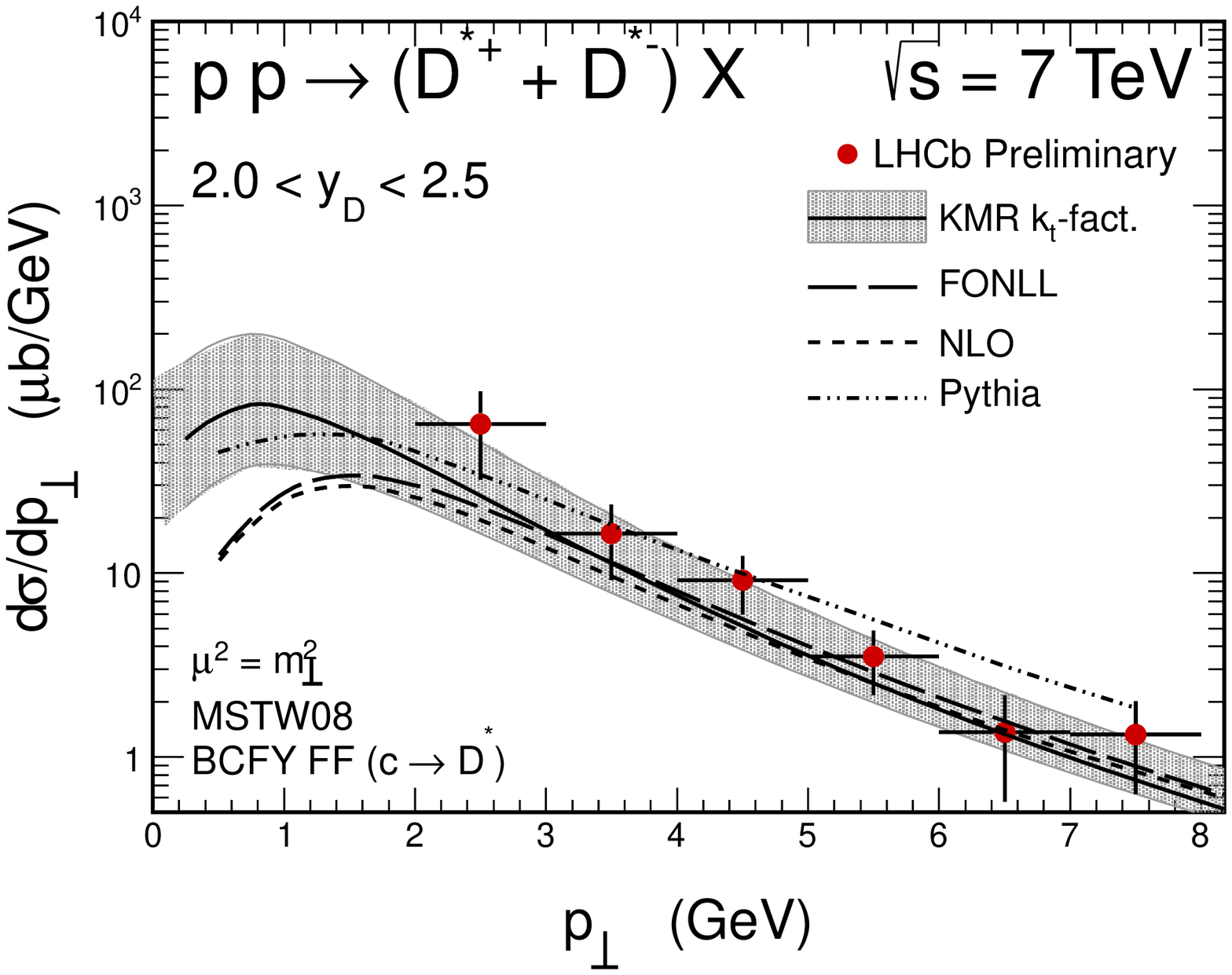}}
\end{minipage}
\hspace{0.5cm}
\begin{minipage}{0.47\textwidth}
 \centerline{\includegraphics[width=1.0\textwidth]{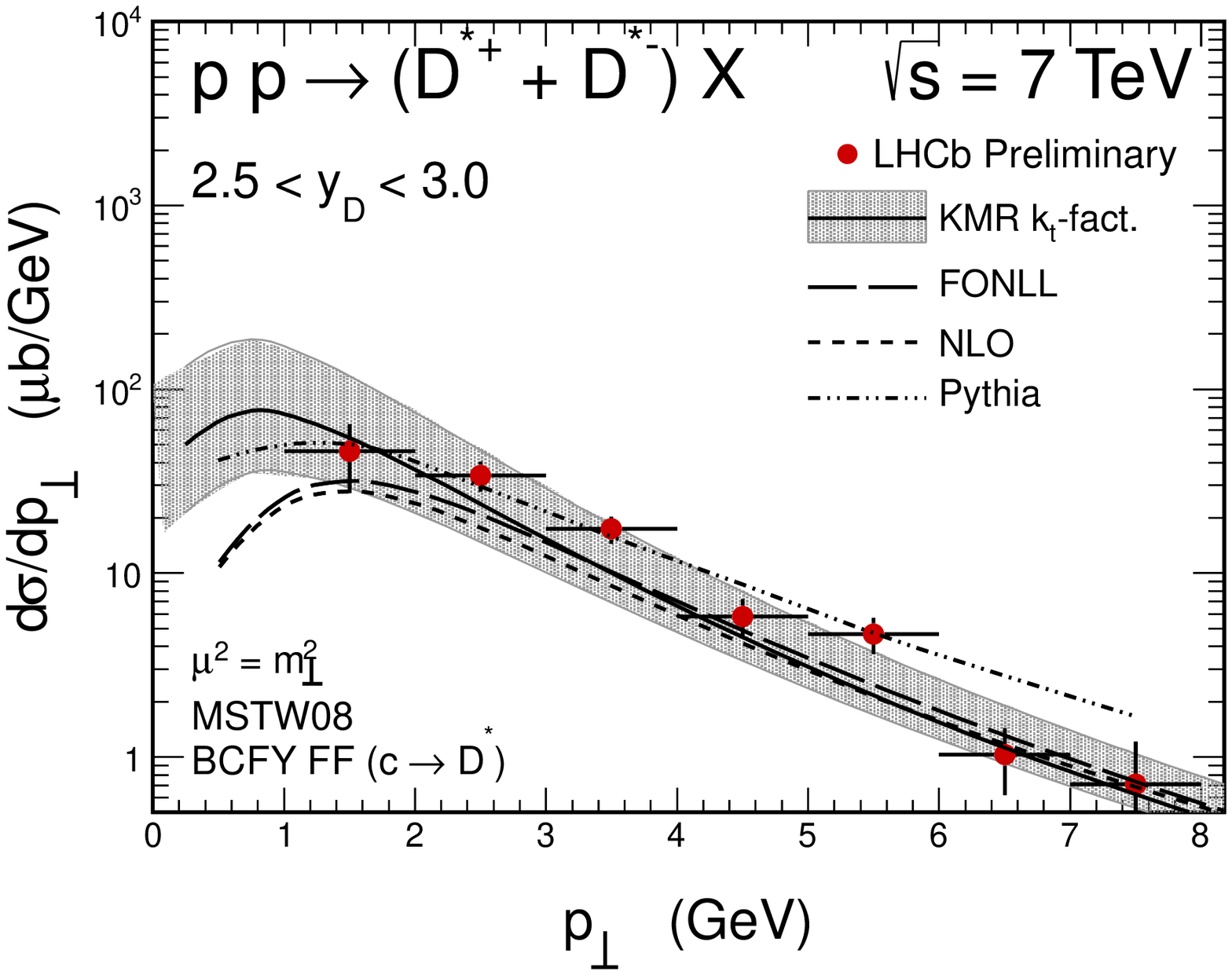}}
\end{minipage}
\begin{minipage}{0.47\textwidth}
 \centerline{\includegraphics[width=1.0\textwidth]{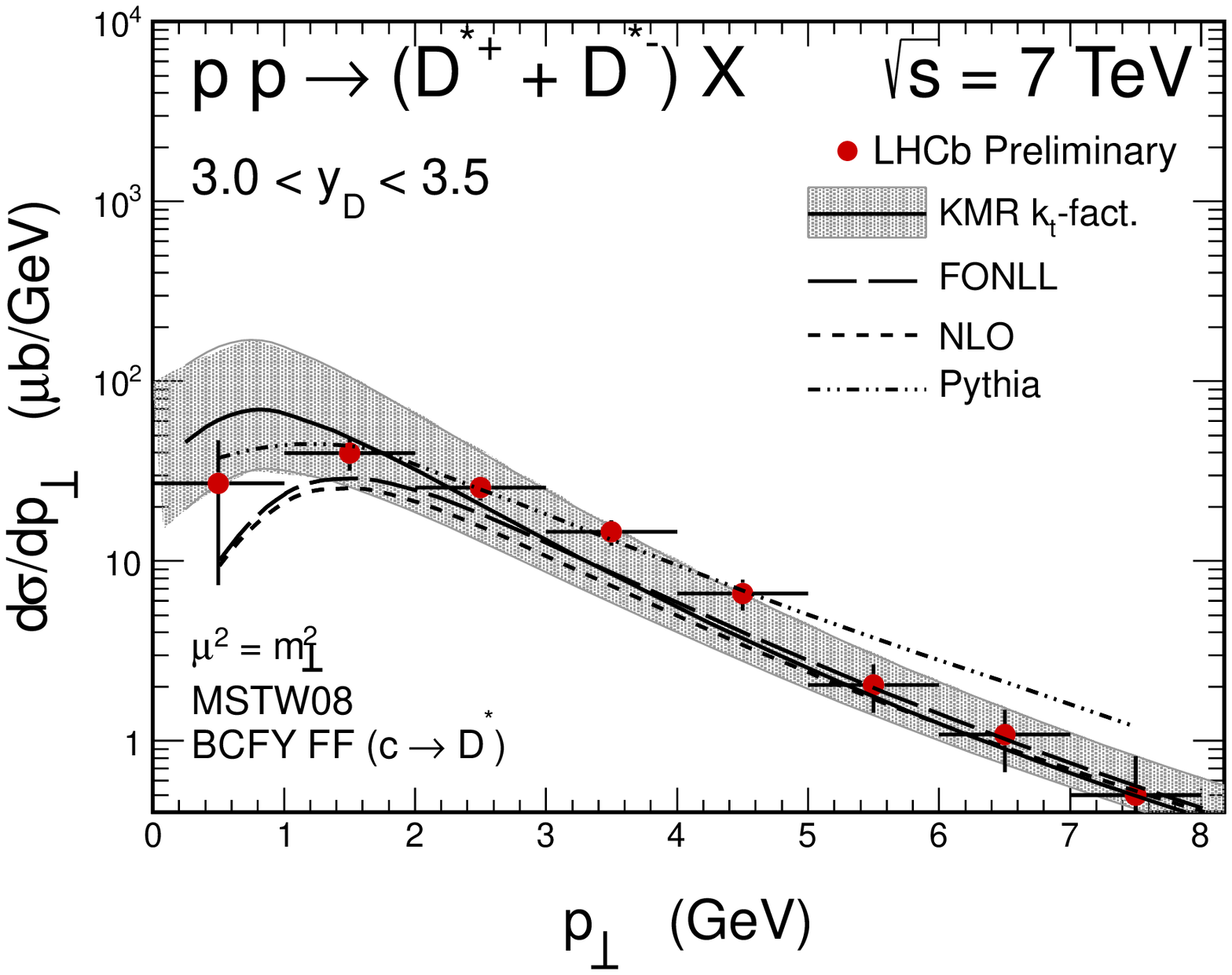}}
\end{minipage}
\hspace{0.5cm}
\begin{minipage}{0.47\textwidth}
 \centerline{\includegraphics[width=1.0\textwidth]{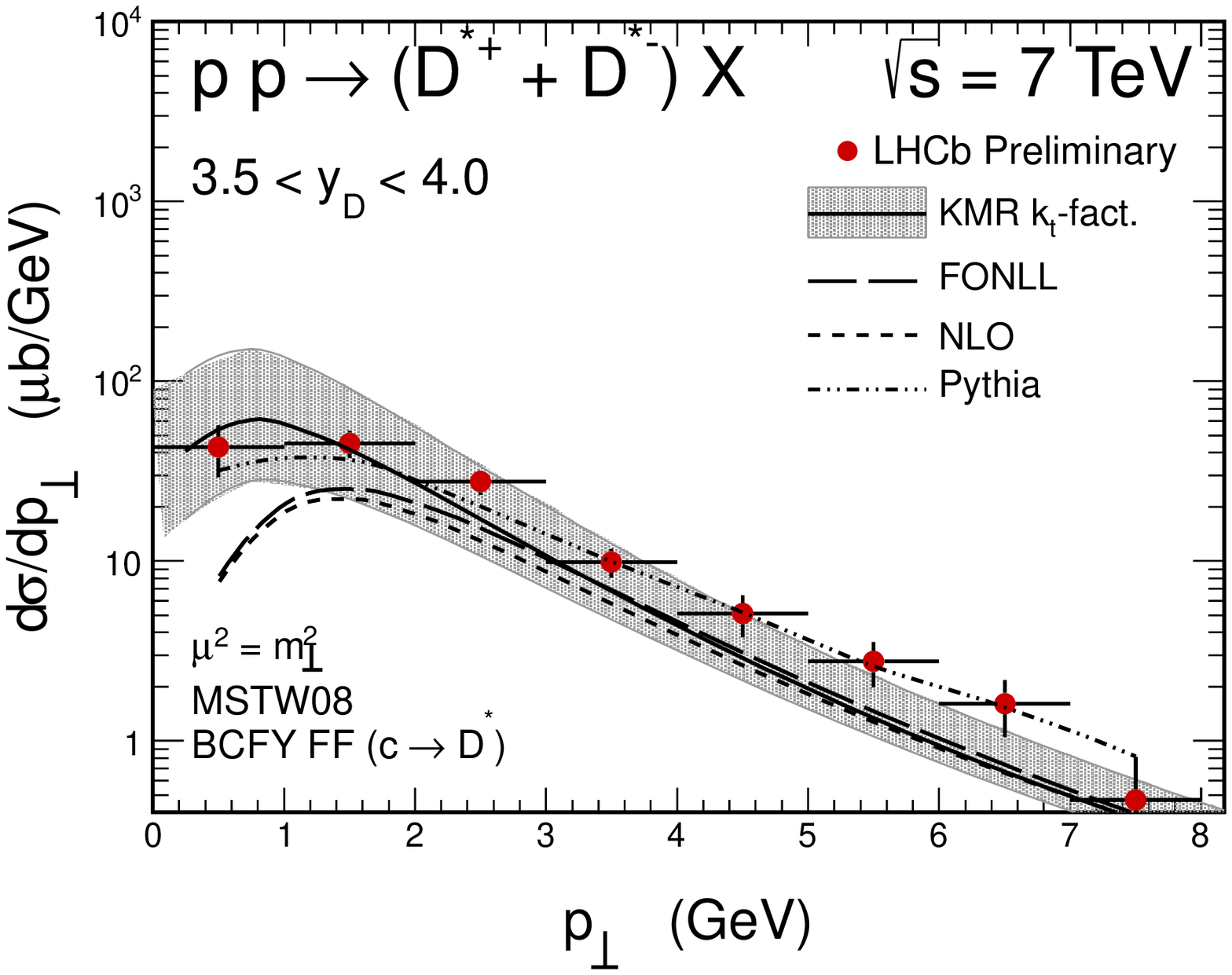}}
\end{minipage}
\begin{minipage}{0.47\textwidth}
 \centerline{\includegraphics[width=1.0\textwidth]{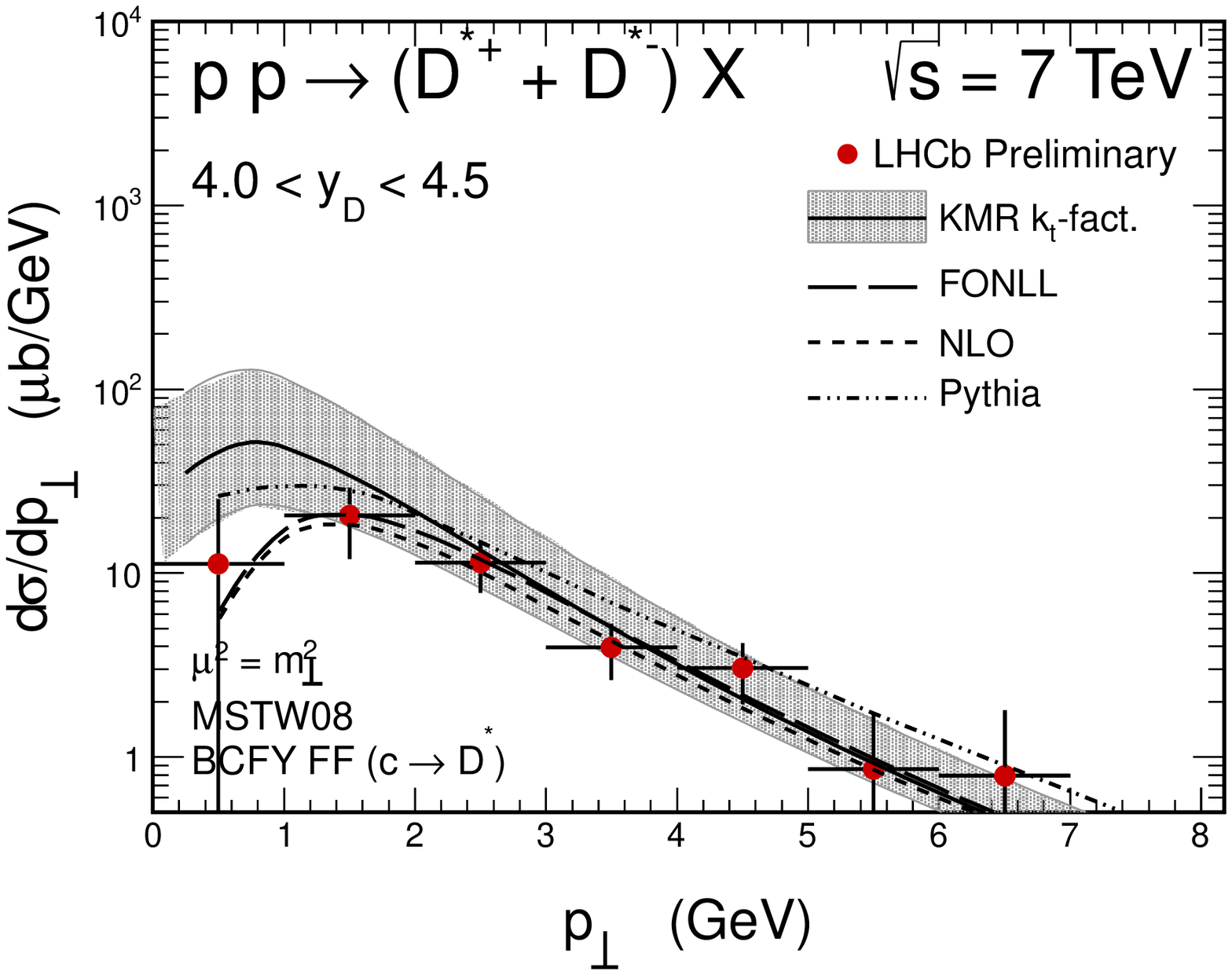}}
\end{minipage}
\hspace{8.2cm}
\begin{minipage}{0.47\textwidth}
\end{minipage}

   \caption{
\small Transverse momentum distribution of $D^{*+}$ mesons for different ranges of rapidities specified in the figures. 
We compare results of the $k_t$-factorization approach with the KMR UGDF
and those obtained within other approaches known from the literature.
}
 \label{fig:pt-lhcb-D-7}
\end{figure}

The LHCb collaboration was able to measure transverse momentum distributions
of mesons in many narrow bins of (pseudo)rapidity. Below 
(Figs.~\ref{fig:pt-lhcb-D-5},~\ref{fig:pt-lhcb-D-6} and 
~\ref{fig:pt-lhcb-D-7}) we show such distributions for 
$D^0$, $D^+$ and $D^{*+}$, respectively. In general, different bins are
sensitive to different regions of longitudinal momentum fractions carried by gluons.
However, we do not observe any interesting trend in the quality of the description
of the LHCb data. Our results with the KMR UGDF within uncertainties are consistent
with the experimental data and with the FONLL and NLO PM predictions.
Corresponding distributions for PYTHIA are taken from Ref.~\cite{LHCbincD}
and have slightly different $p_t$-slope than the other ones.

\clearpage

\clearpage

\section{Production of $D\bar{D}$ pairs}

Most of the calculations in the literature concentrates on single meson distributions.
We wish to focus now on correlation observables for $D$ and $\bar{D}$ mesons.
In order to calculate correlation observables for two mesons we follow here, similar as in the single meson case,
the fragmentation function technique for hadronization process:
\begin{equation}
\frac{d \sigma(pp \to D \bar{D} X)}{d y_1 d y_{2} d^2 p_{1t}^{D} d^2 p_{2t}^{\bar{D}}}
 \approx
\int \frac{D_{c \to D}(z_{1})}{z_{1}}\cdot \frac{D_{\bar c \to \bar D}(z_{2})}{z_{2}}\cdot
\frac{d \sigma(pp \to c \bar{c} X)}{d y_1 d y_{2} d^2
  p_{1t}^{c} d^2 p_{2t}^{\bar{c}}} d z_{1} d z_{2} \; ,
\end{equation}
where: 
$p_{1t}^{c} = \frac{p_{1,t}^{D}}{z_{1}}$, $p_{2,t}^{\bar{c}} =
  \frac{p_{2t}^{\bar{D}}}{z_{2}}$ and
meson longitudinal fractions  $z_{1}, z_{2}\in (0,1)$.
The multidimensional distribution for $c$ quark and $\bar c$ antiquark is convoluted with respective
fragmentation functions simultaneously. As a
result of the hadronization one obtains corresponding two-meson 
multidimensional distribution. In the last step experimental kinematical
cuts on the distributions can be imposed. Then the resulting
distributions can be compared with experimental ones. 
For numerical calculations here we again apply the Peterson model of fragmentation
function \cite{Peterson}.

\begin{table}[tb]%
\caption{Integrated cross sections for the two mesons modes specified in the table below within the LHCb detector.}
\newcolumntype{Z}{>{\centering\arraybackslash}X}
\label{table-cor}
\centering %
\begin{tabularx}{16.5cm}{ZZZZZZZZZZ}
\toprule[0.1em] %
\\[-2.4ex] 

 \multirow{3}*{Mode} & &  & \multicolumn{7}{c}{$\sigma_{tot}^{THEORY} \;\;$[nb]} \\ [+0.4ex]
                     & \multicolumn{2}{c}{$\sigma_{tot}^{EXP} \;\;$[nb]}  & \multicolumn{3}{c}{KMR $^{+}_{-}(\mu)$ $^{+}_{-}(m_{c})$} & \multicolumn{2}{c}{Jung setA$+$} & \multicolumn{2}{c}{KMS} \\[+0.4ex] 
                     &                                             &  & \multicolumn{2}{c}{$\varepsilon_{c} = 0.05$}  &  $\varepsilon_{c} = 0.02$  & $\varepsilon_{c} = 0.05$ & $\varepsilon_{c} = 0.02$ & $\varepsilon_{c} = 0.05$ & $\varepsilon_{c} = 0.02$ \\[+0.4ex]

\toprule[0.1em]\\[-2.4ex] 

 $D^{0}\bar{D^{0}}$    & \multicolumn{2}{c}{$6230\pm120\pm630$} & \multicolumn{2}{c}{$5193$ $^{+1346}_{-879}$ $^{+654}_{-576}$} & 6971 & 4532 & 5814 & 2895 & 3894  \\ [+0.4ex]
 $D^{0}D^{-}$          & \multicolumn{2}{c}{$3990\pm90\pm500$ } & \multicolumn{2}{c}{$4155$ $^{+1076}_{-704}$ $^{+523}_{-461}$}  & 5577  & 3626 & 4652 & 2316 & 3115 \\ [+0.4ex]
 $D^{0}D^{-}_{S}$      & \multicolumn{2}{c}{$1680\pm110\pm240$} & \multicolumn{2}{c}{$1471$ $^{+381}_{-249}$ $^{+185}_{-163}$}  & 1974   & 1284 & 1647 & 820  & 1103 \\  [+0.4ex]
 $D^{+}D^{-}$          & \multicolumn{2}{c}{$780\pm40\pm130$  } & \multicolumn{2}{c}{$831$  $^{+215}_{-141}$ $^{+105}_{-92}$}   & 1115  & 725  & 930  & 463  & 623 \\[+0.4ex] 
 $D^{+}D^{-}_{S}$      & \multicolumn{2}{c}{$550\pm60\pm90$   } & \multicolumn{2}{c}{$588$  $^{+152}_{-99}$ $^{+74}_{-65}$}   & 790 & 513  & 659  & 328  & 441 \\[+0.4ex]
 $D^{+}_{S}D^{-}_{S}$  & \multicolumn{2}{c}{$-$               } & \multicolumn{2}{c}{$104$   $^{+27}_{-17}$ $^{+13}_{-11}$}    & 139  & 91   & 117  & 59   & 78 \\[+0.4ex]
\hline
\bottomrule[0.1em]

\end{tabularx}

\end{table}

\begin{figure}[!h]
\begin{center}
\begin{minipage}{0.5\textwidth}
 \centerline{\includegraphics[width=1.0\textwidth]{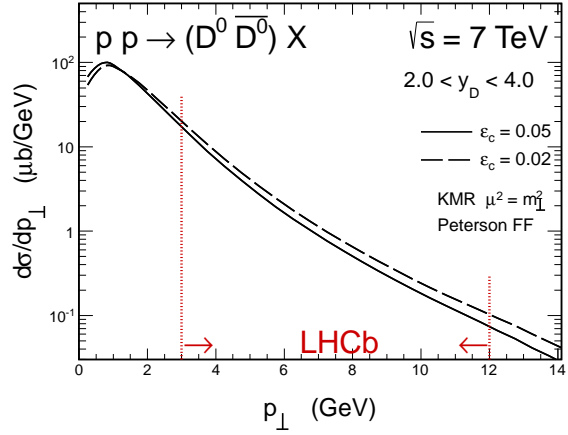}}
\end{minipage}
\end{center}
   \caption{
\small Transverse momentum of $D^0$ meson within the LHCb acceptance
provided that $\bar D^0$ was registered too.
Here the KMR UGDF was used. We show results for different values of the
Peterson fragmentation function parameter $\varepsilon_c$. }
 \label{fig:pt-lhcb-DDbar-FF}
\end{figure}

The experimental cross sections for the production of two mesons
are also (or even a bit more) sensitive to the details of hadronization as it was in the cases of 
the inclusive single $D$ meson production discussed
in the previous section. For example in Fig.~\ref{fig:pt-lhcb-DDbar-FF}
we compare transverse momentum distribution of $D^{0}$ meson provided that $\bar{D^{0}}$ is also measured for two different values of the $\varepsilon_c$ parameter
of the Peterson fragmentation function. The larger meson transverse
momentum, the larger sensitivity to the value of $\varepsilon_c$.
For illustration we show the range of transverse momenta relevant
for the recent experiments of the LHCb collaboration \cite{LHCb-DPS-2012}.
The effect of the modification of the $\varepsilon_c$ from $0.05$ to $0.02$ is quite sizeable.
In the LHCb acceptance, it does not really affect the shape of calculated $p_t$ distribution
but has an important effect for the predictions of integrated cross sections.
In Table~\ref{table-cor} we compare measured by the LHCb collaboration cross sections
for different $D\bar{D}$ modes with our theoretical results. Calculated values for three different
UGDFs are consistent with the measured ones, taking into account rather large experimental
and theoretical uncertainties. In particular, this is true only when $\varepsilon_c = 0.02$
is taken in the calculation of the fragmentation process. Results obtained with the
KMR UGDF are the closest to the experimental numbers.

\begin{figure}[!h]
\begin{minipage}{0.47\textwidth}
 \centerline{\includegraphics[width=1.0\textwidth]{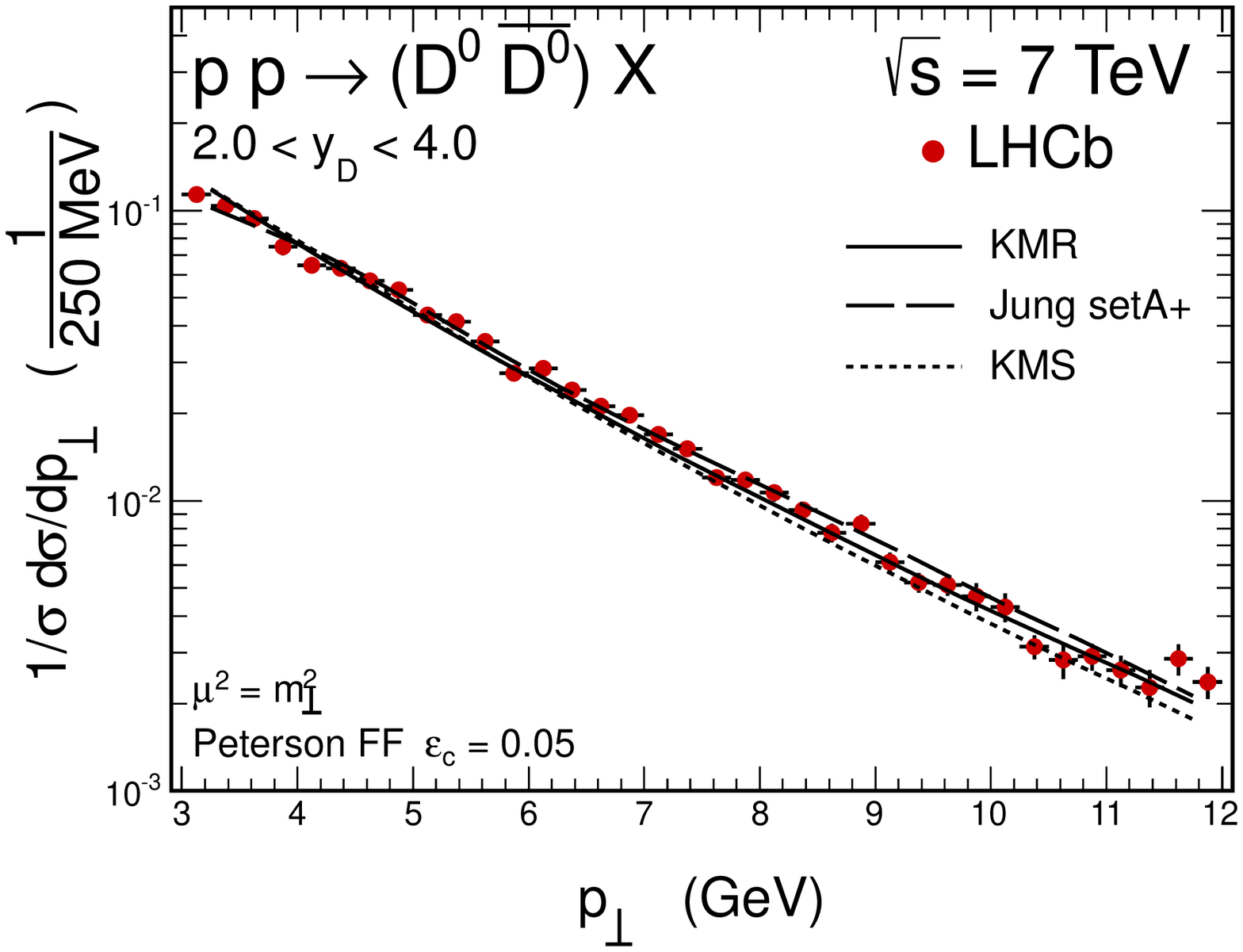}}
\end{minipage}
\hspace{0.5cm}
\begin{minipage}{0.47\textwidth}
 \centerline{\includegraphics[width=1.0\textwidth]{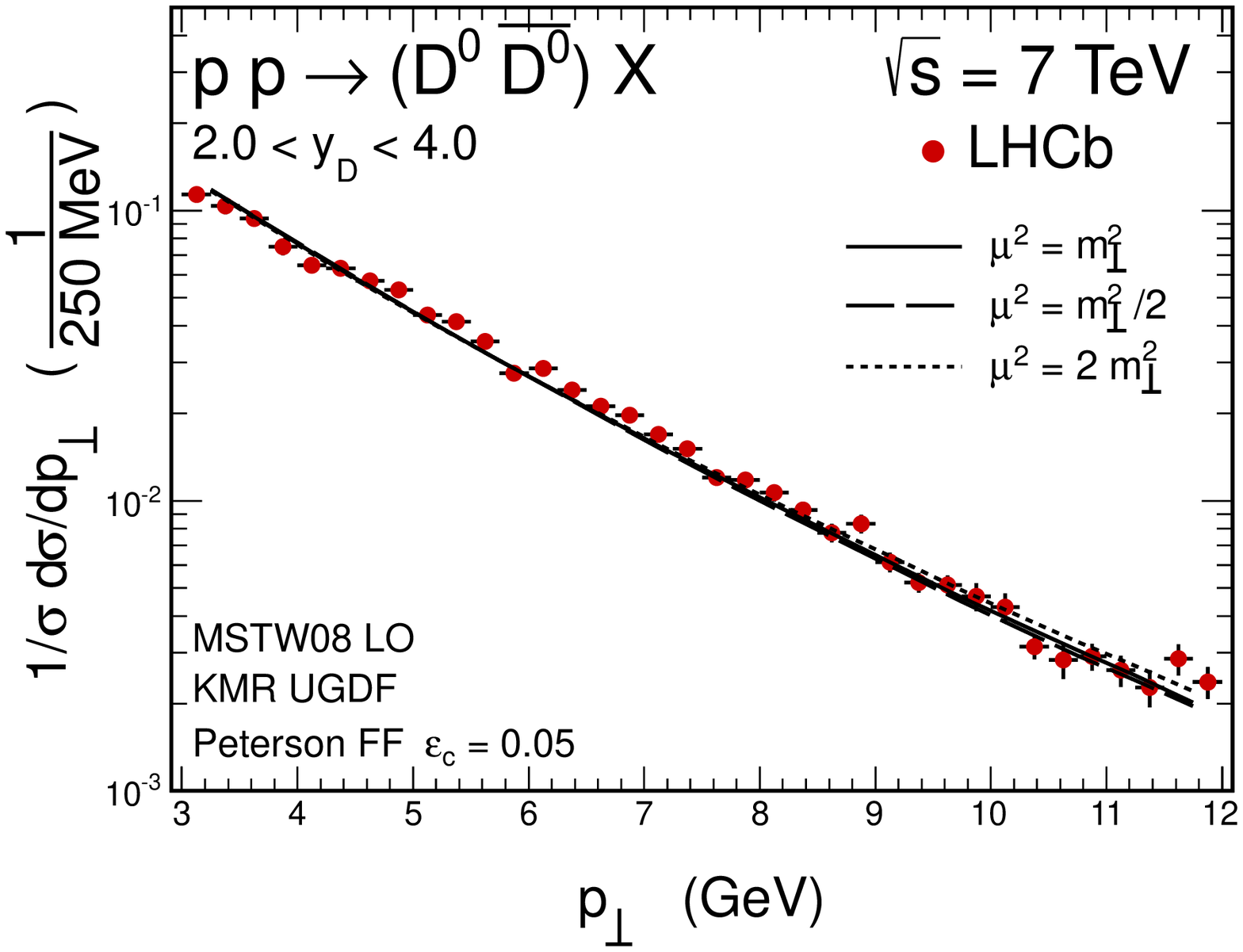}}
\end{minipage}
   \caption{
\small Uncertainties for the conditional transverse momentum
distribution due to the choice of UGDF model (left) and due to the
choice of scales for the KMR UGDF (right). The experimental data
of the LHCb collaboration are from Ref.\cite{LHCb-DPS-2012}.
}
 \label{fig:pt-lhcb-DDbar-1}
\end{figure}

In Fig.~\ref{fig:pt-lhcb-DDbar-1} we present transverse momentum distributions of $D^{0}$
meson for the case when $D^{0}\bar{D^{0}}$ pairs are counted.
We compare theoretical distributions for different
UGDFs (left panel) as well as discuss effect
of the scale dependence (right panel) on the shape of the $p_t$ distribution.
The experimental data points are normalized by a factor $1/\sigma$. The shape of the
transverse momentum distribution is rather well reproduced by all used UGDFs. The
normalization, as discussed already in Table \ref{table-cor}, is less consistent.

\begin{figure}[!h]
\begin{minipage}{0.47\textwidth}
 \centerline{\includegraphics[width=1.0\textwidth]{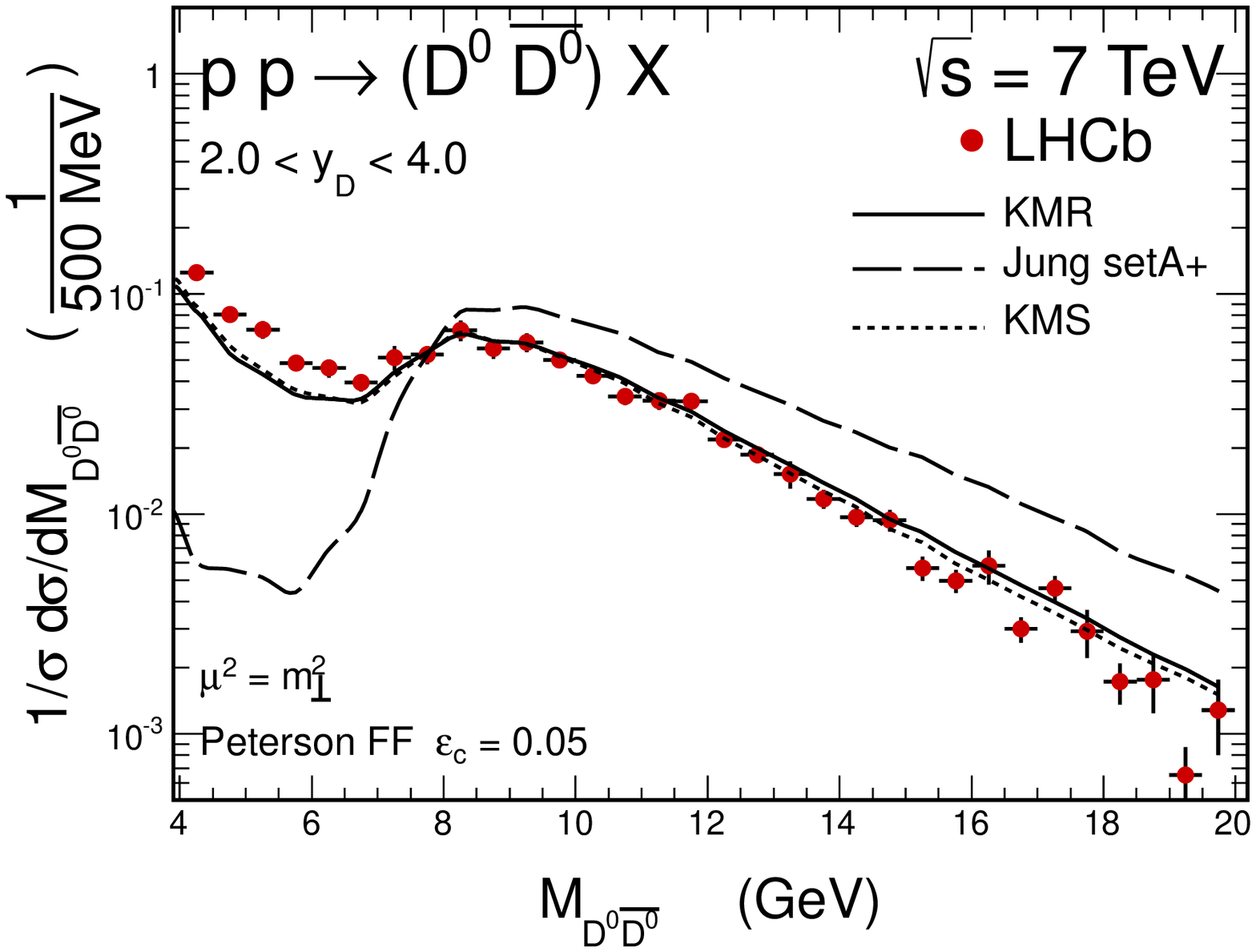}}
\end{minipage}
\hspace{0.5cm}
\begin{minipage}{0.47\textwidth}
 \centerline{\includegraphics[width=1.0\textwidth]{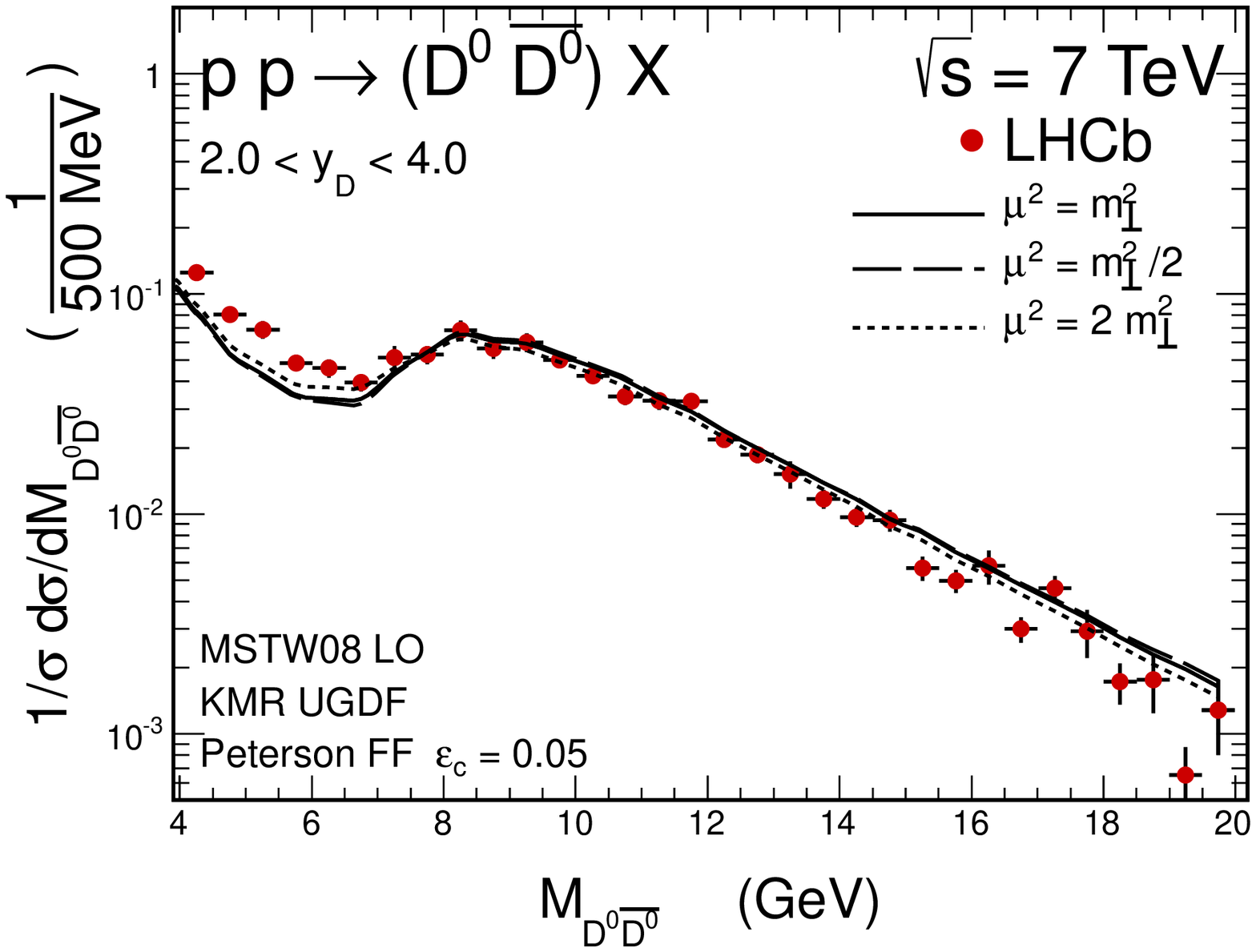}}
\end{minipage}
   \caption{
\small Invariant mass distribution of the $D^0 \bar D^0$ system for
different UGDFs (left) and uncertainties due to the choice of the scale
for the KMR UGDF (right). The experimental data of the LHCb collaboration
are taken from Ref.~\cite{LHCb-DPS-2012}.
}
 \label{fig:minv-lhcb-DDbar-2}
\end{figure}

The LHCb collaboration presented also distribution in the $D^0 \bar D^0$
invariant mass $M_{D^0 \bar D^0}$. In Fig.~\ref{fig:minv-lhcb-DDbar-2} we show the
corresponding theoretical result for different UGDFs. Both, the KMR and KMS UGDFs provide the
right shape of the distribution. The dip at small invariant masses
is due to specific LHCb cuts on kinematical variables. On the other hand the shape of the
distribution almost does not depend on the choice of the scales for the KMR
UGDF (see right panel).

\begin{figure}[!h]
\begin{minipage}{0.47\textwidth}
 \centerline{\includegraphics[width=1.0\textwidth]{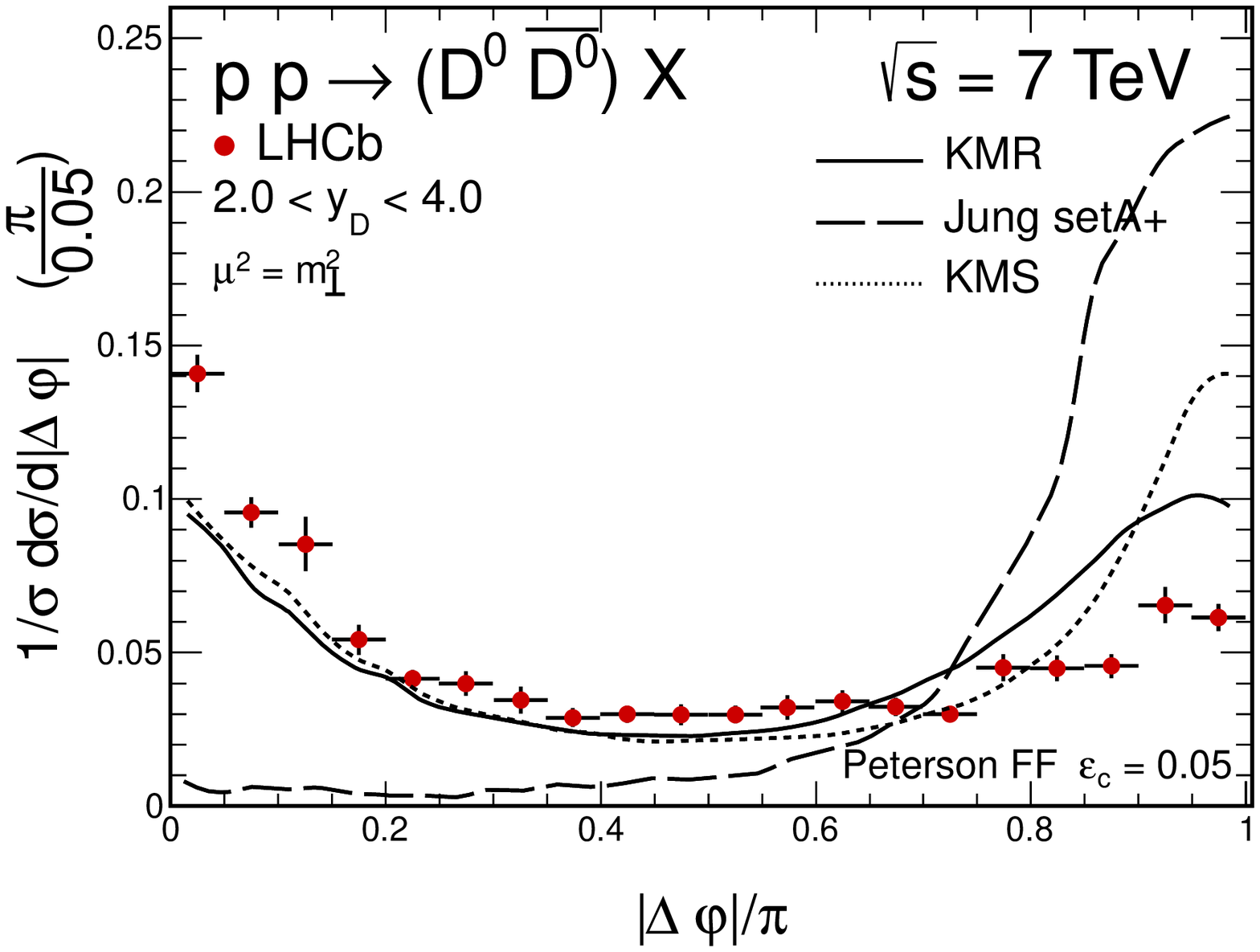}}
\end{minipage}
\hspace{0.5cm}
\begin{minipage}{0.47\textwidth}
 \centerline{\includegraphics[width=1.0\textwidth]{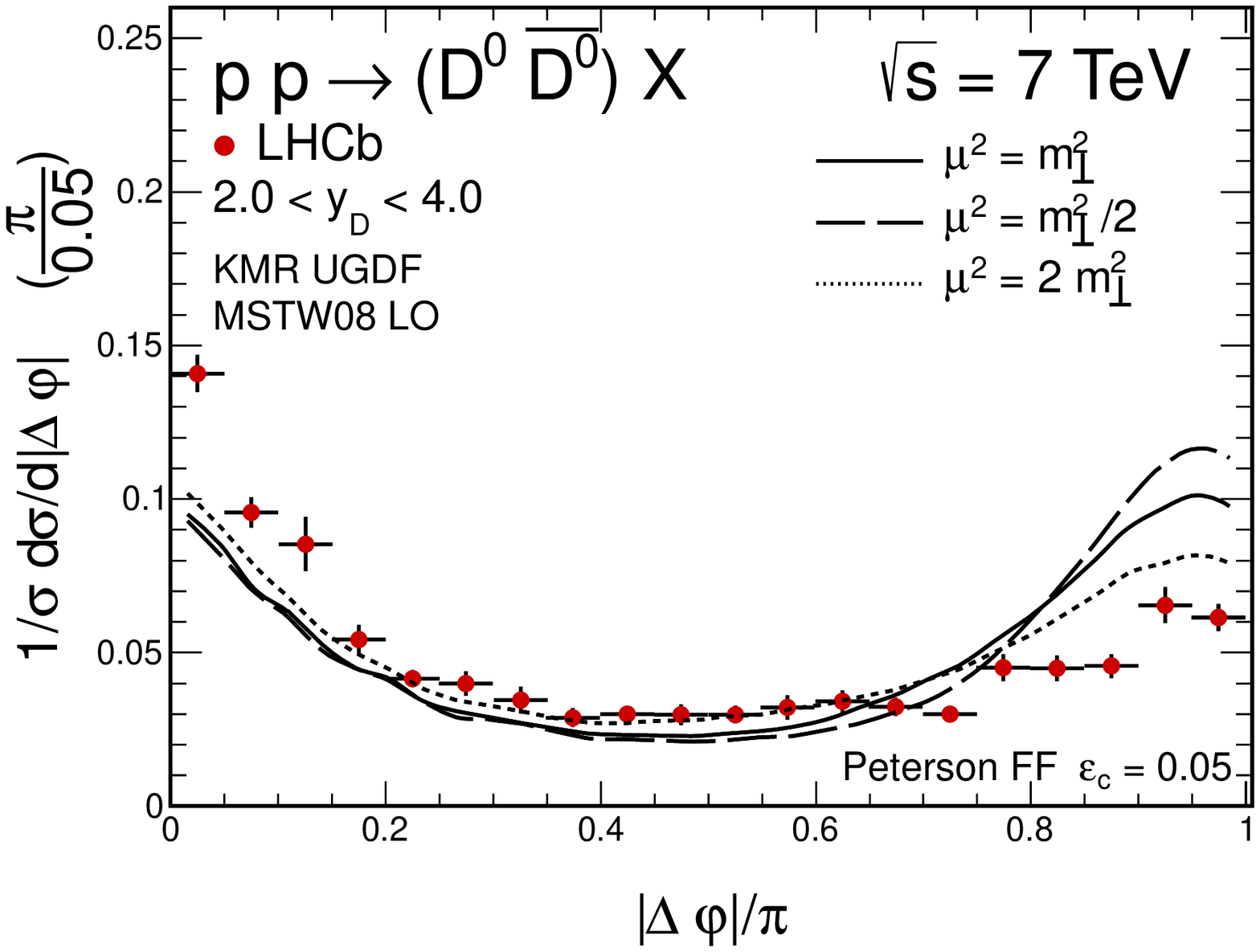}}
\end{minipage}
   \caption{
\small Distribution in relative azimuthal angle between $D^0$ and 
$\bar D^0$ for different UGDFs (left) and uncertainties due to 
the choice of the scale for the KMR UGDF (right).}
 \label{fig:phid-lhcb-DDbar-3}
\end{figure}

The LHCb detector has almost full coverage in azimuthal angle.
In Fig.~\ref{fig:phid-lhcb-DDbar-3} we discuss distribution in azimuthal
angle between the $D^0$ and $\bar D^0$ mesons $\varphi_{D^0 \bar D^0}$. Again the KMR and KMS
distributions give quite reasonable description of the shape of the
measured distribution. Both of them, give the enhancement of the cross section
at $\phi_{D \bar D} \sim$ 0. This is due to the fact that these
approaches include effectively gluon splitting contribution, not included
in the case of the Jung UGDFs. This was also discussed in Ref.~\cite{JKLZ2011} where additional calculations of the $g^{*}g^{*} \rightarrow g g \rightarrow g c \bar{c}$ subprocess
in the case of the Jung UGDFs were performed to describe azimuthal angle correlation between $D$ and $\bar D$ mesons measured at Tevatron. Some dependence of the shape on the choice
of the factorization/renormalization scale in the case of the KMR UGDF can be also observed (see the right panel).
However, still even with the KMR UGDF, one can observe some small missing strenght at small angles. It may suggest that within the KMR model the gluon splitting contribution is not fully
included.

\begin{figure}[!h]
\begin{minipage}{0.3\textwidth}
 \centerline{\includegraphics[width=1.0\textwidth]{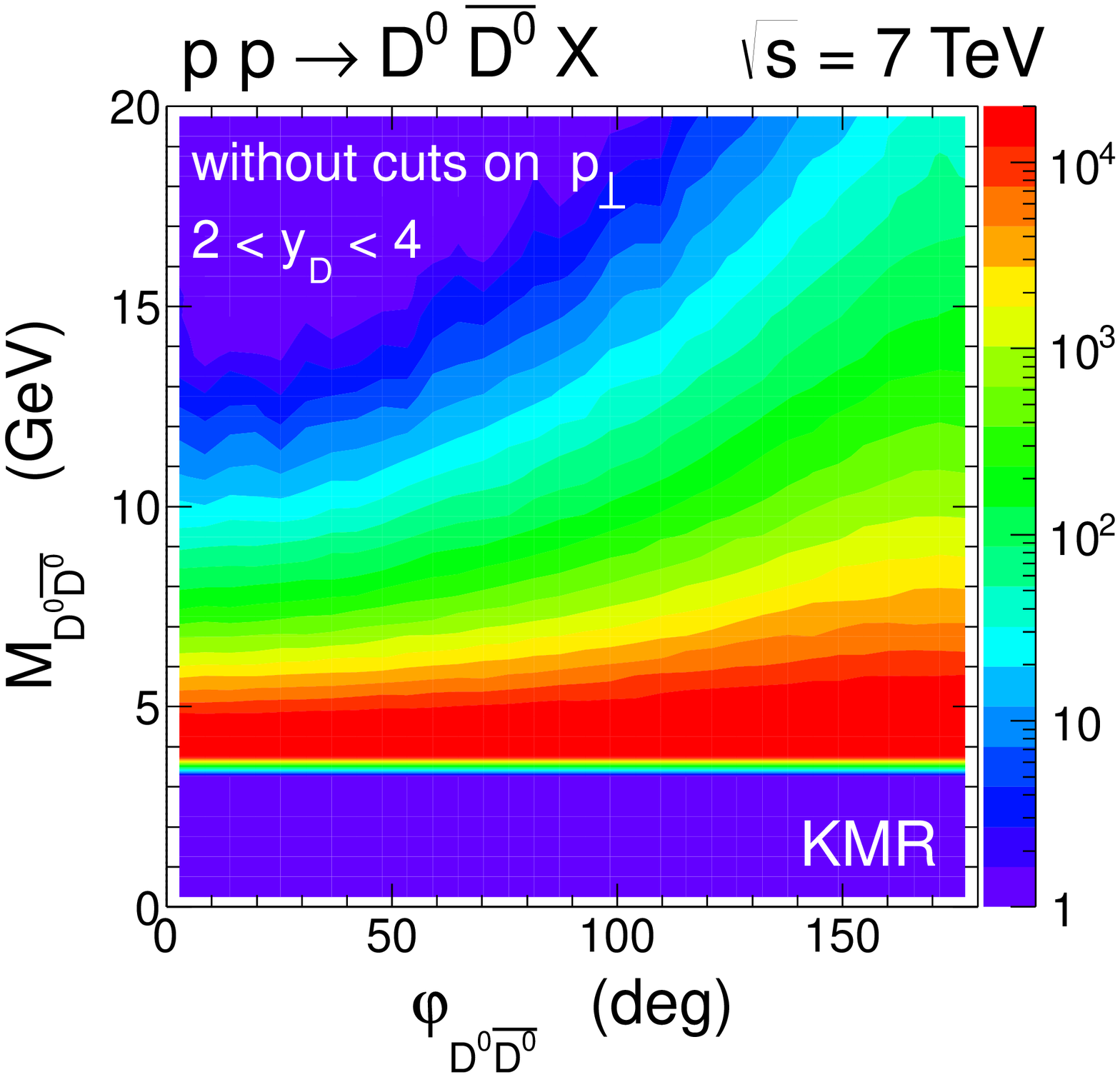}}
\end{minipage}
\hspace{0.5cm}
\begin{minipage}{0.3\textwidth}
 \centerline{\includegraphics[width=1.0\textwidth]{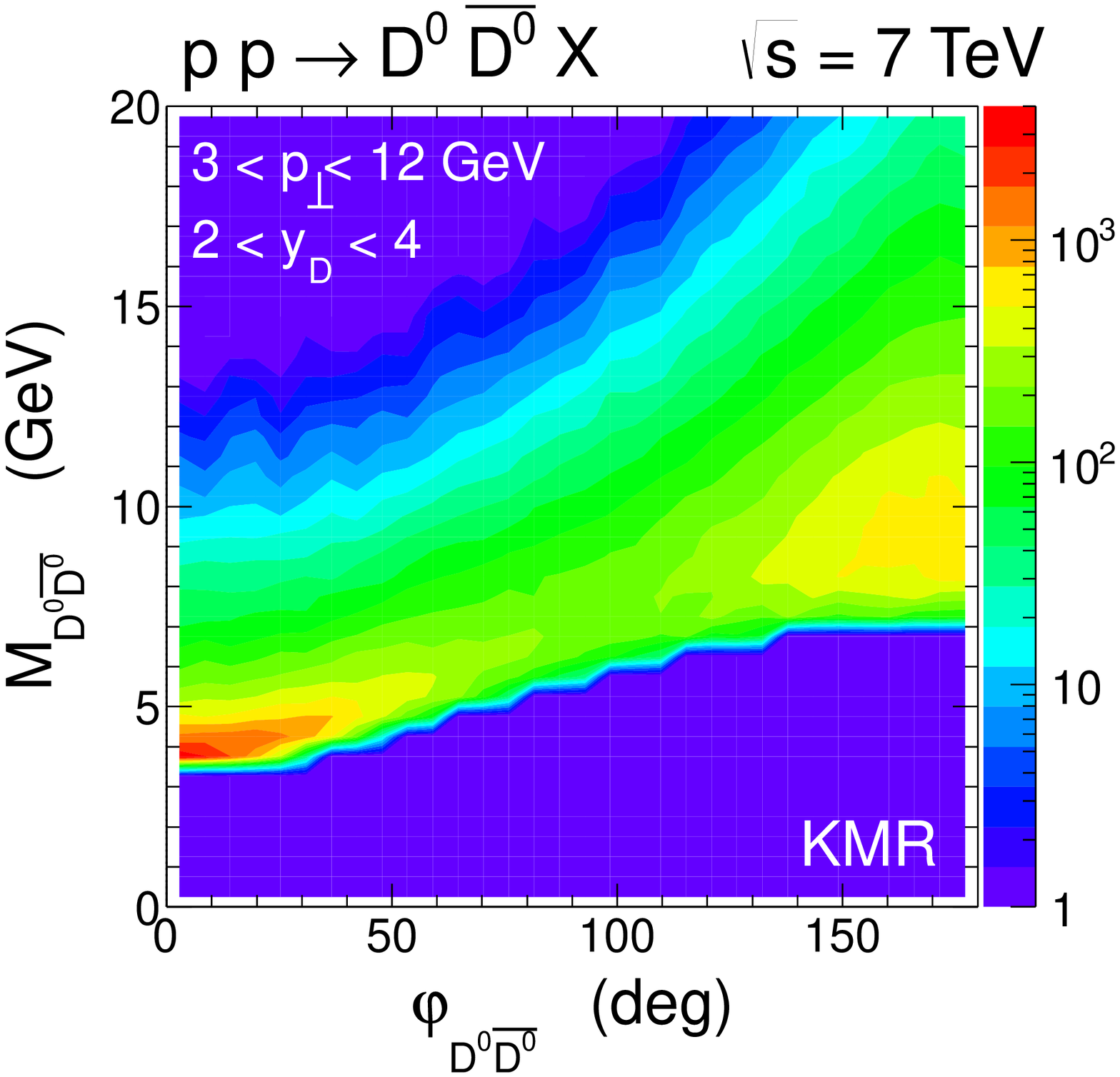}}
\end{minipage}
\hspace{0.5cm}
\begin{minipage}{0.3\textwidth}
 \centerline{\includegraphics[width=1.0\textwidth]{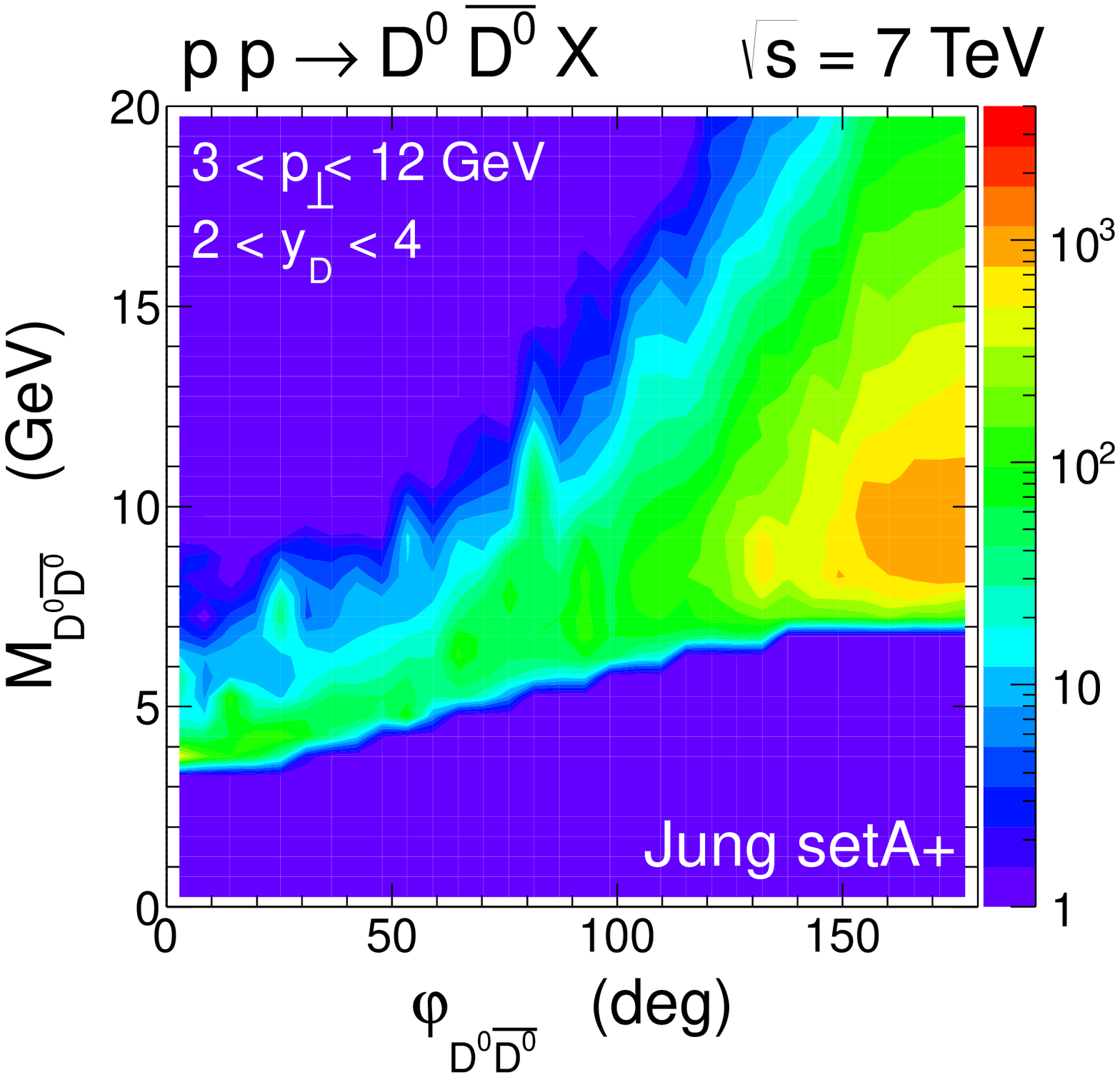}}
\end{minipage}
   \caption{
\small Two-dimensional distribution in $D \bar D$ invariant mass and
relative azimuthal angle between $D$ and $\bar D$ for the KMR and Jung setA+
UGDFs.}
 \label{fig:Minv-phid-lhcb-DDbar-1}
\end{figure}

In order to better understand the result for the azimuthal correlations
in Fig.~\ref{fig:Minv-phid-lhcb-DDbar-1} we show two-dimensional
distributions in invariant mass $M_{D^0 \bar D^0}$ and azimuthal angle
between mesons $\varphi_{D^0 \bar D^0}$. The maximum obtained for the KMR UGDF for small relative 
azimuthal angle between $D$ and $\bar{D}$ mesons corresponds to small invariant masses of the $D^0 \bar D^0$
system. This strongly supports the interpretation of the effect as the gluon
splitting into $c \bar c$ pair. However, one can also see that these interesting shapes
of the correlation observable are the consequence of the specific LHCb kinematical cuts.
 
\section{Summary}

First we have discussed the general situation with the $c\bar{c}$ production at LHC energies.
We have argued that the $c \bar c$ production is related with small-$x$ physics. Therefore it has a good
potential to test different models of unintegrated gluon distributions.

In the present paper we have focused on production of $D$ mesons at the LHC
within the $k_{t}$-factorization formalism with unintegrated gluon distributions. Only 
the Kimber-Martin-Ryskin unintegrated gluon distribution gives transverse momentum
distributions of charmed mesons similar to the recently measured ones by the ATLAS, ALICE 
and LHCb collaborations. Our inclusive theoretical distributions with
the KMR UGDFs are very similar to those obtained within FONLL or MC@NLO 
approaches. All other unintegrated gluon distributions strongly
underpredict the experimental results. This may suggest that some
mechanism of charm production is still missing. In a following paper 
we shall discuss double parton scattering effects as a potential missing
mechanism \cite{MS2013}.

Recently the LHCb collaboration has presented first results for $D$ and 
$\bar D$ meson two-particle distributions.
We have presented first theoretical results for such observables. 
Our model calculation with the KMR UGDF relatively well describes both $D \bar D$
meson invariant mass distributions as well as $D \bar D$ correlations 
in relative azimuthal angle between meson and antimeson. This shows that 
the $k_t$-factorization approach is very efficient in describing the 
two-particle distributions. In contrast the NLO QCD approach can be used
only in a limited region of the phase space but no real results have been
presented so far in the literature.

\vspace{1cm}

{\bf Acknowledgments}

We are indebted to Vanya Belyaev and Marek Szczekowski for the discussion
concerning recent results at LHC. This study was partially supported by 
the Polish Grants DEC-2011/01/B/ST2/04535 and N202 237040.


\end{document}